\documentclass[openany]{nature}
\usepackage[T1]{fontenc}
\usepackage[left]{lineno}
\usepackage{amsthm,amsmath,amssymb}
\usepackage{mathrsfs}
\usepackage{overpic}
\usepackage{multirow}
\usepackage[T1]{fontenc}
\usepackage{graphicx}
\usepackage{pdflscape}
\usepackage{hyperref}
\usepackage{threeparttable}
\usepackage{xcolor}
\usepackage{ulem}
\usepackage{cancel}
\usepackage[figuresleft]{rotating}
\usepackage{lineno}
\usepackage{caption}
\usepackage{subfigure}
\usepackage{tablefootnote}
\usepackage{amsmath, amssymb}
\usepackage{hyperref}
\usepackage{graphicx}
\usepackage{longtable}
\usepackage{supertabular}
\usepackage{float} 
\usepackage{authblk}
\usepackage{multibib}
\usepackage{subfiles}

\makeatletter
 
\def\emph#1 {\textit{ #1 } }

\let\saved@includegraphics\includegraphics
\AtBeginDocument{\let\includegraphics\saved@includegraphics}
\renewenvironment*{figure}{\@float{figure}}{\end@float}
\makeatother

\newcommand{\apj}{Astrophys. J.}

\newcommand{\pasp}{Publ. Astron. Soc. Pac.}
\newcommand{\apjs}{Astrophys. J. Supp.}
\newcommand{\araa}{Annu. Rev. Astron. Astrophys.}
\newcommand{\mnras}{Mon. Not. R. Astron. Soc.}
\newcommand{\apjl}{Astrophys. J. Let.}
\newcommand{\aap}{Astron. Astrophys.}
\newcommand{\aj}{Astron. J.}
\newcommand{\nat}{Nature}

\newcommand{\prl}{Phys. Rev. Lett.}
\newcommand{\prd}{Phys. Rev. D}

\newcommand{\ssr}{Space. Sci. Reviews.}
\newcommand{\actaa}{Acta Astron.}
\newcommand{\jcap}{Journal of Cosmology and Astroparticle Physics}

\def\be{\begin{eqnarray}}
\def\ee{\end{eqnarray}}

\makeatletter

\def\@fnsymbol#1{\ensuremath{\ifcase#1\or \dagger\or \ddagger\or
 \mathsection\or \mathparagraph\or \|\or **\or \dagger\dagger
 \or \ddagger\ddagger \else\@ctrerr\fi}}
\makeatother

\definecolor{dkblue}{RGB}{54, 86, 169}

\title{A long-duration gamma-ray burst with a peculiar origin}

\author{
Jun Yang$^{1,2}$, 
Shunke Ai$^{3,4}$,
Bin-Bin Zhang$^{1,2}$\thanks{E-mail: bbzhang@nju.edu.cn},
Bing Zhang$^{3, 4}$\thanks{E-mail: bing.zhang@unlv.edu},
Zi-Ke Liu$^{1,2}$,
Xiangyu Ivy Wang$^{1,2}$,
Yu-Han Yang$^{1,2}$,
Yi-Han Yin$^{5}$,
Ye Li$^{6}$, 
Hou-Jun L\"{u}$^{7}$}

\begin{document}

\maketitle

\begin{affiliations}
 \item School of Astronomy and Space Science, Nanjing University, Nanjing 210093, China
 \item Key Laboratory of Modern Astronomy and Astrophysics (Nanjing University), Ministry of Education, China
 \item Nevada Center for Astrophysics, University of Nevada Las Vegas, NV 89154, USA
 \item Department of Physics and Astronomy, University of Nevada Las Vegas, NV 89154, USA
 \item School of Physics, Nanjing University, Nanjing 210093, China
 \item Purple Mountain Observatory, Chinese Academy of Sciences, Nanjing 210023, China
 \item Guangxi Key Laboratory for Relativistic Astrophysics, School of Physical Science and Technology, Guangxi University, Nanning 530004, China
\end{affiliations}
\bigskip

\begin{abstract}

It is generally believed that long-duration gamma-ray bursts (GRBs) are associated with massive star core-collapse\cite{Woosley_2006ARAA}, whereas short-duration GRBs are associated with mergers of compact star binaries\cite{Berger_2014ARAA}. However, growing observations\cite{Gehrels_2006Natur, Della_2006Natur, Zhang_2021NatAs, Ahumada_2021NatAs} have suggested that oddball GRBs do exist, and multiple criteria (prompt emission properties, supernova/kilonova associations, and host galaxy properties) rather than burst duration only are needed to classify GRBs physically\cite{Zhang_2009ApJ}. A previously reported long-duration burst, GRB 060614\cite{Gehrels_2006Natur}, could be viewed as a short GRB with extended emission if it were observed at a larger distance\cite{Zhang_2007ApJ} and was associated with a kilonova-like feature\cite{Yang_2015NatCo}. As a result, it belongs to the Type-I (compact star merger) GRB category and is likely of the binary neutron star merger origin. Here we report a peculiar long-duration gamma-ray burst, GRB 211211A, whose prompt emission properties in many aspects differ from all known Type-I GRBs, yet its multi-band observations suggest a non-massive-star origin. In particular, significant excess emission in both optical and near-infrared wavelengths has been discovered (see also Ref.\cite{Rastinejad_2022arXiv}), which resembles kilonova emission as observed in some Type-I GRBs. These observations point towards a new progenitor type of GRBs. A scenario invoking a white dwarf-neutron star merger with a post-merger magnetar engine provides a self-consistent interpretation for all the observations, including prompt gamma-rays, early X-ray afterglow, as well as the engine-fed \cite{Yu_2013ApJ, Ai_2022arXiv} kilonova emission.

\end{abstract}

\begin{figure*}
\begin{center}
\includegraphics[width=110 mm]{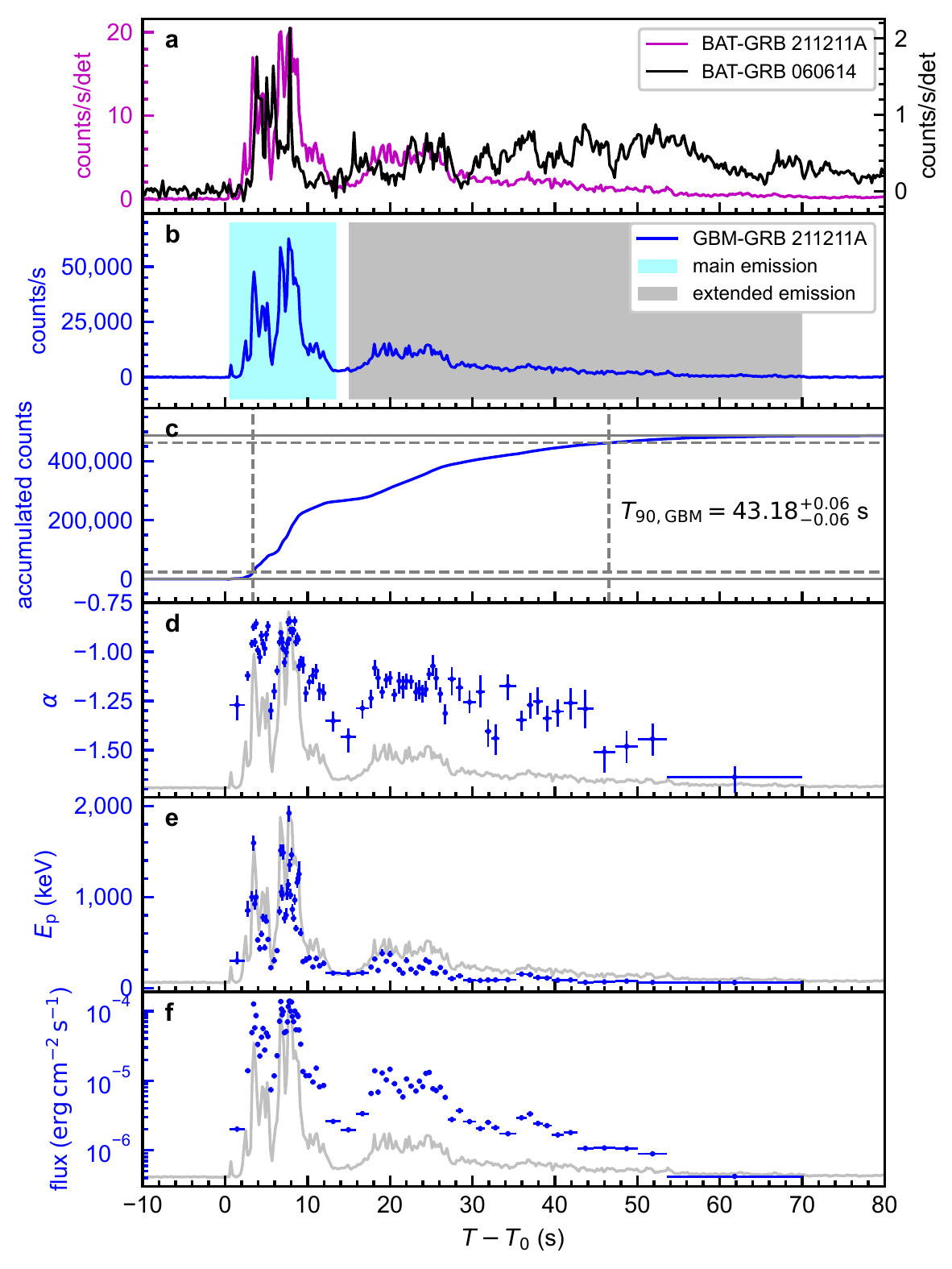}
\end{center}
\caption{\noindent\textbf{The temporal and spectral behaviors of GRB 211211A.} \textbf{a}, The net light curves of GRB 211211A (magenta line) and GRB 060614 (black line) obtained from {\it Swift}/BAT data. For comparison, we reduce the trigger time of GRB 060614 by 5 seconds. \textbf{b}, The net light curve (blue line) of GRB 211211A obtained from {\it Fermi}/GBM data. The cyan and grey shaded areas represent main emission (ME) and extended emission (EE) phases, respectively. \textbf{c}: The accumulated counts (blue line) of the {\it Fermi}/GBM net light curve. The grey horizontal dashed (solid) lines are drawn at $5\%$ ($0\%$) and $95\%$ ($100\%$) of the total accumulated counts. The $T_{\rm90,GBM}$ interval is marked by the grey vertical dashed lines. \textbf{d}, \textbf{e} and \textbf{f} show the evolution of spectral index $\alpha$, peak energy $E_{\rm p}$ and energy flux. The {\it Fermi}/GBM net light curve (grey line) is plotted in the background as a reference. All error bars represent the 1$\sigma$ confidence level.}
\label{fig:anares}
\end{figure*}

GRB 211211A triggered the Gamma-ray Burst Monitor (GBM \cite{Meegan_2009ApJ}) onboard {\it The Fermi Gamma-Ray Space Telescope}, the Burst Alert Telescope (BAT \cite{Barthelmy_2005SSRv}) onboard {\it The Neil Gehrels Swift Observatory}, and the high energy X-ray Telescope onboard {\it Insight}-HXMT \cite{Xiao2022arXiv} at 13:09:59 Coordinated Universal Time on 11 December 2021 (hereafter $T_{\rm 0}$). The burst is characterized by a spiky main emission (ME) phase lasting $\sim 13$ seconds and a longer, weaker extended emission (EE) phase lasting $\sim$ 55 seconds (Figure \ref{fig:anares} and Table \ref{tab:summary}). Our detailed spectral analysis (Methods and Table \ref{tab:summary}) shows that the time-integrated spectrum of the ME phase can be fitted by a Band \cite{Band_1993ApJ} model with spectral indices $\alpha=-0.996_{-0.005}^{+0.005}$, $\beta=-2.36_{-0.02}^{+0.02}$, and peak energy $E_{\rm p}=687.1_{-11.0}^{+12.5}$, and that of the EE phase can be fitted by a Band model with spectral indices $\alpha=-0.97_{-0.04}^{+0.03}$, $\beta=-2.02_{-0.02}^{+0.01}$, and peak energy $E_{\rm p}=82.0_{-2.3}^{+3.8}$. Follow-up observations revealed a candidate host galaxy at redshift $z=0.076$ (Ref.\cite{Rastinejad_2022arXiv}).

\begin{figure}
\centering
\includegraphics[width=160 mm]{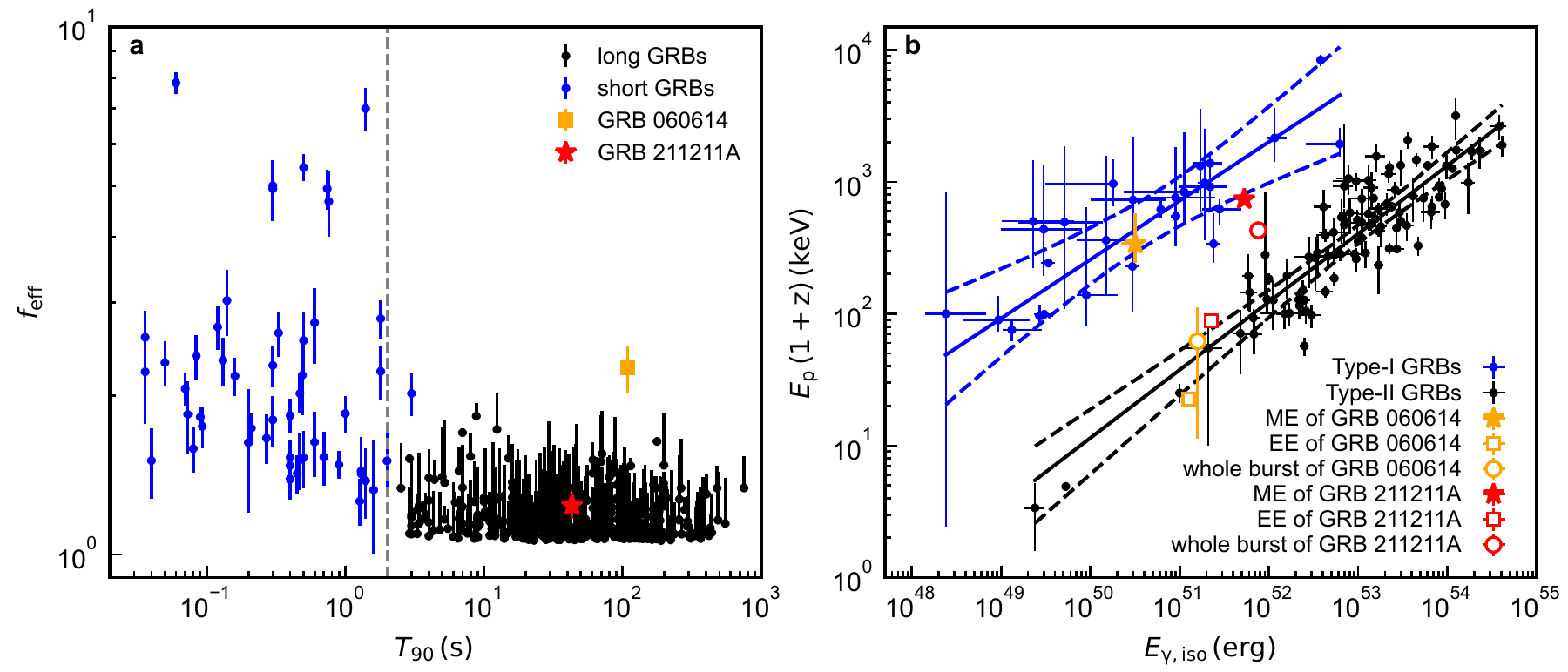}
\caption{\noindent\textbf{GRB 211211A in already classified clusters.} \textbf{a}, The $f_{\rm eff}$ and $T_{\rm 90}$ diagram. Long and short GRBs are presented by black and blue circles, respectively. The grey vertical line is the division line at 2 s. GRB 211211A and GRB 060614 are highlighted by the red star and orange square, respectively. \textbf{b}, The $E_{\rm p,z}$ (or $E_{\rm p}(1+z)$) and $E_{\rm\gamma,iso}$ correlation diagram. The best-fit correlations (solid lines) and corresponding 3$\sigma$ confidence bands (dashed lines) are presented for Type-I (blue circles) and Type-II (black circles) GRB populations, respectively. The main emission (ME), extended emission (EE), and the whole burst for GRB 211211A (red) and GRB 060614 (orange) are highlighted by stars, squares, and circles, respectively. All error bars on data points represent their 1$\sigma$ confidence level.}
\label{fig:pvp}
\end{figure}

\begin{figure}
\centering
\includegraphics[width=160 mm]{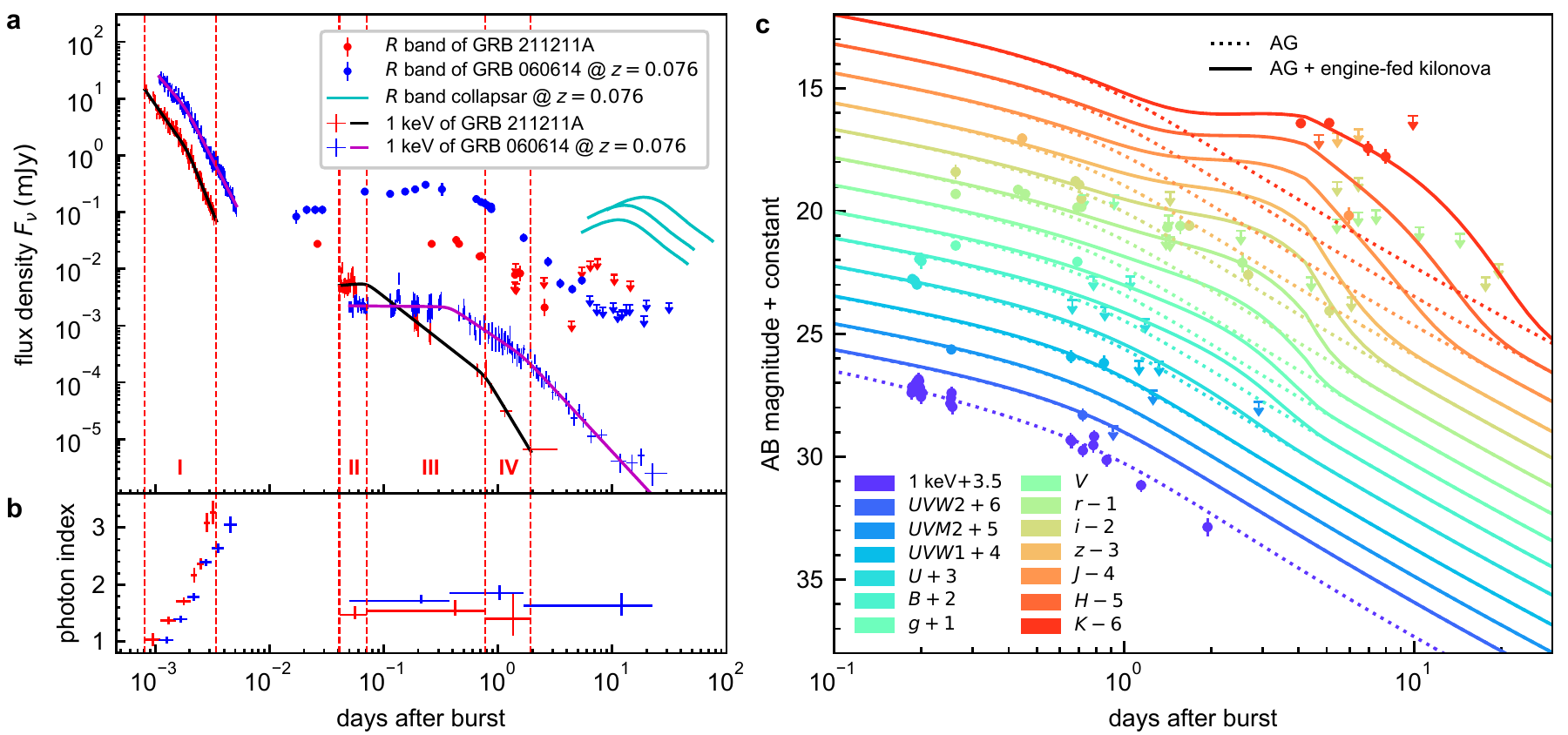}
\caption{\noindent\textbf{The multi-wavelength observations and fitting models.} \textbf{a}, The X-ray (1 keV) and $R$ band afterglows, and the $R$ band collapsars. The observed (fitted with the smoothed broken-power-law model) X-ray light curves for GRB 211211A and GRB 060614 are shown with red pluses (black lines) and blue pluses (magenta lines), respectively. The $R$ band observations for GRB 211211A and GRB 060614 are shown with red and blue solid circles (downward arrows for upper limits), respectively. The cyan lines represent the $R$ band collapsars, including SN 1998bw/GRB 980425, SN 2006aj/GRB 060218A, SN 2010bh/GRB 100316D. The vertical red dashed lines are drawn at the edges of each phase in the X-ray light curve of GRB 211211A. \textbf{b}, The evolution of X-ray photon index of GRB 211211A (red pluses) and GRB 060614 (blue pluses). \textbf{c}, The fittings to multi-wavelength observations. The detections and upper limits of the multi-wavelength data of GRB 211211A are shown with solid circles and downward arrows, respectively. The best-fit afterglow (AG) and AG plus engine-fed kilonova models are shown with dotted and solid lines, respectively.}
\label{fig:agmnmodeling}
\end{figure}

Despite its long duration, GRB 211211A does not fit into the Type-II (massive star core-collapse) GRB category, to which most long-duration GRBs belong. Instead, multiple pieces of observational evidence place GRB 211211A squarely into the Type-I GRB category. First, GRB 211211A shows tiny spectral lags between the multi-band light curves (Methods and Table \ref{tab:summary}). Such a characteristic is typically observed in Type-I GRBs. Second, the offset between the GRB location and its host galaxy is calculated to be 8.19 kpc, corresponding to a normalized offset value of 9.02 (Methods and Table \ref{tab:summary}). Such a large offset would be very peculiar if GRB 211211A were a Type-II GRB, but it is in agreement with the prediction of a compact object merger origin of the burst, since a supernova associated with the birth of a neutron star member in the binary system likely ``kicks'' the system away from the star-forming region in the host galaxy. Third, the ME phase of GRB 211211A follows the Type-I (rather than Type-II) GRB track in the Amati relation diagram \cite{Amati_2002A&A} (Methods and Figure \ref{fig:pvp}b), even though the EE phase follows the track of Type-II GRBs. Fourth, there is an apparent absence of a supernova (Methods and Figure \ref{fig:agmnmodeling}a) in the long-term optical data, which rules out the possibility of GRB 211211A being a typical Type-II GRB caused by the core-collapse of a massive star. Lastly, the standard external shock afterglow model, while fitting the overall trend of the multi-wavelength afterglow (especially in X-rays), significantly under-fits the optical and near-infrared band data around 1--10 days (Methods and Figure \ref{fig:agmnmodeling}c). Such kind of excess has long been considered as evidence of the existence of an underlying kilonova emission, typically produced from neutron star mergers \cite{Li_1998ApJ, Metzger_2010MNRAS}. Indeed, Ref.\cite{Rastinejad_2022arXiv} has claimed the detection of a kilonova based on the multi-wavelength excess data, and our analysis independently confirms such an association (Methods and Figure \ref{fig:agmnmodeling}c).

Such a GRB has never been observed before. The closest analogy was GRB 060614, a previously-discovered long-duration Type-I burst\cite{Gehrels_2006Natur}. GRB 211211A, however, distinguishes itself by its significantly longer ME duration and much harder EE. This is evident from the following observational facts: (1) As listed in Table \ref{tab:summary}, GRB 211211A exhibits a twice longer ME phase and a half shorter EE phase in comparison with GRB 060614 (Figure \ref{fig:anares}a). This makes GRB 211211A very different from the short main burst + extended emission pattern as observed in some Type-I GRBs; (2) Most Type-I GRBs have a relatively large ``effective amplitude parameter'' $f_{\rm eff}$, which suggests that the short duration is not due to a ``tip-of-iceberg'' effect\cite{Lu_2014MNRAS}. GRB 211211A, on the other hand, has a much smaller $f_{\rm eff}$ (Table \ref{tab:summary}), making it fall into the distribution of typical long-duration GRBs in the $f_{\rm eff}-T_{\rm 90}$ diagram (Methods and Figure \ref{fig:pvp}a), confirming its genuinely long-duration nature. In contrast, GRB 060614 had a relatively large $f_{\rm eff}$ (Table \ref{tab:summary}), making it more resemble short-duration Type-I GRB \cite{Zhang_2007ApJ}; (3) Comparatively speaking, the spectra of GRB 211211A in both the ME and EE phases are much harder than those of GRB 060614 (Table \ref{tab:summary}). The fact that the ME of GRB 211211A is both longer and harder than that of GRB 060614 rules out any possibility that the two events are intrinsically similar but appear differently because of jet structure or viewing angle effects. To summarize, GRB 211211A does not belong to an extreme version of short GRBs with extended emission (as GRB 060614 does). It pushes the envelope of Type-I GRBs to the genuinely long-duration regime.

The unique properties of GRB 211211A call for a new progenitor system that was not invoked to interpret standard short or long GRBs. First, the lack of a supernova explosion signature directly rules out the massive star core-collapse interpretation. Second, whereas the $\sim 6$-s ME duration for GRB 060614 may still allow it to belong to the short GRB category based on the $T_{90}$ distribution, the $\sim 13$-s ME duration of GRB 211211A rejects it from the short GRB category beyond the $3\sigma$ level (Methods). For an accretion-powered engine, the duration of a burst should scale with the density of the progenitor star\cite{zhang_2018book}. The longer-than-short ME duration and the lack of a supernova signature suggest that the progenitor star should have a lower density than a neutron star (NS, which is typically used to interpret short GRBs) but cannot be a massive star. A white dwarf (WD) star naturally falls into such a category. Because the GRB engine needs to be either an NS or a black hole (BH) to reproduce the short variability timescale observed\cite{zhang_2018book} (Methods and Table \ref{tab:summary}), the WD needs to evolve to a more compact engine, so the most likely progenitor system would be a WD-NS merger system. Indeed, all other compact star merger scenarios suffer difficulties in interpreting this event (Methods).

Observations also suggest that the merger product is not a BH but rather a rapidly spinning magnetar. First, the 55-s EE with distinct emission properties from the ME (Methods and Table \ref{tab:summary}) requires that the engine is not an accreting black hole but is rather a proto-magnetar (Methods). After the initial accretion phase to power the ME, the proto-magnetar goes through a differential rotation phase during which magnetic bubbles are launched to power the EE\cite{Kluzniak_1998ApJ, Ruderman_2000ApJ, Dai_2006Sci} (Methods). A magnetar with an initial differential kinetic energy of $\sim10^{51}$ erg is capable of producing tens of similar eruptions, with the interval between two adjacent bubble eruptions being $\tau_{\rm b} \simeq (1.6 ~ {\rm s}) (B_{\rm r}/10^{13} ~ {\rm G})^{-1} (P_0/{\rm ms})$, where $B_{\rm r}$ and $P_0$ are the internal radial magnetic field and the initial spin period of the proto-magnetar \cite{Kluzniak_1998ApJ}. Second, the early X-ray afterglow has a plateau lasting up to $\sim$ 6,000 s (Methods), consistent with the existence of a long-lasting central engine. Assuming isotropic emission and $P_0 \sim 1$ ms, we obtain the surface dipolar magnetic field at the pole being $B_{\rm p} \approx 3\times10^{14}$ G (Methods). Third, the kilonova light curve can be well modeled by an engine-fed kilonova \cite{Yu_2013ApJ, Ai_2022arXiv} with the following parameters: $P_0 = 1$ ms, $B_{\rm p}=1.9_{-0.4}^{+1.0}\times10^{13}$ G, radioactive mass ratio $f_{\rm n}=0.80_{-0.24}^{+0.03}$, opacity $\kappa=0.73_{-0.06}^{+0.53}$ ${\rm cm^2\,g^{-1}}$, total ejecta mass $M_{\rm ej}=0.037_{-0.004}^{+0.008}$ ${\rm M_{\odot}}$, and initial ejecta velocity $v_{\rm ej}=0.24_{-0.02}^{+0.06}\,{\rm c}$ (Methods). The smaller fitted $B_{\rm p}$ compared with the X-ray constraint suggests that the X-ray emission is collimated with a beaming factor of 0.004, corresponding to an opening angle of $\sim 5$ degrees. One can see that a WD-NS merger progenitor with a millisecond post-merger magnetar engine provides a self-consistent interpretation for all the data.

Within such a picture, the ``kilonova'' differs from the traditional picture that only invokes r-process and $\beta$-decay nuclear heating. The ejected materials include a neutron-poor component and a neutron-rich component so that only a fraction of the ejecta mass contributes to nuclear heating. In addition, energy injection from the magnetar central engine also provides additional heating to the ejecta \cite{Yu_2013ApJ, Ai_2022arXiv}. Our engine-fed kilonova model fitting suggests that about $0.037$ ${\rm M_{\odot}}$ total mass is ejected, with a fraction of $\sim 0.8$ being neutron-rich materials partially powering the kilonova emission. The rarity of the GRB 211211A and the fact that WD-NS mergers are more common than NS-NS mergers suggest that the majority of WD-NS mergers must not produce GRBs. Indeed, only a very small fraction of WD-NS mergers with a mass ratio close to unity could produce the necessary conditions for a magnetar-powered GRB (Methods).

The observations of GRB 211211A suggest that Type-I GRBs can have diverse sub-categories with different progenitor types. WD-NS mergers could be another type of progenitor to power Type-I GRBs. The gravitational-wave (GW) signals from these mergers have a typical frequency of around 0.1 Hz\cite{Toonen_2018AA}, which is below the sensitivity of ground-based GW detectors LIGO/Virgo/KAGRA\cite{Buikema_2020PhRvD, Bersanetti_2021Univ, Kagra_2019NatAs}. They can, however, potentially be detected by future space missions such as LISA\cite{Amaro_2017arXiv}, Taiji\cite{Luo_2020ResPh}, and Tianqin\cite{Luo_2016CQGra}.

\bigskip
\bigskip

\clearpage

\begin{table*}
\centering
\renewcommand\arraystretch{0.8}
\caption{\textbf{Summary of the observed properties of GRB 211211A in comparison with GRB 060614\cite{GCN5264, Della_2006Natur, Amati_2007AA, Yang_2015NatCo, Blanchard_2016ApJ, Lu_2014MNRAS}.}}
\label{tab:summary}
\begin{tabular}{lcc}
\hline
\hline
Observed Properties & GRB 211211A & GRB 060614 \\
\hline
\textbf{Main Emission:} & & \\
Duration ($\rm s$) & 13 & 6 \\ 
Averaged variability ($\rm ms$) & 16 & ... \\
Spectral lag ($\rm ms$) & $10_{-4}^{+3}$ & $3\pm6$ \\
Spectral index $\alpha$ & $-0.996_{-0.005}^{+0.005}$ & $-1.57_{-0.14}^{+0.12}$ \\
Spectral index $\beta$ & $-2.36_{-0.02}^{+0.02}$ & ... \\
Peak energy ($\rm keV$) & $687.1_{-11.0}^{+12.5}$ & $302_{-85}^{+214}$ \\
Energy fluence ($\rm erg\,cm^{-2}$) & $3.77_{-0.01}^{+0.01}\times10^{-4}$ & $8.19_{-2.52}^{+0.56}\times10^{-6}$ \\
Isotropic energy ($\rm erg$) & $5.30_{-0.01}^{+0.01}\times10^{51}$ & $3.18_{-0.98}^{+0.22}\times10^{50}$ \\
\hline
\textbf{Extended Emission:} & & \\
Duration ($\rm s$) & 55 & 100 \\ 
Averaged variability ($\rm ms$) & 48 & ... \\
Spectral lag ($\rm ms$) & $5_{-5}^{+5}$ & $3\pm9$ \\
Spectral index $\alpha$ & $-0.97_{-0.04}^{+0.03}$ & $-2.13\pm{0.05}$ \\
Spectral index $\beta$ & $-2.02_{-0.02}^{+0.01}$ & ... \\
Peak energy ($\rm keV$) & $82.0_{-2.3}^{+3.8}$ & $\lesssim20$ \\
Peak luminosity ($\rm erg\,s^{-1}$) & $2.05_{-0.06}^{+0.06}\times10^{50}$ & ... \\
Isotropic energy ($\rm erg$) & $2.26_{-0.01}^{+0.01}\times10^{51}$ & $1.27_{-0.09}^{+0.06}\times10^{51}$ \\
\hline
\textbf{Whole Burst:} & & \\
$T_{90}$ ($\rm s$) & $43.18_{-0.06}^{+0.06}$ & $102\pm5$ \\
Spectral lag ($\rm ms$) & $12\pm10$ & ... \\
$f_{\rm eff}$ parameter & $1.24\pm0.07$ & $2.26\pm0.23$ \\
Spectral index $\alpha$ & $-1.20_{-0.01}^{+0.01}$ & ... \\
Spectral index $\beta$ & $-2.05_{-0.02}^{+0.02}$ & ... \\
Peak energy ($\rm keV$) & $399.3_{-16.1}^{+14.0}$ & $10-100$ \\
Peak flux ($\rm erg\,cm^{-2}\,s^{-1}$) & $1.38_{-0.02}^{+0.02}\times10^{-4}$ & $4.50_{-1.53}^{+0.72}\times10^{-6}$ \\
Total fluence ($\rm erg\,cm^{-2}$) & $(5.42\pm0.08)\times10^{-4}$ & $4.09_{-0.34}^{+0.18}\times10^{-5}$ \\
Peak luminosity ($\rm erg\,s^{-1}$) & $1.94_{-0.03}^{+0.03}\times10^{51}$ & $1.74_{-0.59}^{+0.28}\times10^{50}$ \\
Isotropic energy ($\rm erg$) & $(7.61\pm0.11)\times10^{51}$ & $1.59_{-0.13}^{+0.07}\times10^{51}$ \\
\hline
\textbf{Host Galaxy:} & & \\
Redshift & 0.076 & 0.125 \\
Half-light radius ($\rm kpc$) & 0.91 & 0.87    \\
Offset ($\rm kpc$) & 8.19 & 0.80\\
Normalized offset & 9.02 & 1.09 \\
\hline
\textbf{Associations:} & & \\
Kilonova & Yes & Yes \\ 
Supernova & No & No \\
\hline
\hline
\end{tabular}
\end{table*}

\clearpage


\clearpage

\clearpage

\section*{Methods}

\subsection{Prompt emission.\\}

We retrieved the time-tagged event dataset covering the time range of GRB 211211A from the {\it Fermi}/GBM public data archive (\url{https://heasarc.gsfc.nasa.gov/FTP/fermi/data/gbm/daily/}). Two sodium iodide (NaI) detectors, namely n2 and na, with the smallest viewing angles in respect to the GRB source direction, are selected for our analysis. Additionally, one bismuth germanium oxide (BGO) detector, b0, closest to the GRB direction, was also selected for spectral analysis. Since GRB 211211A is also detected by {\it Swift}/BAT, we employ those BAT data in our analysis. {\it Swift}/BAT data was downloaded from the {\it Swift} archive website (\url{https://www.swift.ac.uk/archive/ql.php}).

\textit{Light curves}: The BAT light curve is obtained following the standard analysis threads (\url{https://www.swift.ac.uk/analysis/bat/}) using the \emph{HEASoft} tools (version 6.28). We first check the energy conversion of BAT event data using \emph{bateconvert}. Next, \emph{batbinevt}, \emph{bathotpix}, and \emph{batmaskwtevt} are utilized to generate the detector plane image, remove the hot pixels and apply mask-weighting, respectively. Then the BAT light curve (Figure \ref{fig:anares}a) is created with a bin size of 0.2 s and with an energy range of 15--350 keV using \emph{batbinevt}. The $T_{\rm 90, BAT}$ of GRB 211211A is determined by \emph{battblocks} to be $51.00\pm1.02$ s. 

The GBM light curve (Figure \ref{fig:anares}b) is derived by binning the photons with a bin size of 0.2 s in the energy range of 10--1,000 keV collected from the selected two NaI detectors (namely, n2 and na). The background level is modeled by the baseline algorithm in the \emph{pybaseline} package (\url{https://github.com/derb12/pybaselines}). The burst duration, $T_{\rm 90, GBM}$, determined by the time difference between the epochs when the total accumulated net photon counts within 10--1,000 (50--300) keV reaches the $5\%$ level and the $95\%$ level, is calculated to be $43.18_{-0.06}^{+0.06}$ ($36.58_{-0.02}^{+0.02}$) s (Table \ref{tab:summary} and Figure \ref{fig:anares}c). 

The light curves in Figure \ref{fig:anares} exhibit two distinct phases, namely, main emission (ME) and extended emission (EE). The ME is a sharply changing phase (which lasts for approximately 13 s) that is characterized by several hard spikes. Following a relatively quiescent time of $\sim$2 s, the EE phase appears as a decaying tail extending to $T_{0}+70$ s. Such ME+EE feature was also observed in the light curve of GRB 060614 (Figure \ref{fig:anares}a). However, GRB 211211A exhibits a twice longer ME and a half shorter EE in comparison with GRB 060614, as listed in Table \ref{tab:summary}. We further explore their prompt light curves in a multi-band manner by extracting them separately within several energy bands from the event data (Extended Data Figure \ref{fig:wavelag}a). The temporal profiles of both bursts appear not to evolve with energy. Contrary to GRB 060614, there is significant emission above 350 keV during the ME phase of GRB 211211A, thanks to the broad energy coverage of {\it Fermi}/GBM detectors. We also note that the EE phase of GRB 060614 is dominated by photons softer than 50 keV.

\textit{$T_{90}$ distributions}: To quantify how likely GRB 211211A and GRB 060614 are to belong to the short GRB population, we place their ME phases in the duration distributions of the GRB sample from the fourth {\it Fermi}/GBM catalog \cite{von_2020ApJ}. In the first case, as shown in Extended Data Figure \ref{fig:t90pdf}a, we fit the $T_{\rm 90}$ distribution of the short GRB ($T_{\rm 90} < 2~{\rm s}$) sample with a single log-normal distribution. Here we denote the normal distribution as $N(\mu,~\sigma^2)$, where $\mu$ and $\sigma$ are the mean and standard deviation of the normal distribution, respectively, all in logarithmic space. Based on the fitted normal distribution of $N(-0.27,~0.36^2)$, the p-values corresponding to 6 s and 13 s are $1.73\times10^{-3}$ and $5.61\times10^{-5}$, respectively. This implies that GRB 060614 and GRB 211211A are rejected from being a short GRB with a confidence level of $2.92\sigma$ and $3.86\sigma$, respectively. In the second case, as shown in Extended Data Figure \ref{fig:t90pdf}b, we fit the $T_{\rm 90}$ distribution of the whole sample with a two-component log-normal mixture model. The two components, namely $N(-0.24,~0.40^2)$ and $N(1.40,~0.50^2)$, are responsible for short and long GRB populations, respectively. Based on the fitted model, one can evaluate the probabilities for GRB 060614 and GRB 211211A to belong to the short GRB class (denoted as $P_{\rm sGRB}$), which are $1.78\times10^{-2}$ and $7.68\times10^{-4}$, respectively. The confidence levels of rejecting GRB 060614 and GRB 211211A from the short GRB population are $2.57\sigma$ and $3.42\sigma$, respectively. In both cases, the probability of a GRB with $T_{90} = 13$ s (the case of ME of GRB 211211A) is almost two orders of magnitude smaller than a GRB with $T_{90} = 6$ s (the case of ME of GRB 060614), even though the difference in duration is only by a factor of two. And the $\sim 13$-s ME duration of GRB 211211A firmly rejects it from the short GRB category beyond the $3\sigma$ level.

\textit{Variability timescale}: To quantify the fluctuations of temporal profiles, we calculate the averaged variability time scales of the ME and EE phases. We apply the Bayesian block method \cite{Scargle_2013ApJ} to the un-binned event data of both ME and EE phases and take half of the averaged block size as the mean variability time scale \cite{Vianello_2018ApJ}. Our results yield $\tau_{\rm var}\sim 16$ ms for the ME phase and $\tau_{\rm var}\sim48$ ms for the EE phase (Table \ref{tab:summary}), suggesting a much smoother time profile in the latter phase. The ME phase exhibits a three times smaller averaged variability time scale than that of the EE phase, implying two distinct physical origins of radiation.

\textit{Amplitude parameter}: The amplitude parameter\cite{Lu_2014MNRAS} is defined as $f=F_{\rm p}/F_{\rm b}$, where $F_{\rm p}$ and $F_{\rm b}$ stand for the peak flux and background flux at the same epoch, respectively. In order to distinguish intrinsically short Type-I GRBs from disguised short-duration GRBs due to the tip-of-iceberg effect, an effective amplitude parameter, $f_{\rm eff}$, for long-duration GRBs, can be defined as $f_{\rm eff}=F_{\rm p}^{\prime}/F_{\rm b}$, which is the $f$ value of ``pseudo GRBs'' whose signal flux is re-scaled so that $T_{90}$ is shorter than 2 s. For short-duration GRBs, $f$ and $f_{\rm eff}$ are the same by definition. Following the calculation procedure presented in Ref.\cite{Lu_2014MNRAS}, we obtain the effective amplitude as $f_{\rm eff}=1.24\pm0.07$ for GRB 211211A. Such a small $f_{\rm eff}$ places GRB 211211A to be consistent with the long-duration GRBs in the $f_{\rm eff}-T_{\rm 90}$ diagram (Figure \ref{fig:pvp}a), confirming the genuinely long-duration feature of the burst. On the contrary, the value of $f_{\rm eff}$ of GRB 060614 is significantly higher than those of other long-duration GRBs (Figure \ref{fig:pvp}a), supporting that it is an intrinsically short-duration Type-I GRB \cite{Zhang_2007ApJ}.

\textit{Spectral fitting}: We perform detailed time-resolved and time-integrated spectral fitting using the GBM data from detectors n2, na, and b0. According to the criterion that the average number of net photons per energy channel is not less than 20 \cite{Zhang_2018NatAs}, the time interval from $T_{0}+0.5$ to $T_{0}+70$ s is divided into 85 time slices for time-resolved spectral fitting. We additionally bracket three time slices for time-integrated spectral fitting, which span the ME phase, EE phase, and whole burst, respectively. For each time slice, the source and background spectra are acquired by summing the total photons and background photons for each energy channel, where the background photon number is derived by applying the baseline algorithm to each energy channel. The detector response matrix is generated by the \emph{gbm\_drm\_gen} package \cite{Burgess_2018MNRAS, Berlato_2019ApJ}. Then we use a self-developed software package, \emph{MySpecFit}, to perform the spectral fitting. \emph{MySpecFit} employs the nested sampler \emph{Multinest} \cite{Feroz_2008MNRAS, Feroz_2009MNRAS, Buchner_2014A&A, Feroz_2019OJAp} as the fitting engine and PGSTAT \cite{Arnaud_1996ASPC} as the statistic to estimate uncertainties of the best-fit parameters. The cutoff power law (CPL) and band (Band \cite{Band_1993ApJ}) function are adopted to fit time-resolved spectra and time-integrated spectra, respectively. The CPL \& Band models can be expressed as
\begin{equation}
 N(E)=AE^{\alpha}{\rm exp}(-E/E_{\rm c})
\end{equation}
and 
\begin{equation}
 N(E)=\left\{
 \begin{array}{l}
 A(\frac{E}{100\,{\rm keV}})^{\alpha}{\rm exp}(-\frac{E}{E_{\rm c}}),\,E<(\alpha-\beta)E_{\rm c} \\
 A\big[\frac{(\alpha-\beta)E_{\rm c}}{100\,{\rm keV}}\big]^{\alpha-\beta}{\rm exp}(\beta-\alpha)(\frac{E}{100\,{\rm keV}})^{\beta}, E\geq(\alpha-\beta)E_{\rm c} \\
 \end{array}\right.
\end{equation}
respectively, where $\alpha$ and $\beta$ are low-energy and high-energy photon spectral indices. The peak energy $E_{\rm p}$ is related to the cut-off energy, $E_{\rm c}$, through $E_{\rm p}=(2+\alpha)E_{\rm c}$. 


Extended Data Table \ref{tab:specfit} lists the spectral fitting results as well as the corresponding energy fluxes between 10 and 1,000 keV. The time-integrated spectrum of the ME (EE) phase can be fitted by the Band model with spectral indices $\alpha=-0.996_{-0.005}^{+0.005}$ $(-0.97_{-0.04}^{+0.03})$, $\beta=-2.36_{-0.02}^{+0.02}$ $(-2.02_{-0.02}^{+0.01})$, and peak energy $E_{\rm p}=687.1_{-11.0}^{+12.5}$ $(82.0_{-2.3}^{+3.8})$ keV. Comparatively, the spectra in both phases are much harder than those in the ME and EE phases of GRB 060614 (Table \ref{tab:summary}). Thus, a soft extended emission cannot be attributed to the long-duration of GRB 211211A, as had been the case in GRB 060614. As shown in Extended Data Figure \ref{fig:alpha_Ep_flux}f, the time-resolved $E_{\rm p}$ distributes in a wide range between 60 keV and 1.9 MeV. The hardest time-resolved spectrum measured from $T_{0}+7.7$ to $T_{0}+7.8$ s corresponds to the 0.1-s peak flux of $1.38_{-0.02}^{+0.02}\times10^{-4}$ $\rm erg\,cm^{-2}\,s^{-1}$. Summing up all the fluence values derived from time-resolved spectral fitting, we further calculate the total fluence to be $(5.42\pm0.08)\times10^{-4}$ $\rm erg\,cm^{-2}$. 

\textit{Spectral evolution}: In Figure \ref{fig:anares}d-f, both $\alpha$ and $E_{\rm p}$ exhibit the same ``flux-tracking'' \cite{Lu_2012ApJ, Li_2019ApJ} evolution patterns simultaneously (``double-tracking'') across the entire burst. The correlations between any pair of $\alpha$, $E_{\rm p}$, and energy flux $F$ are fitted by using a linear model for both ME and EE phases, as well as the entire burst. We also fit the correlation between time-resolved rest-frame peak energy $E_{\rm p,z}=E_{\rm p}(1+z)$ and time-resolved isotropic energy $E_{\rm\gamma,iso}$. The fitting process is implemented by the python module \emph{emcee}\cite{Foreman_2013PASP}. The fitting results are listed in Extended Data Table \ref{tab:alpha_Ep_flux} and illustrated in Extended Data Figure \ref{fig:alpha_Ep_flux}. As shown in Extended Data Table \ref{tab:alpha_Ep_flux}, the slopes between the ME and EE phases are generally consistent, suggesting that the two phases may share similar radiation mechanisms. Such spectral evolution suggests that the synchrotron model is the most likely explanation of GRB 211211A due to the following reasons: (1). As listed in Extended Data Table \ref{tab:specfit} and plotted in Extended Data Figure \ref{fig:alpha_Ep_flux}e, the values of lower energy photon index, $\alpha$, are typically distributed at $\sim -1.2$ and never exceed the so-called line of death ($-2/3$ \cite{Preece_1998ApJ}) of synchrotron radiation. Such an $\alpha$ distribution is fully consistent with the synchrotron model but not consistent with the simplest photosphere model \cite{Meszaros_2000ApJ}, which predicts a thermal-like spectrum with a much harder $\alpha$ above the synchrotron death line. To reproduce the soft spectrum as observed, one has to introduce some contrived new ingredients within the photosphere model framework \cite{Deng_2014ApJ}; (2). The values of the slopes in $E_{\rm p}\propto F^{k}$ relation are $\sim 0.6$, which is consistent with the predicted relation of $E_{\rm p}\propto L^{1/2}$ by synchrotron models \cite{Zhang_2002ApJ}. Comparably, photosphere models predict $E_{\rm p}\propto L^{1/4}$ for $R_{\rm ph}<R_{\rm s}$, and $E_{\rm p}\propto L^{-5/12}$ for $R_{\rm ph}>R_{\rm s}$, where $R_{\rm ph}$ is photosphere radius and $R_{\rm s}$ is saturation radius \cite{Meszaros_2000ApJ}; (3). The $\alpha$ tracking behavior can be naturally interpreted as synchrotron radiation, as suggested by Ref.\cite{Uhm_2014NatPh}, and specifically, the one-zone Internal-Collision-induced MAgnetic Reconnection and Turbulence (ICMART) model \cite{Zhang_2011ApJ}.

\textit{Spectral lags}: Spectral lag refers to the systematic time delay of the soft-band light curve relative to the hard-band light curve. Type-II GRBs are often associated with significant spectral lags, whereas Type-I GRBs tend to have zero delays or even negative lags \cite{Yi_2006MNRAS, Bernardini_2015MNRAS}. We utilize the discrete cross-correlation function (DCCF \cite{Norris_2000ApJ, Ukwatta_2010ApJ, Zhang_2012ApJ}) to measure the cross-correlation between the light curves of the softest band and each harder band in Extended Data Figure \ref{fig:wavelag}a at different time delays. The DCCF values for different delays are fitted with a 3-order polynomial function to determine the spectral lag, which is defined as the time delay corresponding to the maximum cross-correlation. We calculate spectral lags separately for the ME phase ($T_{0}+[-0.5,\,14.0]$ s; Extended Data Figure \ref{fig:wavelag}b), EE phase ($T_{0}+[14.0,\,70.0]$ s; Extended Data Figure \ref{fig:wavelag}c), and whole burst ($T_{0}+[-0.5,\,70.0]$ s; Extended Data Figure \ref{fig:wavelag}d). Extended Data Figure \ref{fig:wavelag}b-d show that the spectral lags derived from GBM and BAT data are in good agreement with each other, although the former has smaller uncertainties. As a comparison, we obtain the lags of GRB 060614 from Ref.\cite{Gehrels_2006Natur} and also display them in Extended Data Figure \ref{fig:wavelag}b-c. We note that both GRBs exhibit tiny lags, a characteristic typically associated with Type-I GRBs \cite{Yi_2006MNRAS, Bernardini_2015MNRAS, Shao_2017ApJ}.

\textit{Amati relation}: In Figure \ref{fig:pvp}b, we re-plot the $E_{\rm p,z}-E_{\rm\gamma,iso}$ diagram (a.k.a. Amati relation \cite{Amati_2002A&A}) using the Type-I and Type-II GRBs with known redshifts \cite{Amati_2002A&A, Zhang_2009ApJ}, where $E_{\rm p,z}=E_{\rm p}(1+z)$ is the rest-frame peak energy and $E_{\rm\gamma,iso}$ is the isotropic energy of prompt emission. We employ the python module \emph{emcee} \cite{Foreman_2013PASP} to fit the model ${\rm log}E_{\rm p,z}=b + k{\rm log}E_{\rm\gamma,iso}$ to both GRB samples. The best-fitting parameters with 3$\sigma$ uncertainties are $k=0.45_{-0.06}^{+0.05}$ and $b=-19.85_{-2.48}^{+2.87}$ for Type-I GRBs, and $k=0.52_{-0.02}^{+0.03}$ and $b=-24.96_{-1.45}^{+1.30}$ for Type-II GRBs. The best-fit correlations and corresponding 3$\sigma$ confidence bands are presented in Figure \ref{fig:pvp}b. Given a redshift of 0.076, the isotropic energy and peak luminosity of prompt emission of the burst can be calculated as $E_{\rm\gamma,iso}=(7.61\pm0.11)\times10^{51}$ erg and $L_{\rm\gamma,p}=1.94_{-0.03}^{+0.03}\times10^{51}$ $\rm erg\,s^{-1}$, respectively. We over-plot GRB 211211A onto the $E_{\rm p,z}-E_{\rm\gamma,iso}$ diagram. GRBs of different physical origins typically follow different tracks in such a diagram. We find that the whole burst of GRB 211211A resides in the area between the Type-I and Type-II GRB tracks. It is interesting to note that, similar to GRB 060614, the ME and the EE of GRB 211211A are located differently. As can be seen from Figure \ref{fig:pvp}b, the ME phase of GRB 211211A follows the track of Type-I GRBs, while the EE phase follows the track of Type-II GRBs. We note that GRB 211211A systematically resides higher in the Amati relation and poses higher total energy and harder spectra in both the ME and EE phases, suggesting that, although the two bursts may share similar physical origins, GRB 211211A is likely a more intense event. Notice that all the above discussion is based on the assumption that the source is at $z=0.076$ (Ref. \cite{GCN31221, Rastinejad_2022arXiv}). This redshift was indirectly inferred. Even though the possibility that the source is at a much higher redshift is not ruled out, this redshift is consistent with an independent estimate based on the gamma-ray data \cite{GCN31230}. It also provides a self-consistent explanation for the galactic location, rapid optical fading, and red color of the source \cite{GCN31235}.

\subsection{Host galaxy.\\}

A candidate of the optical counterpart of GRB 211211A is reportedly located very close to a Sloan Digital Sky Survey galaxy (SDSS J140910.47+275320.8; \url{http://skyserver.sdss.org/dr12/en/tools/explore/summary.aspx}) with an offset of $R_{\rm off}\simeq$ $5.50^{\prime\prime}$ \cite{GCN31203, GCN31221}. This galaxy has a magnitude of $m$=19.53 mag in the $r$ band and a half-light radius of $R_{50}=0.61^{\prime\prime}$ \cite{Adelman_2008ApJS}. We assume that the surface distributions of galaxies and GRBs are uniform and follow Poisson distributions \cite{Bloom_2002AJ, Stalder_2017ApJ}. The probability of one or more random coincidences is given by
\begin{equation}
 P=1-e^{-\lambda}\approx\lambda\,(\lambda\sim0),
\end{equation}
where $\lambda$ is the expectation number of coincidences. To establish the associations between GRB 211211A, the afterglow candidates, and the SDSS galaxy, the probabilities of chance coincidence (PCC) that the afterglow candidates are unrelated to GRB 211211A (defined as $P_{\rm cc1}$), and that this galaxy is independent of the afterglow candidates (defined as $P_{\rm cc2}$) can be calculated as follows.

\textit{The calculation for $P_{\rm cc1}$}: To derive $P_{\rm cc1}$, $\lambda$ can be regarded as the average number of GRBs occurring within a time window and sky area, and it is the product of the following three quantities: 
\begin{itemize}
 \item $r_1$: the detection rate of GRB over the whole sky. The latest {\it Fermi}/GBM \cite{von_2020ApJ} and {\it Swift}/BAT \cite{Lien_2016ApJ} burst catalogs provide 236 and 92.2 bursts per year detected by the two instruments, respectively. Considering the fact that $13.6\%$ of {\it Fermi}/GBM bursts are co-detected by {\it Swift}/BAT \cite{Bhat_2016ApJS}, we estimate a joint detection rate to be $r_1\sim0.8$ bursts per day over the whole sky.
 
 \item $r_2$: the ratio between the localization error circle and the whole sky. Using the localization uncertainty of $\theta=1^{\circ}$ for GRB 211211A \cite{GCN31201}, one can calculate this ratio to be $r_2=(1-{\rm cos}\theta)/2=7.62\times10^{-5}$.

 \item $r_3$: the time window in which the GRB occurs prior to the optical afterglow. Based on the optical afterglow gallery \cite{Kann_2011ApJ}, we utilize a relatively long time window of one day ($r_3=1$) to avoid underestimating $P_{\rm cc1}$.
\end{itemize}
Based on the above values, one can calculate $P_{\rm cc1}\approx\lambda=\Pi_{\rm i=1}^3r_{\rm i}$ to be $6.09\times10^{-5}$. Such probability allows us to rule out a chance coincidence between the afterglow candidates and the burst at the $4.01\sigma$ confidence level.

\textit{The calculation for $P_{\rm cc2}$}: For $P_{\rm cc2}$, $\lambda$ is the expected number of galaxies brighter than $m$ and within a circle with a radius of $\Delta R$, and it can be derived by the two quantities:
\begin{itemize}
 \item $\sigma(\leqslant m)$: the number density of galaxies brighter than $m$. Following Ref.\cite{Berger_2010ApJ}, we calculate the number density $\sigma(\leqslant m)=1.6\times10^{-4}\,{\rm arcsec}^{-2}$ based on the results of deep optical galaxy surveys \cite{Hogg_1997MNRAS, Beckwith_2006AJ}, i.e.,
 \begin{equation}
 \sigma(\leqslant m)=\frac{10^{0.33(m-24)-2.44}}{0.33\times{\rm ln}(10)}\,{\rm arcsec}^{-2}.
 \end{equation}
 
 \item $\Delta R$: the maximum effective angular radius that the host galaxy is far from the optical counterpart. We take $\Delta R\approx(R_{\rm off}^2+4R_{50}^2)^{1/2}=5.63$ arcsec based on the fact that the optical counterpart is outside of the galaxy \cite{Bloom_2002AJ}.
\end{itemize}
Therefore, we calculate $P_{\rm cc2}$ through $\lambda=\pi(\Delta R)^2\sigma(\leqslant m)$ to be $1.60\times10^{-2}$, which is consistent with the result in Ref.\cite{Rastinejad_2022arXiv}. This value implies a random association between the optical counterpart and the galaxy can be ruled out at the $2.41\sigma$ level. We note that such a PCC is much smaller than those of other galaxies near the position of the GRB. Therefore, we take this galaxy as the host galaxy of GRB 211211A.

\textit{Offset}: Ref.\cite{GCN31221, Rastinejad_2022arXiv} has reported a redshift of $z=0.076$ of the host galaxy. Thus the offset in projection can be calculated as 8.19 kpc. And the normalized offset, defined by $R_{\rm off}/R_{50}$, is 9.02. The large offset of GRB 211211A would be very peculiar if it were a Type-II GRB. On the other hand, Type-I GRBs have substantially larger offsets relative to the host galaxies than Type-II GRBs \cite{Bloom_2002AJ, Fong_2013ApJ, Berger_2014ARAA}, which is in agreement with the prediction of the merger origin of compact binary where the neutron star member undergoes a supernova ``kick'' as it forms. The offset (normalized offset) of GRB 211211A is larger than $\sim$70\% ($\sim$85\%) of the offsets (normalized offsets) of Type-I GRBs \cite{Fong_2013ApJ}. Therefore, GRB 211211A is more likely to originate from a compact star merger event based on its host galaxy information.

\subsection{Multi-wavelength afterglow.\\}

\textit{X-ray afterglow}: The \emph{Swift} X-ray Telescope (XRT \cite{Burrows_2005SSRv}) began to observe the BAT field from 69 s after the BAT trigger and identified a bright, uncatalogued X-ray source located at RA=$14^{\rm h}09^{\rm m}10.09^{\rm s}$ and DEC=$+27^{\circ}53^{\prime}18.8^{\prime\prime}$ (J2000) with an uncertainty of $2.1^{\prime\prime}$ \cite{GCN31205}. Figure \ref{fig:agmnmodeling}a and \ref{fig:agmnmodeling}b shows the X-ray light curve and photon index evolution of GRB 211211A obtained from the {\it Swift}/XRT GRB light curve repository (\url{https://www.swift.ac.uk/xrt_curves/}) and {\it Swift}/XRT GRB spectrum repository (\url{https://www.swift.ac.uk/xrt_spectra/}), respectively. Note that we convert the 0.3--10 keV flux into the 1 keV flux density using the time-averaged photon index for each of both the Windowed Timing (WT) mode and Photon Counting (PC) mode. A convention of $f_\nu\propto t^{-\alpha_{\rm X}}v^{-\Gamma_{\rm X}+1}$ is employed here, where $\alpha_{\rm X}$ is the decay slope of the light curve and $\Gamma_{\rm X}$ is defined as the photon index of the X-ray spectrum. Due to the long gap between the WT mode and PC mode, we fit the light curves in WT and PC modes separately using a multi-segment smoothed broken-power-law (SBPL\cite{Liang_2007ApJ, Xiao_2018ApJ}) model. The fitting process is implemented by the python module \emph{emcee}\cite{Foreman_2013PASP}. We perform the model comparison based on the Bayesian information criterion\cite{Schwarz_1978AnSta} (BIC, defined as ${\rm BIC}=-2{\rm ln}\mathcal{L}+k{\rm ln} N$, where $\mathcal{L}$ is the maximum likelihood, $k$ is the number of parameters of the model, and $N$ is the number of data points) and Akaike information criterion \cite{Akaike1974, sugiura1978} (here we use the variant AICc corrected for small sample sizes, defined as ${\rm AICc}=-2{\rm ln}\mathcal{L}+2k+2k(k+1)/(N-k-1)$). Generally speaking, the model with the lower BIC or AICc is the preferred model. The 2-segment SBPL can model the light curve in WT mode well. For the light curve in PC mode, the 3-segment SBPL model (stat/dof=40.81/46, BIC=64.52, AICc=54.68) provides a better fit than the 2-segment SBPL model (stat/dof=51.25/48, BIC=67.06, AICc=60.10) as shown in Extended Data Figure \ref{fig:xrtfit}. The differences in BIC and AICc between the 2-segment and 3-segment SBPL models are $\Delta{\rm BIC}=2.54$ and $\Delta{\rm AICc}=5.42$, respectively. Such ``strength of evidence'' (i.e., $\Delta{\rm BIC}$ and $\Delta{\rm AICc}$) positively supports the 3-segment SBPL model. Based on our fits, the X-ray light curve of GRB 211211A can be divided into four phases (Figure \ref{fig:agmnmodeling}a). The time range, power-law decay index of the light curve ($\alpha_{\rm X}$), and photon index of the spectrum ($\Gamma_{\rm X}$) of each phase are listed in Extended Data Table \ref{tab:xprofile}. These four phases match up well with a canonical GRB X-ray light curve\cite{Zhang_2006ApJ, Nousek_2006ApJ}. Details of each phase are further itemized below.

\begin{itemize}
 \item Phase I -- the steep decay. The sharp decay slopes ($\alpha_{\rm X}\ge3$) and strong spectral evolution can be interpreted by the higher latitude effect \cite{Kumar_2000ApJ, Dermer_2004ApJ} of the tail of the prompt emission with an intrinsically curved spectral shape \cite{BBZhang_2007ApJ, BBZhang_2009ApJ}. The spectra in Phase I were claimed to be consistent with the fast-cooling synchrotron emission\cite{gompertz2022}.

 \item Phase II -- the plateau. With a decay index $\alpha_{\rm X}\sim0$, Phase II is a well-defined plateau, as seen in many previous GRB X-ray light curves. Physically speaking, X-ray plateaus can be divided into two categories, namely, external plateaus and internal plateaus. The external plateau is followed by a normal decay phase and can be interpreted by invoking a continuous energy injection caused by a long-lasting central engine \cite{Dai_1998PhRvL, Zhang_2001ApJ} or stratified ejecta \cite{Rees_1998ApJ, Sari_2000ApJ} within the external forward shock model \cite{Zhang_2006ApJ, Nousek_2006ApJ}. The internal plateaus are usually followed by a late-time steep decay segment and are inconsistent with the standard external shock model. One has to introduce an internal dissipation process (e.g., the formation of a black hole from a magnetar \cite{Troja_2007ApJ, Lyons_2010MNRAS}) to explain the sudden change of decay slopes. The plateau of GRB 211211A fits the former case as it is followed by a normal decay phase, and no significant spectral evolution is observed between the two phases (Figure \ref{fig:agmnmodeling}b). This suggests that Phase II is likely powered by the continuous energy injection in an external shock. The leading scenario is that the energy injection is from a long-lasting central engine\cite{Zhang_2006ApJ, Nousek_2006ApJ}.

 \item Phase III -- the normal decay. With $\alpha_{\rm X}=1.57$ and $\Gamma_{\rm X}=1.54$, the normal decay phase is well consistent with the prediction of the standard external shock afterglow model \cite{Zhang_2006ApJ, Nousek_2006ApJ}.

 \item Phase IV -- post jet-break decay. A break appears at $\sim$ 0.8 days, a typical time as a jet break \cite{Racusin_2009ApJ}. The break is followed by a slightly steeper decay phase with $\alpha_X\sim 3.33$ that lasts until $\sim $ 2 days. As modeled with the afterglow model (see below), we found a slightly off-axis afterglow from a structured jet succeeds in interpreting the observed data.
\end{itemize}

Comparatively, we analyze the {\it Swift}/XRT light curve of GRB 060614 and list the fitting results in Extended Data Table \ref{tab:xprofile}. We also re-scale the X-ray afterglow of GRB 060614 to the redshift of 0.076 and over-plot it in Figure \ref{fig:agmnmodeling}a. We note that both bursts exhibit similar temporal and spectral profiles in the X-ray bands. Nevertheless, the flux of GRB 211211A is generally weaker than that of GRB 060614, except during the plateau phase in the X-ray band, when the energy injection of GRB 211211A is stronger but lasts for a shorter period of time.

\textit{Ultraviolet, optical, and near-infrared afterglows}: We collect all ultraviolet, optical, and near-infrared observations at the {\it Swift}/XRT coordinate as listed in Extended Data Table \ref{tab:uvoniobs}. The \emph{Swift} Ultraviolet and Optical Telescope (UVOT \cite{Roming_2005SSRv}) performed five rounds of observations in the field of GRB 211211A from 92 seconds to $\sim3$ days after the BAT trigger \cite{GCN31222}. The UVOT data are obtained from the \emph{Swift} archive website (\url{https://www.swift.ac.uk/archive/ql.php}) and listed in Extended Data Table \ref{tab:uvoniobs}. The white filter data in UVOT observations are excluded from this study due to its flat transmission function. Additionally, we only employ the tightest upper limit when several time bins are very close together. The detections and upper limits reported in GCN circulars (\url{https://gcn.gsfc.nasa.gov/other/211211A.gcn3}) by multiple facilities, including KAIT\cite{GCN31203}, MITSuME/Akeno\cite{GCN31217}, HCT\cite{GCN31227}, LCO/Sinistro\cite{GCN31214}, GMG\cite{GCN31232}, DOT\cite{GCN31299}, GIT\cite{GCN31227}, AbAO\cite{GCN31233}, Zeiss-1000\cite{GCN31234}, and TNG/NICS\cite{GCN31242}, are collected in Extended Data Table \ref{tab:uvoniobs}. We also collect additional optical and near-infrared data from Ref.\cite{Rastinejad_2022arXiv, Xiao2022arXiv} and list the magnitudes uncorrected for foreground Galactic extinction in Extended Data Table \ref{tab:uvoniobs}.

Figure \ref{fig:agmnmodeling}a shows the $R$ band light curve of GRB 211211A. The initial 11-hour photometric data does not show significant decay. Afterward, it decays as a power law with an index of 1.00 until 1.56 days. The final two detections in the $R$ band indicate a fast power-law decay with an index of 2.80. We also collected the $R$ band light curve of GRB 060614 from Ref.\cite{Gal-Yam_2006Natur, Fynbo_2006Natur} and over-plot it in Figure \ref{fig:agmnmodeling}a by re-scaling it to the redshift of 0.076. In comparison with GRB 060614, the optical afterglow of GRB 211211A has a similar temporal profile but is generally weaker by about one order of magnitude. Regarding the multi-band behavior, GRB 060614 appears to behave normally, whereas GRB 211211A shows a significant chromatic feature, that is, the light curves of different bands break at different times. Given that the breaks themselves are not of spectral origins (Figure \ref{fig:agmnmodeling}b), the chromatic behavior cannot be interpreted by the standard external shock afterglow model, which predicts achromatic breaks. We will further show the deviation between the afterglow data and the standard model by modeling the multi-wavelength afterglows (see below).

\subsection{Non-detection of supernova.\\}

A robust association between Type-II GRBs and supernovae (SNe) has been established, and the smoking gun signature is the detection of characteristic spectral features of the supernova in the optical afterglow \cite{Galama_1998Natur, Reeves_2002Natur, Hjorth_2003Natur}. With a redshift of 0.076, if GRB 211211A were a Type-II GRB, a significant supernova component would be expected approximately ten days later. To examine the possibility of an association between GRB 211211A and a supernova, we obtain the quasi-bolometric unreddened rest-frame light curves of SN 1998bw from Ref.\cite{Clocchiatti_2011AJ}. We then apply a scale factor, $k$, and a stretch factor, $s$, to the template supernova, SN 1998bw, in order to obtain a specific supernova light curve. The observed flux density at redshift $z$ of the specific supernova can be described as
\begin{equation}
 F_{\nu}^{\rm SN}(t_{\rm obs}, \nu_{\rm obs})=\frac{(1+z)k}{4\pi d_{\rm L}^2} L_{\nu}^{\rm SN} \bigg(\frac{t_{\rm obs}}{(1+z)s},(1+z)\nu_{\rm obs}\bigg),
\end{equation}
where $d_{\rm L}$ is luminosity distance and $L_{\nu}^{\rm SN}$ is specific luminosity at frequency $\nu$ of template supernova. In Figure \ref{fig:agmnmodeling}a, we demonstrate the $R$ band light curves of three well-studied supernovae associated with Type-II GRBs, including the template SN 1998bw/GRB 980425, SN 2006aj/GRB 060218A ($s=0.68$ and $k=0.72$; see Ref.\cite{Cano_2013MNRAS}) and SN 2010bh/GRB 100316D ($s=0.60$ and $k=0.40$; see Ref.\cite{Cano_2013MNRAS}), by putting them at the redshift of GRB 211211A. The three supernovae cover a considerable range of luminosity, with SN 1998bw and SN 2010bh ranked the brightest and faintest GRB-SNe, respectively. The $R$ band afterglow of GRB 211211A decays so rapidly at 1--5 days that it is about two orders of magnitude fainter than the supernovae at this time, and there is still no sign of a supernova bump. The $R$ band upper limits at $\sim 10$ days further confirm the absence of supernova. The absence of a supernova associated with GRB 211211A rules out it being a typical Type-II GRB produced by the core collapse of a massive star.

\subsection{Identification of an engine-fed kilonova.\\}

\textit{Fit with afterglow-only model}: The theoretical afterglow light curves are generated using the public python package \emph{Afterglowpy}\cite{Ryan_2020ApJ}, which can compute external forward shock synchrotron radiation and handle both the structured jet and off-axis-observer cases. We apply a Gaussian structured jet for our calculation. A model-dependent set of input parameters includes the isotropic kinetic energy ($E_{\rm k,iso}$), viewing angle ($\theta_{\rm v}$), Gaussian width of jet core ($\theta_{\rm c}$), circumburst density ($n$), electron energy distribution index ($p$), and the fractions of shock energy that go to electrons ($\epsilon_{\rm e}$) and magnetic fields ($\epsilon_{\rm B}$). For a Gaussian structured jet, we provide a sensible value of 4$\theta_{\rm c}$ to truncate the outer regions of the jet. To correct the data in Extended Data Table \ref{tab:uvoniobs} for foreground Galactic extinction on the line of sight, we applied the F99 \cite{Fitzpatrick_1999PASP} model with a visual extinction to reddening ratio $R_{\rm V}=3.1$ to extinguish the flux density of the afterglow model, which is implemented with the \emph{dust-extinction} python package \cite{Astropy_2013AA}. We leave the color excess $E(B-V)$ as a free parameter during the procedure. 

The fitting of the observed data is based on the nested sampling algorithm by applying the Bayesian computation python package \emph{PyMultinest} \cite{Buchner_2014A&A}, with the log-likelihood function written as
\begin{equation}
 {\rm ln}\mathcal{L}=-\frac{1}{2}\sum_{i=1}^{n}\Bigg[\frac{(O_{i}-M_{i})^2}{\sigma_{i}^2+v^2}+{\rm ln}\Big[2\pi(\sigma_{i}^2+v^2)\Big]\Bigg],
\end{equation}
where $O$, $M$, and $\sigma$ stand for the observed magnitudes, modeled magnitudes, and uncertainties of observed magnitudes, respectively. The subscript $i$ represents the serial number for each observed and modeled value. An extra variance parameter $v$ is introduced to account for the variance of Gaussian distribution, which encompasses additional uncertainty in observed data. The log-likelihood function can be described as a one-sided Gaussian penalty term for upper limits. We utilize $-2{\rm ln}\mathcal{L}$ as a statistic to measure the goodness of modeling. To avoid the effect of energy injection, we only consider the afterglow after the X-ray plateau. Then we allow all model parameters to vary with typical but broad priors to search for the best fit for observed data.

Fitting the afterglow-only (AG-only) model with all the observed data, from X-ray to near-infrared, we find that significant residual errors exist even with the best-fit parameters, especially in the optical and near-infrared bands. Then, we exclude the observed data in the $r$, $i$, $J$, and $K$ bands and fit the rest again with the AG-only model. The best-fit AG-only model can explain the overall trends of the multi-wavelength light curves, in particular in the X-ray band. Nonetheless, significant disagreement between the observational data and the afterglow model at $t\gtrsim$ 1 day is observed in both optical and near-infrared wavelengths, as plotted in Figure \ref{fig:agmnmodeling}c. These significant excesses indicate that other emission components are required to explain the observed data.

\textit{Fits with afterglow plus kilonova model}: Such kind of excess has long been considered as the existence of an underlying kilonova emission. Therefore, we investigate the afterglow plus kilonova (AG + KN) model to explain the significant excess observed in the optical and near-infrared bands of GRB 211211A.

In a WD-NS system, if the masses of the two components are comparable, the WD might not be totally disrupted and thrown around the NS right before the merger like a tidal disruption event. Instead, the NS might merge into the center of the WD and induce the collapse of the WD. During the collapsing, materials from the WD would be neutronized and mixed with the materials from the pre-merger NS. Later, a proto-magnetar would form, which is surrounded by a disk with neutron-rich materials. Therefore, the ejecta from the WD-NS merger consists of two components: a neutron-poor component from the dynamical process in the front and a neutron-rich component from the disk wind in the rear. Here we define a parameter $f_n = M_{\rm ej,disk}/(M_{\rm ej,dyn} + M_{\rm ej,disk})$ to describe the fraction of neutron-rich ejecta, where $M_{\rm ej,disk}$ and $M_{\rm ej,dyn}$ represent the masses of ejecta from the disk wind and dynamical process, respectively. R-process nucleosynthesis can happen inside the neutron-rich ejecta and power a kilonova emission\cite{Li_1998ApJ,Metzger_2010MNRAS,Kasen_2013ApJ}. 

Since a post-merger magnetar has been introduced to explain the EE phase and the long-lasting plateau in the X-ray afterglow, a considerable amount of spin-down energy from the magnetar should be injected into both the dynamical and disk-wind ejecta. Therefore, instead of the r-process powered kilonova model, we employ an engine-fed kilonova model based on Ref.\cite{Yu_2013ApJ}. In this model, the spin-down energy is injected with an efficiency $\xi$ while only a fraction $\xi_{\rm t}$ can be directly deposited as the ejecta's internal energy. The detailed energy injection mechanism has been studied recently \cite{Ai_2022arXiv} in the framework of a shock system produced through the interaction between the magnetar wind and the ejecta, and the efficiencies have been quantitatively determined. In this work, we adopt $\xi = 0.5$ and $\xi_{\rm t} = 0.03 e^{-1/\tau}$, where $\tau$ stands for the mean optical depth of the ejecta.

The light curve of the engine-fed kilonova depends on the properties of both the ejecta and magnetar. The parameters related to the ejecta include the total ejecta mass ($M_{\rm ej}$), opacity ($\kappa$), initial velocity ($v_{\rm ej}$), and the fraction of neutron-rich ejecta ($f_{\rm n}$). The total ejecta mass and the fraction of neutron-rich ejecta determine how much radioactive energy would be generated. The ejecta mass would also highly influences the dynamics of the kilonova's radiation region. The opacity and velocity significantly influence the radiation transfer process and the subsequent thermal emission. Fixing the collapsing time of the newly born magnetar as 6,000 s (corresponding to the end time of the X-ray plateau), the energy injection from the central engine would only depend on the spin-down energy, in other words, the surface magnetic strength ($B$) and the initial spin period of the magnetar ($P_0$, fixing 1 ms).

Again, the AG + KN model is fitted to the full-band observed data with the python package \emph{PyMultinest} \cite{Buchner_2014A&A}. The best-fitting parameter values and their 1$\sigma$ uncertainties for the afterglow and engine-fed kilonova are presented in Extended Data Table \ref{tab:agmnmodeling}. The corner plot of posterior probability distributions of the parameters is exhibited in Extended Data Figure \ref{fig:mnmodeling}a. The modeled light curves of AG and AG + KN are plotted as dotted and solid lines in Figure \ref{fig:agmnmodeling}c, respectively. As shown in Extended Data Figure \ref{fig:mnmodeling}b, the engine-fed kilonova model is successful in fitting the excesses in both optical and near-infrared bands.

As listed in Extended Data Table \ref{tab:agmnmodeling}, the parameters of afterglow are generally consistent with the statistical results in Ref.\cite{Wang_2015ApJS}. With the aid of a clearly visible jet break in the X-ray afterglow, the jet opening angle $\theta_{\rm c}$ can be well constrained. The viewing angle $\theta_{\rm v}\sim2\theta_{\rm c}$ indicates that this burst was seen in a slightly off-axis manner. From the theoretical side, the properties of the merger ejecta are still poorly known because of the lack of numerical simulations on WD-NS mergers. According to our best-fitting values, the opacity of the ejecta ($\kappa \sim 0.73~{\rm g~cm^{-2}}$) is higher than the typical opacity due to the electron scattering ($\kappa = 0.2~{\rm g~cm^{-2}}$), but much lower than that for the neutron-rich materials ($\kappa \sim 10~{\rm g~cm^{-2}}$). It is consistent with our scenario that the whole ejecta is a mixture of neutron-rich and neutron-poor materials. Our best-fitting opacity can be regarded as an effective value.

\subsection{Interpretation to the extended emission.\\}

Since the EE has distinct temporal and spectral properties from the ME, a different physical mechanism is needed. For compact object mergers, if the merger product is a black hole, it is essentially impossible to maintain a roughly constant jet luminosity because of its small disk mass, short viscous timescale, and decaying accretion rate\cite{rosswog07, lu2022}. A rapid-spin-powered BH engine scenario was mentioned in Ref.\cite{vanputten14}, but the specific mechanism has been only developed for long GRBs\cite{vanputten12}.

A natural mechanism to interpret EE for Type-I GRBs is to invoke a magnetar central engine\cite{metzger08}. The specific mechanism may be through invoking differential-rotation-induced magnetic bubbles as the mechanism for producing GRB prompt emission \cite{Kluzniak_1998ApJ, Ruderman_2000ApJ, Dai_2006Sci, Zhang_2021NatAs}. In such a model, the duration of a GRB is dependent on the number of bubbles and the interval between two bubble eruptions. Each magnetic bubble eruption follows the same process as described below. We consider a differentially rotating magnetar with an internal poloidal magnetic field $B_{\rm r}=10^{13}\rm\,G$ and an initial spin period $P_{0}\sim1$ ms. Here we take the convention that $Q_x = Q/10^x$ and all the quantities are in the \emph{cgs} unit system. Due to differential rotation, $B_{\rm r}$ will be wound up into a toroidal magnetic field $B_{\rm\phi}$ which increases as $dB_{\rm\phi}/dt=\Delta\Omega B_{\rm r}$ \cite{Dai_2006Sci}, where $\Delta\Omega$, with an initial value of $2\pi/P_{0}$, is the differential angular velocity between the inner and outer parts of the magnetar. Continually amplified $B_{\rm\phi}$ forms a magnetically confined toroid inside the magnetar to enclose some matter. When $B_{\rm\phi}$ is amplified to a critical magnetic field\cite{Kluzniak_1998ApJ} $B_{\rm b}\approx10^{17}\rm\,G$, the magnetic buoyant force can balance the antibuoyant stratification in magnetar composition, which corresponds to a critical time scale $\tau_{\rm b}=B_{\rm b}/(\Delta\Omega B_{\rm r})\simeq1.6\times B_{\rm r,13}^{-1}P_{\rm 0,-3}$ s. Such a time scale corresponds to the interval between two bubble eruptions. As $B_{\rm\phi}$ continues to increase after $\tau_{\rm b}$, the buoyancy force will exceed the antibuoyant force. The magnetic toroid becomes unstable and floats up toward the magnetar surface due to excess buoyancy. The buoyancy time scale, during which the magnetic toroid floats up and penetrates through the magnetar surface, can be calculated as $\Delta\tau_{\rm b}\simeq8.4\times B_{\rm r,13}^{-1/3}P_{\rm 0,-3}^{1/3}$ ms. Upon penetration through the magnetar surface, the toroidal fields may reconnect and give rise to an explosive event with energy $E_{\rm b}=B_{\rm b}^2V_{\rm b}/8\pi\simeq1.6\times10^{50}\,\xi_{\rm b,-1}$ erg, where $\xi_{\rm b}=V_{\rm b}/V_{\rm s}$ is the ratio of the toroid's volume to the stellar volume. Such a reconnection process typically occurs within $\sim 10^{-4}$ s, corresponding to the large fluctuations as observed in the observations. A magnetar with an initial differential kinetic energy of $\sim3\times10^{51}$ erg is capable of producing $\sim 30$ similar eruptions. It yields a temporal profile consisting of several sub-bursts with a total duration of $\sim 50$ s, which is in agreement with the observed EE phase of GRB 211211A. Once this event is over, a similar process described above will lead to another explosion as long as $\Delta\Omega$ remains large enough. Once the differential kinetic is exhausted or no longer supports another wind-up of the toroidal magnetic field, the millisecond magnetar can continue to lose its rotation energy through magnetic dipole radiation. Such a model predicts that the different pulses of the EE are locally generated in small emission regions and are independent of one another, so the spectral lag of the EE will be small. Such a prediction is consistent with that observed in GRB 211211A.

\subsection{Spin-down luminosity.\\}
The spin-down power of magnetar from magnetic dipole radiation has successfully explained the energy injection in the plateau phase \cite{Zhang_2006ApJ, Nousek_2006ApJ}. For a millisecond magnetar, the spin-down luminosity due to the magnetic dipole radiation can be calculated as \cite{shapiro83, Zhang_2001ApJ}
\begin{eqnarray}
L_{\rm sd,m} = \frac{\eta B_{\rm p}^2 R_{\rm s}^6 \Omega^4}{6c^3} \approx 10^{46} {\,\rm erg~s^{-1}}~\eta_{-3} B_{\rm p, 15}^2 R_{\rm s,6}^6 P_{0,-3}^{-4},
\end{eqnarray}
where $B_{\rm p}$ represents the strength of the surface dipolar magnetic field of the magnetar, $R_{\rm s}$ stands for the radius of the magnetar, $\eta$ is the efficiency of converting the magnetar wind energy into X-ray radiation \cite{Xiao_2019ApJ, Xiao_2019ApJL}, and $c$ is the speed of light. To explain the observed luminosity of $\sim 10^{45}$ ${\rm erg\,s^{-1}}$ during the X-ray plateau of GRB 211211A, a surface dipolar magnetic field $B_{\rm p} \approx 3\times10^{14}$ G is required.

\subsection{Special properties of the WD-NS merger.\\}
Since the WD-NS merger rate is much higher than the NS-NS merger rate and since the rate density of GRB 211211A is much smaller than that of short GRBs, an inevitable conclusion is that only a rare type of WD-NS mergers can make GRBs. We suggest that the preferred systems that can make GRBs are those with a mass ratio close to unity. In other words, the WD needs to be massive and close to the Chandrasekhar limit. Such a WD is compact and is the progenitor of the accretion-induced collapse (AIC) model for GRBs\cite{fryer99}. Before the merger, a small amount of WD mass (which is neutron poor) would be tidally ejected. The majority of the mass would undergo collapse and neutralization after the NS enters the interior of the WD because of the huge excess gravity provided by the NS. Because of the large orbital angular momentum, the WD undergoes AIC to form a massive NS surrounded by a massive accretion disk. A neutron-rich wind is subsequently launched from such a system, similar to the AIC model for GRBs\cite{fryer99}, which provides the necessary neutron-rich ejecta to power the kilonova. The product of such a merger carries a significant angular momentum. The magnetic field of the post-merger product depends on the magnetic field strengths of the NS and the WD, but additional amplification is possible due to the $\alpha-\Omega$ dynamo. A millisecond magnetar is a likely outcome.

\subsection{Other progenitor models.\\}
NS-NS\cite{Paczynski1986, Eichler1989} and BH-NS\cite{Paczynski1991} mergers have been proposed to power short GRBs. Both types of engines have difficulties interpreting GRB 211211A. The product of a BH-NS merger system must be a BH. Since the NS matter has a very high density, it is very difficult to power a GRB with a duration longer than 2 s, let alone to produce both a 13-s ME and a 55-s EE with distinct emission properties, as well as an X-ray plateau lasting up to $\sim$ 6,000 s. An NS-NS merger can produce a magnetar central engine, but it is difficult to interpret the much longer ME than all other short GRBs or short GRBs with EE, which are believed to be due to NS-NS mergers. A WD-WD merger model was proposed to interpret the kilonova emission associated with GRB 170817A\cite{Rueda2018}. However, this model cannot interpret the gravitational wave data, which pointed toward an NS-NS merger. For GRB 211211A, the WD-WD merger model predicts too high an event rate, and it is unclear whether a rapidly spinning magnetar can be formed as the consequence of the merger. Another interesting possibility is the r-process enhanced collapsar model\cite{Siegel2019}. However, such a collapsar event is expected to be associated with a bright supernova, which is not observed in association with GRB 211211A. The same objection applies to the model that interprets the infrared hump as evidence of dust echo of a standard long GRB\cite{waxman22}. In order to avoid this apparent issue, Ref.\cite{waxman22} invokes another galaxy, SDSS J140909.60+275325.8 (hereafter G2) at redshift $z=0.459$, in the GRB field of view as the GRB host galaxy. However, the chance coincidence probability $P_{\rm cc, G2}=0.15$ between GRB 211211A and G2 is much higher than that between GRB 211211A and SDSS J140910.47+275320.8 (G1), the host galaxy candidate adopted in our paper and all other papers on GRB 211211A. Furthermore, if G2 is the host galaxy, the location of GRB 211211A has a physical offset of 61.81 kpc and a normalized offset of 36.78 from the host. These numbers are inconsistent with long GRBs and even short GRBs. At such a large offset, it is essentially impossible to have a massive star and a dusty environment to support the suggested dust echo model. We, therefore, conclude that a WD-NS merger provides the best interpretation for the broad-band observations of GRB 211211A.

\section*{Data Availability}

The processed data are presented in the tables and figures of the paper, which are available upon reasonable request. The authors point out that the data used in the paper are publicly available, whether through {\it Fermi}/GBM data archive, {\it Swift} data archive, or GCN circulars.

\section*{Code Availability}
Upon reasonable requests, the code (mostly in python) used to produce the results and figures will be provided.

\clearpage


\begin{addendum}

\item This work is supported by the National Key Research and Development Programs of China (2018YFA0404204), the National Natural Science Foundation of China (Grant Nos. 11833003, U2038105, 12121003, 11922301, 12041306, 12103089), the science research grants from the China Manned Space Project with NO.CMS-CSST-2021-B11, the Natural Science Foundation of Jiangsu Province (Grant No. BK20211000), and the Program for Innovative Talents, Entrepreneur in Jiangsu. S.-K.A. and B.Z. acknowledge support of the Top Tier Doctoral Graduate Research Assistantship (TTDGRA) and Nevada Center for Astrophysics at the University of Nevada, Las Vegas. We acknowledge the use of public data from the {\it Fermi} Science Support Center, {\it Swift} Science Data Centre, and GCN circulars reported by multiple facilities. We thank Yan-Zhi Meng and Zong-Kai Peng for helpful comments.

\item[Author Contributions] 

B.-B.Z. and J.Y. initiated the study. B.-B.Z., B.Z., J.Y., and S.A. coordinated the scientific investigations of the event. J.Y. processed and analyzed the {\it Fermi}/GBM and {\it Swift} data. J.Y. and Z.-K.L. calculated the spectral lags. J.Y., X.I.W., and H.-J.L. calculated the amplitude parameter. J.Y., Y.-H.Yang, and Y.-H.Yin fitted the Amati relation. J.Y. and Y.L. contributed to the information about the host galaxy. J.Y. contributed to the afterglow analysis and modeling. S.A. provided the engine-fed kilonova model. J.Y. and S.A. performed theoretical modeling of the engine-fed kilonova. J.Y. and S.A. contributed to the magnetic bubble and spin-down models. B.Z. and S.A. developed the WD-NS merger scenario. B.Z., S.A., and J.Y. investigated other progenitor models. J.Y., B.-B.Z., B.Z., and S.A. wrote the manuscript with contributions from all authors.
 
\item[Competing Interests] The authors declare that they have no competing financial interests.
 
\item[Additional information] Correspondence and requests for materials should be addressed to B.-B.Z. and B.Z.

\end{addendum}

\clearpage
\setcounter{figure}{0}
\setcounter{table}{0}

\captionsetup[table]{name={\bf Extended Data Table}}
\captionsetup[figure]{name={\bf Extended Data Figure}}

\begin{table*}
\tiny
\caption{\textbf{Spectral fitting results and corresponding energy flux in each time interval of GRB 211211A.} The energy fluxes are calculated between 10 and 1,000 keV. All errors represent the 1$\sigma$ uncertainties.}
\label{tab:specfit}
\hspace{-1.2 cm}
\begin{tabular}{ccclcc|ccclcc}
\hline
t1 (s) & t2 (s) & $\alpha$ ($\beta$) & $E_{\rm p}$ (keV) & flux ($\rm erg\,cm^{-2}\,s^{-1}$) & pgstat/dof & t1 (s) & t2 (s) & $\alpha$ & $E_{\rm p}$ (keV) & flux ($\rm erg\,cm^{-2}\,s^{-1}$) & pgstat/dof \\
\hline
\multirow{2}{*}{0.500}&\multirow{2}{*}{13.000}&$-0.996_{-0.005}^{+0.005}$&\multirow{2}{*}{$687.1_{-11.0}^{+12.5}$}&\multirow{2}{*}{$3.02_{-0.01}^{+0.01}\times10^{-5}$}&\multirow{2}{*}{2,399.59/361}&\multirow{2}{*}{10.900}&\multirow{2}{*}{11.200}&$-1.10_{-0.04}^{+0.04}$&\multirow{2}{*}{$323.9_{-22.0}^{+33.2}$}&\multirow{2}{*}{$1.52_{-0.04}^{+0.06}\times10^{-5}$}&\multirow{2}{*}{332.01/362} \\
&&($-2.36_{-0.02}^{+0.02}$)&&&&&&&&& \\
\multirow{2}{*}{15.000}&\multirow{2}{*}{70.000}&$-0.97_{-0.04}^{+0.03}$&\multirow{2}{*}{$82.0_{-2.3}^{+3.8}$}&\multirow{2}{*}{$2.92_{-0.02}^{+0.01}\times10^{-6}$}&\multirow{2}{*}{883.32/361}&\multirow{2}{*}{11.200}&\multirow{2}{*}{11.700}&$-1.20_{-0.05}^{+0.04}$&\multirow{2}{*}{$242.8_{-20.4}^{+30.9}$}&\multirow{2}{*}{$8.16_{-0.30}^{+0.35}\times10^{-6}$}&\multirow{2}{*}{317.85/362} \\
&&($-2.02_{-0.02}^{+0.01}$)&&&&&&&&& \\
\multirow{2}{*}{0.500}&\multirow{2}{*}{70.000}&$-1.20_{-0.01}^{+0.01}$&\multirow{2}{*}{$399.3_{-16.1}^{+14.0}$}&\multirow{2}{*}{$7.24_{-0.03}^{+0.04}\times10^{-6}$}&\multirow{2}{*}{2,648.05/361}&\multirow{2}{*}{11.700}&\multirow{2}{*}{12.200}&$-1.21_{-0.04}^{+0.05}$&\multirow{2}{*}{$269.4_{-24.5}^{+30.7}$}&\multirow{2}{*}{$8.54_{-0.31}^{+0.33}\times10^{-6}$}&\multirow{2}{*}{297.89/362} \\
&&($-2.05_{-0.02}^{+0.02}$)&&&&&&&&& \\
\cline{1-6}
0.500&2.400&$-1.27_{-0.08}^{+0.05}$&$300.3_{-39.3}^{+102.9}$&$2.01_{-0.11}^{+0.16}\times10^{-6}$&204.38/362&12.200&14.000&$-1.35_{-0.05}^{+0.05}$&$164.3_{-13.6}^{+24.5}$&$2.60_{-0.10}^{+0.14}\times10^{-6}$&291.13/362 \\
2.400&3.100&$-1.12_{-0.03}^{+0.02}$&$851.4_{-65.4}^{+108.2}$&$1.39_{-0.02}^{+0.02}\times10^{-5}$&303.82/362&14.000&15.900&$-1.43_{-0.08}^{+0.04}$&$155.1_{-12.4}^{+40.2}$&$1.97_{-0.07}^{+0.16}\times10^{-6}$&259.65/362 \\
3.100&3.400&$-0.96_{-0.02}^{+0.02}$&$997.7_{-49.4}^{+66.7}$&$4.91_{-0.06}^{+0.07}\times10^{-5}$&393.99/362&15.900&17.400&$-1.29_{-0.05}^{+0.04}$&$167.4_{-12.2}^{+22.1}$&$3.36_{-0.11}^{+0.17}\times10^{-6}$&336.08/362 \\
3.400&3.500&$-0.87_{-0.02}^{+0.02}$&1,593.6$_{-84.1}^{+85.8}$&$1.26_{-0.02}^{+0.02}\times10^{-4}$&440.78/362&17.400&18.000&$-1.24_{-0.05}^{+0.04}$&$230.5_{-21.4}^{+28.8}$&$6.56_{-0.25}^{+0.29}\times10^{-6}$&289.01/362 \\
3.500&3.700&$-0.95_{-0.02}^{+0.02}$&$923.1_{-60.0}^{+60.8}$&$5.74_{-0.10}^{+0.07}\times10^{-5}$&377.34/362&18.000&18.300&$-1.08_{-0.04}^{+0.04}$&$319.0_{-22.4}^{+32.2}$&$1.39_{-0.04}^{+0.05}\times10^{-5}$&301.40/362 \\
3.700&3.800&$-0.86_{-0.03}^{+0.02}$&$999.2_{-56.5}^{+82.2}$&$8.60_{-0.15}^{+0.16}\times10^{-5}$&379.22/362&18.300&18.800&$-1.13_{-0.05}^{+0.05}$&$193.5_{-13.8}^{+18.4}$&$6.85_{-0.23}^{+0.30}\times10^{-6}$&285.09/362 \\
3.800&4.100&$-0.99_{-0.03}^{+0.02}$&$528.0_{-27.9}^{+32.6}$&$3.31_{-0.07}^{+0.05}\times10^{-5}$&402.86/362&18.800&19.300&$-1.20_{-0.03}^{+0.03}$&$380.7_{-28.9}^{+43.8}$&$1.29_{-0.04}^{+0.04}\times10^{-5}$&365.17/362 \\
4.100&4.315&$-1.03_{-0.03}^{+0.04}$&$433.8_{-30.9}^{+35.0}$&$2.26_{-0.06}^{+0.06}\times10^{-5}$&323.71/362&19.300&19.800&$-1.14_{-0.04}^{+0.03}$&$294.3_{-22.2}^{+30.0}$&$1.03_{-0.04}^{+0.03}\times10^{-5}$&363.09/362 \\
4.315&4.500&$-0.92_{-0.03}^{+0.02}$&$590.8_{-30.6}^{+41.7}$&$4.21_{-0.10}^{+0.08}\times10^{-5}$&337.54/362&19.800&20.200&$-1.13_{-0.04}^{+0.03}$&$370.7_{-29.6}^{+36.8}$&$1.46_{-0.04}^{+0.04}\times10^{-5}$&366.96/362 \\
4.500&4.700&$-0.96_{-0.02}^{+0.02}$&$775.6_{-46.8}^{+45.4}$&$5.61_{-0.09}^{+0.09}\times10^{-5}$&386.44/362&20.200&20.900&$-1.22_{-0.04}^{+0.03}$&$260.3_{-16.5}^{+24.6}$&$9.06_{-0.22}^{+0.31}\times10^{-6}$&303.95/362 \\
4.700&4.900&$-0.98_{-0.04}^{+0.03}$&$442.3_{-25.5}^{+35.6}$&$2.75_{-0.06}^{+0.07}\times10^{-5}$&336.71/362&20.900&21.330&$-1.15_{-0.07}^{+0.05}$&$201.6_{-13.6}^{+28.5}$&$7.05_{-0.24}^{+0.42}\times10^{-6}$&296.20/362 \\
4.900&5.100&$-0.91_{-0.04}^{+0.02}$&$734.9_{-26.5}^{+70.6}$&$4.83_{-0.08}^{+0.09}\times10^{-5}$&365.44/362&21.330&21.800&$-1.18_{-0.05}^{+0.06}$&$163.4_{-13.2}^{+15.6}$&$5.84_{-0.24}^{+0.26}\times10^{-6}$&312.70/362 \\
5.100&5.300&$-0.87_{-0.03}^{+0.03}$&$534.1_{-28.8}^{+29.4}$&$4.32_{-0.10}^{+0.06}\times10^{-5}$&323.32/362&21.800&22.200&$-1.15_{-0.04}^{+0.04}$&$302.3_{-23.4}^{+35.7}$&$1.06_{-0.03}^{+0.04}\times10^{-5}$&334.77/362 \\
5.300&5.800&$-1.30_{-0.04}^{+0.04}$&$223.4_{-18.9}^{+28.6}$&$7.40_{-0.27}^{+0.32}\times10^{-6}$&326.06/362&22.200&22.900&$-1.15_{-0.04}^{+0.03}$&$207.3_{-11.8}^{+14.7}$&$8.36_{-0.23}^{+0.25}\times10^{-6}$&355.29/362 \\
5.800&6.105&$-1.20_{-0.04}^{+0.04}$&$300.8_{-27.1}^{+37.0}$&$1.17_{-0.04}^{+0.05}\times10^{-5}$&326.81/362&22.900&23.400&$-1.21_{-0.04}^{+0.05}$&$166.7_{-12.1}^{+13.5}$&$7.11_{-0.24}^{+0.27}\times10^{-6}$&312.39/362 \\
6.105&6.500&$-1.10_{-0.02}^{+0.03}$&$411.4_{-25.1}^{+24.9}$&$2.28_{-0.05}^{+0.05}\times10^{-5}$&363.45/362&23.400&23.730&$-1.19_{-0.05}^{+0.04}$&$231.7_{-19.2}^{+26.3}$&$9.85_{-0.36}^{+0.42}\times10^{-6}$&298.87/362 \\
6.500&6.700&$-0.95_{-0.02}^{+0.02}$&$841.7_{-46.3}^{+52.4}$&$7.13_{-0.09}^{+0.11}\times10^{-5}$&458.43/362&23.730&24.200&$-1.21_{-0.05}^{+0.04}$&$217.1_{-16.8}^{+23.6}$&$8.22_{-0.25}^{+0.36}\times10^{-6}$&322.31/362 \\
6.700&6.800&$-0.90_{-0.02}^{+0.01}$&1,511.6$_{-67.1}^{+64.6}$&$1.37_{-0.02}^{+0.02}\times10^{-4}$&469.41/362&24.200&24.500&$-1.19_{-0.04}^{+0.04}$&$305.1_{-27.4}^{+32.6}$&$1.27_{-0.04}^{+0.04}\times10^{-5}$&291.18/362 \\
6.800&6.875&$-0.93_{-0.03}^{+0.02}$&1,055.5$_{-65.7}^{+84.0}$&$1.07_{-0.02}^{+0.02}\times10^{-4}$&396.71/362&24.500&25.000&$-1.11_{-0.03}^{+0.03}$&$285.6_{-15.4}^{+23.0}$&$1.31_{-0.03}^{+0.04}\times10^{-5}$&337.06/362 \\
6.875&6.965&$-0.95_{-0.03}^{+0.03}$&1,021.6$_{-67.2}^{+86.6}$&$8.84_{-0.16}^{+0.17}\times10^{-5}$&362.51/362&25.000&25.400&$-1.07_{-0.05}^{+0.06}$&$167.5_{-11.1}^{+13.4}$&$7.66_{-0.27}^{+0.30}\times10^{-6}$&295.48/362 \\
6.965&7.100&$-0.98_{-0.02}^{+0.02}$&1,486.8$_{-80.1}^{+89.0}$&$9.83_{-0.12}^{+0.17}\times10^{-5}$&502.24/362&25.400&25.800&$-1.13_{-0.06}^{+0.05}$&$160.5_{-10.6}^{+15.3}$&$7.21_{-0.24}^{+0.32}\times10^{-6}$&285.70/362 \\
7.100&7.300&$-1.05_{-0.02}^{+0.03}$&$765.9_{-58.1}^{+65.1}$&$4.89_{-0.08}^{+0.08}\times10^{-5}$&465.77/362&25.800&26.400&$-1.21_{-0.04}^{+0.04}$&$225.7_{-15.0}^{+24.3}$&$8.05_{-0.22}^{+0.34}\times10^{-6}$&304.57/362 \\
7.300&7.500&$-1.00_{-0.03}^{+0.02}$&$800.3_{-49.5}^{+75.0}$&$5.08_{-0.07}^{+0.09}\times10^{-5}$&406.72/362&26.400&27.000&$-1.31_{-0.05}^{+0.05}$&$174.9_{-14.4}^{+23.6}$&$5.76_{-0.20}^{+0.30}\times10^{-6}$&279.77/362 \\
7.500&7.610&$-0.96_{-0.03}^{+0.02}$&1,035.8$_{-67.5}^{+83.9}$&$6.98_{-0.17}^{+0.11}\times10^{-5}$&431.29/362&27.000&28.000&$-1.14_{-0.08}^{+0.06}$&$103.7_{-5.8}^{+8.5}$&$2.77_{-0.09}^{+0.12}\times10^{-6}$&268.42/362 \\
7.610&7.700&$-0.85_{-0.02}^{+0.02}$&1,136.8$_{-59.8}^{+61.4}$&$1.16_{-0.02}^{+0.02}\times10^{-4}$&487.18/362&28.000&28.900&$-1.18_{-0.07}^{+0.05}$&$133.9_{-8.0}^{+13.8}$&$3.71_{-0.11}^{+0.17}\times10^{-6}$&293.07/362 \\
7.700&7.800&$-0.94_{-0.01}^{+0.02}$&1,920.8$_{-93.8}^{+77.6}$&$1.38_{-0.02}^{+0.02}\times10^{-4}$&484.62/362&28.900&30.400&$-1.26_{-0.06}^{+0.06}$&$85.9_{-4.2}^{+5.2}$&$2.58_{-0.07}^{+0.08}\times10^{-6}$&267.26/362 \\
7.800&7.870&$-0.84_{-0.02}^{+0.02}$&1,352.4$_{-79.0}^{+69.5}$&$1.38_{-0.02}^{+0.03}\times10^{-4}$&420.36/362&30.400&31.500&$-1.20_{-0.08}^{+0.09}$&$83.6_{-5.4}^{+7.0}$&$2.05_{-0.07}^{+0.09}\times10^{-6}$&280.33/362 \\
7.870&8.000&$-0.89_{-0.02}^{+0.02}$&1,021.6$_{-43.4}^{+67.0}$&$1.00_{-0.01}^{+0.01}\times10^{-4}$&371.07/362&31.500&32.305&$-1.40_{-0.08}^{+0.06}$&$88.90_{-6.8}^{+9.8}$&$2.51_{-0.09}^{+0.13}\times10^{-6}$&254.40/362 \\
8.000&8.100&$-0.89_{-0.02}^{+0.02}$&1,464.1$_{-60.2}^{+78.3}$&$1.35_{-0.02}^{+0.02}\times10^{-4}$&451.51/362&32.305&33.300&$-1.44_{-0.08}^{+0.07}$&$90.1_{-6.6}^{+12.2}$&$2.11_{-0.08}^{+0.13}\times10^{-6}$&245.51/362 \\
8.100&8.200&$-0.90_{-0.03}^{+0.03}$&$864.6_{-53.2}^{+56.2}$&$8.36_{-0.16}^{+0.14}\times10^{-5}$&386.92/362&33.300&35.300&$-1.17_{-0.07}^{+0.06}$&$91.4_{-4.3}^{+7.1}$&$1.73_{-0.05}^{+0.07}\times10^{-6}$&255.59/362 \\
8.200&8.400&$-0.89_{-0.02}^{+0.02}$&$766.2_{-34.1}^{+45.3}$&$7.05_{-0.11}^{+0.09}\times10^{-5}$&453.50/362&35.300&36.600&$-1.35_{-0.05}^{+0.05}$&$153.8_{-12.4}^{+21.3}$&$2.93_{-0.10}^{+0.15}\times10^{-6}$&298.30/362 \\
8.400&8.500&$-0.84_{-0.03}^{+0.02}$&$965.8_{-47.3}^{+60.5}$&$9.99_{-0.18}^{+0.16}\times10^{-5}$&354.00/362&36.600&37.380&$-1.27_{-0.07}^{+0.06}$&$147.7_{-12.1}^{+21.9}$&$3.33_{-0.13}^{+0.20}\times10^{-6}$&272.55/362 \\
8.500&8.700&$-0.95_{-0.02}^{+0.02}$&$654.4_{-33.4}^{+41.7}$&$5.38_{-0.09}^{+0.09}\times10^{-5}$&402.11/362&37.380&38.500&$-1.25_{-0.07}^{+0.07}$&$115.3_{-9.4}^{+12.4}$&$2.43_{-0.11}^{+0.11}\times10^{-6}$&241.96/362 \\
8.700&8.795&$-0.92_{-0.02}^{+0.02}$&1,159.4$_{-61.6}^{+83.6}$&$8.85_{-0.16}^{+0.16}\times10^{-5}$&369.54/362&38.500&39.600&$-1.34_{-0.07}^{+0.07}$&$110.2_{-7.5}^{+13.6}$&$2.26_{-0.08}^{+0.12}\times10^{-6}$&254.22/362 \\
8.795&8.900&$-0.94_{-0.02}^{+0.02}$&1,204.8$_{-63.1}^{+109.1}$&$8.37_{-0.15}^{+0.14}\times10^{-5}$&375.64/362&39.600&41.100&$-1.30_{-0.08}^{+0.06}$&$86.4_{-5.2}^{+7.4}$&$1.66_{-0.06}^{+0.07}\times10^{-6}$&256.25/362 \\
8.900&9.030&$-1.07_{-0.02}^{+0.02}$&1,252.2$_{-83.8}^{+136.7}$&$5.35_{-0.11}^{+0.10}\times10^{-5}$&433.73/362&41.100&42.700&$-1.26_{-0.09}^{+0.07}$&$88.8_{-5.9}^{+8.9}$&$1.80_{-0.06}^{+0.09}\times10^{-6}$&240.01/362 \\
9.030&9.300&$-1.04_{-0.03}^{+0.02}$&$603.6_{-33.2}^{+55.2}$&$3.35_{-0.06}^{+0.06}\times10^{-5}$&300.74/362&42.700&44.700&$-1.29_{-0.10}^{+0.10}$&$62.4_{-4.1}^{+4.5}$&$1.07_{-0.04}^{+0.04}\times10^{-6}$&203.05/362 \\
9.300&9.600&$-1.07_{-0.04}^{+0.04}$&$291.4_{-23.2}^{+26.2}$&$1.35_{-0.05}^{+0.05}\times10^{-5}$&344.43/362&44.700&47.300&$-1.51_{-0.11}^{+0.03}$&$68.1_{-2.8}^{+9.9}$&$1.08_{-0.02}^{+0.07}\times10^{-6}$&234.63/362 \\
9.600&10.000&$-1.21_{-0.05}^{+0.03}$&$307.4_{-24.0}^{+42.0}$&$1.18_{-0.04}^{+0.05}\times10^{-5}$&312.69/362&47.300&50.100&$-1.48_{-0.09}^{+0.08}$&$76.2_{-6.0}^{+9.0}$&$1.06_{-0.04}^{+0.06}\times10^{-6}$&208.91/362 \\
10.000&10.400&$-1.15_{-0.04}^{+0.04}$&$331.0_{-25.3}^{+39.0}$&$1.18_{-0.03}^{+0.05}\times10^{-5}$&330.12/362&50.100&53.600&$-1.44_{-0.09}^{+0.08}$&$63.1_{-3.9}^{+5.4}$&$8.91_{-0.33}^{+0.40}\times10^{-7}$&247.34/362 \\
10.400&10.900&$-1.12_{-0.04}^{+0.04}$&$232.2_{-16.5}^{+22.1}$&$9.57_{-0.30}^{+0.40}\times10^{-6}$&346.45/362&53.600&70.000&$-1.64_{-0.08}^{+0.05}$&$60.8_{-4.6}^{+7.4}$&$4.21_{-0.15}^{+0.24}\times10^{-7}$&218.64/362 \\
\hline
\end{tabular}
\end{table*}

\clearpage

\begin{table*}
\footnotesize
\centering
\caption{\textbf{The best-fit parameters of linear models for $\alpha$--log$F$, log$E_{\rm p}$--log$F$, log$E_{\rm p}$--$\alpha$, and log$E_{\rm p,z}$--log$E_{\rm \gamma, iso}$ correlations.} All errors represent the 1$\sigma$ uncertainties.}
\label{tab:alpha_Ep_flux}
\begin{tabular}{ccccccccc}
\hline
\multirow{2}{*}{Linear model}&\multicolumn{2}{c}{Main emission}&&\multicolumn{2}{c}{Extended emission}&&\multicolumn{2}{c}{Whole burst}\\
\cline{2-3}
\cline{5-6}
\cline{8-9}
&$k$&$b$&&$k$&$b$&&$k$&$b$\\
\hline
$\alpha=k{\rm log}F+b$&$0.26_{-0.02}^{+0.02}$&$0.14_{-0.08}^{+0.07}$&&$0.27_{-0.02}^{+0.03}$&$0.21_{-0.11}^{+0.17}$&&$0.26_{-0.01}^{+0.01}$&$0.15_{-0.05}^{+0.05}$ \\ 
${\rm log}E_{\rm p}=k{\rm log}F+b$&$0.55_{-0.03}^{+0.04}$&$5.26_{-0.15}^{+0.16}$&&$0.55_{-0.02}^{+0.05}$&$5.15_{-0.10}^{+0.25}$&&$0.62_{-0.02}^{+0.02}$&$5.53_{-0.08}^{+0.08}$ \\ 
${\rm log}E_{\rm p}=k\alpha+b$&$1.69_{-0.18}^{+0.20}$&$4.53_{-0.19}^{+0.20}$&&$1.38_{-0.32}^{+0.15}$&$3.92_{-0.41}^{+0.20}$&&$2.08_{-0.14}^{+0.09}$&$4.87_{-0.15}^{+0.11}$ \\ 
${\rm log}E_{\rm p,z}=k{\rm log}E_{\gamma,{\rm iso}}+b$&$1.22_{-0.12}^{+0.12}$&$-58.44_{-6.07}^{+6.04}$&&$0.95_{-0.17}^{+0.22}$&$-44.96_{-10.76}^{+8.42}$&&$1.50_{-0.08}^{+0.09}$&$-72.26_{-4.26}^{+4.03}$ \\ 
\hline
\end{tabular}
\end{table*}

\clearpage

\begin{table*}
\scriptsize
\centering
\caption{\textbf{The temporal and spectral profiles of X-ray afterglows.} All errors represent the 1$\sigma$ uncertainties.}
\label{tab:xprofile}
\begin{tabular}{lcccccccccc}
\hline
\multirow{2}{*}{Phase}&\multicolumn{4}{c}{GRB 211211A}&&\multicolumn{4}{c}{GRB 060614} \\
\cline{2-5}
\cline{7-10}
& start (s) & stop (s) & $\alpha_{\rm X}$ & $\Gamma_{\rm X}$ & & start (s) & stop (s) & $\alpha_{\rm X}$ & $\Gamma_{\rm X}$ \\
\hline
\multirow{2}{*}{I: steep decays} & 70 & 177 & $2.98_{-0.06}^{+0.03}$ & $1.42_{-0.05}^{+0.05}$ & & 97 & 173 & $2.47_{-0.06}^{+0.06}$ & $1.21_{-0.05}^{+0.05}$ \\
& 177 & 296 & $5.16_{-0.13}^{+0.06}$ & $2.48_{-0.06}^{+0.06}$ & & 173 & 464 & $3.90_{-0.03}^{+0.03}$ & $2.28_{-0.04}^{+0.04}$ \\
II: plateau & 3,526 & 6,140 & $-0.17_{-0.55}^{+0.52}$ & $1.47_{-0.08}^{+0.15}$ & & 4,538 & 34,006 & $0.02_{-0.05}^{+0.05}$ & $1.71_{-0.04}^{+0.11}$ \\
III: normal decay & 6,140 & 66,490 & $1.57_{-0.21}^{+0.15}$ & $1.54_{-0.09}^{+0.19}$ & & 34,006 & 152,033 & $1.34_{-0.09}^{+0.10}$ & $1.85_{-0.12}^{+0.13}$ \\
IV: post jet-break decay & 66,490 & 167,615 & $3.33_{-0.42}^{+0.73}$ & $1.40_{-0.30}^{+0.44}$ & & 152,033 & 2,027,831 & $2.17_{-0.12}^{+0.11}$ & $1.63_{-0.18}^{+0.22}$ \\
\hline
\end{tabular}
\end{table*}

\clearpage

\begin{table*}
\tiny
\centering
\caption{\textbf{Ultraviolet, optical, and near-infrared observations of GRB 211211A.}}
\label{tab:uvoniobs}
\renewcommand\arraystretch{0.8}
\begin{tabular}{ccccr}
\hline
$T-T_{0}$ (day) & Telescope & Band & AB Magnitude & Ref. \\
\hline
0.026 & KAIT & Clear ($\sim$R) & $20.30\pm0.20$ & 64 \\
0.042 & Swift/UVOT & U & $19.86\pm0.15$ & 98 \\
0.044 & Swift/UVOT & B & $19.74\pm0.23$ & 98 \\
0.049 & Swift/UVOT & UVW2 & $19.74\pm0.15$ & 98 \\
0.051 & Swift/UVOT & V & $19.27\pm0.30$ & 98 \\
0.053 & Swift/UVOT & UVM2 & $19.79\pm0.18$ & 98 \\
0.056 & Swift/UVOT & UVW1 & $19.55\pm0.14$ & 98 \\
0.057 & Swift/UVOT & U & $19.56\pm0.19$ & 98 \\
0.186 & Swift/UVOT & U & $19.76\pm0.12$ & 98 \\
0.190 & Swift/UVOT & U & $19.83\pm0.12$ & 98 \\
0.193 & Swift/UVOT & U & $19.99\pm0.14$ & 98 \\
0.196 & Swift/UVOT & B & $19.95\pm0.26$ & 98 \\
0.200 & Swift/UVOT & B & $20.03\pm0.33$ & 98 \\
0.254 & Swift/UVOT & UVM2 & $20.63\pm0.17$ & 98 \\
0.263 & MITSuME/Akeno & Ic & $20.40\pm0.30$ & 99 \\
0.263 & MITSuME/Akeno & Rc & $20.30\pm0.10$ & 99 \\
0.263 & MITSuME/Akeno & $\rm g^{\prime}$ & $20.40\pm0.20$ & 99 \\
0.431 & Nanshan/NEXT & r & $20.13\pm0.06$ & 15 \\
0.445 & Nanshan/NEXT & z & $20.04\pm0.23$ & 15 \\
0.456 & HCT & R & $20.30\pm0.13$ & 100 \\
0.656 & Swift/UVOT & UVW1 & $21.93\pm0.27$ & 98 \\
0.663 & Swift/UVOT & U & $>20.62$ & 98 \\
0.680 & CAHA/CAFOS & i & $20.78\pm0.08$ & 10 \\
0.690 & NOT/ALFOSC & g & $21.06\pm0.04$ & 10 \\
0.690 & NOT/ALFOSC & r & $20.85\pm0.05$ & 10 \\
0.700 & NOT/ALFOSC & i & $20.92\pm0.06$ & 10 \\
0.713 & LCO/Sinistro & R & $20.83\pm0.09$ & 101 \\
0.713 & LCO/Sinistro & I & $21.49\pm0.25$ & 101 \\
0.720 & Swift/UVOT & UVW2 & $22.30\pm0.28$ & 98 \\
0.726 & Swift/UVOT & V & $>19.34$ & 98 \\
0.784 & Swift/UVOT & B & $>20.66$ & 98 \\
0.852 & Swift/UVOT & UVW1 & $22.18\pm0.31$ & 98 \\
0.858 & Swift/UVOT & U & $>20.89$ & 98 \\
0.917 & Swift/UVOT & UVW2 & $>22.75$ & 98 \\
0.923 & Swift/UVOT & V & $>19.37$ & 98 \\
1.049 & Swift/UVOT & B & $>20.66$ & 98 \\
1.126 & Swift/UVOT & UVW1 & $>22.10$ & 98 \\
1.182 & Swift/UVOT & U & $>21.39$ & 98 \\
1.259 & Swift/UVOT & UVM2 & $>22.31$ & 98 \\
1.321 & Swift/UVOT & UVW1 & $>22.13$ & 98 \\
1.396 & GMG & r & $>22.00$ & 102 \\
1.410 & DOT & R & $21.66\pm0.07$ & 103 \\
1.425 & GIT & $\rm r^{\prime}$ & $>21.19$ & 100 \\
1.440 & Nanshan/NEXT & r & $>22.06$ & 15 \\
1.441 & Nanshan/NEXT & i & $>21.22$ & 15 \\
1.562 & AbAO & R & $21.60\pm0.33$ & 104 \\
1.680 & CAHA/CAFOS & i & $22.59\pm0.13$ & 10 \\
2.527 & AbAO & R & $>21.80$ & 104 \\
2.558 & Zeiss-1000 & Rc & $23.10\pm0.40$ & 105 \\
2.680 & CAHA/CAFOS & i & $24.59\pm0.34$ & 10 \\
2.912 & Swift/UVOT & UVM2 & $>22.77$ & 98 \\
4.070 & Gemini/NIRI & K & $22.42\pm0.14$ & 10 \\
4.420 & DOT & R & $>23.70$ & 103 \\
4.700 & TNG/NICS & H & $>21.90$ & 106 \\
5.100 & Gemini/NIRI & K & $22.41\pm0.17$ & 10 \\
5.110 & Gemini/GMOS & i & $26.06\pm0.30$ & 10 \\
5.419 & Nanshan/NEXT & r & $>21.34$ & 15 \\
5.428 & Nanshan/NEXT & i & $>20.81$ & 15 \\
5.440 & Nanshan/NEXT & z & $>19.83$ & 15 \\
5.960 & MMT/MMIRS & J & $24.18\pm0.35$ & 10 \\
6.080 & Gemini/GMOS & i & $>25.52$ & 10 \\
6.419 & Nanshan/NEXT & r & $>21.05$ & 15 \\
6.428 & Nanshan/NEXT & i & $>20.64$ & 15 \\
6.440 & Nanshan/NEXT & z & $>19.67$ & 15 \\
6.940 & MMT/MMIRS & K & $23.44\pm0.31$ & 10 \\
7.403 & Nanshan/NEXT & r & $>20.97$ & 15 \\
7.980 & MMT/MMIRS & K & $23.78\pm0.30$ & 10 \\
9.920 & MMT/MMIRS & K & $>22.12$ & 10 \\
10.434 & Nanshan/NEXT & r & $>21.66$ & 15 \\
14.412 & Nanshan/NEXT & r & $>21.93$ & 15 \\
17.650 & NOT/ALFOSC & i & $>24.70$ & 10 \\
19.570 & CAHA/CAFOS & i & $>24.17$ & 10 \\
\hline
\end{tabular}
\end{table*}

\clearpage

\begin{table*}
\centering
\caption{\textbf{The best-fit parameters of the afterglow and engine-fed kilonova models.} All errors represent the 1$\sigma$ uncertainties.}
\label{tab:agmnmodeling}
\begin{tabular}{lcc}
\hline
Parameter & Posterior & Prior bounds \\
\hline
{\bf Afterglow:} & & \\
log$(E_{\rm k,iso}/{\rm erg})$ & $53.2_{-1.2}^{+1.0}$ & $(50, 55)$ \\
log$(\theta_{\rm v}/{\rm rad})$ & $-1.6_{-0.4}^{+0.2}$ & $(-2.2, -1.0)$ \\
log$(\theta_{\rm c}/{\rm rad})$ & $-1.8_{-0.2}^{+0.5}$ & $(-2.5, -1.0)$ \\
log$(n/{\rm cm^{-3}})$ & $-2.8_{-1.2}^{+2.3}$ & $(-6.0, 1.0)$ \\
$p$ & $2.24_{-0.05}^{+0.10}$ & $(2.0, 2.5)$ \\
log$\epsilon_{\rm e}$ & $-2.3_{-0.7}^{+1.2}$ & $(-4.0, -0.5)$ \\
log$\epsilon_{\rm B}$ & $-2.0_{-2.0}^{+0.5}$ & $(-5.0, -0.5)$ \\
\hline
{\bf Kilonova:} & & \\
$P_{0}/{\rm ms}$ & $(1)$ & \\
log$(B/{\rm G})$ & $13.3_{-0.1}^{+0.2}$ & $(13, 14)$ \\
$f_{\rm n}$ & $0.80_{-0.24}^{+0.03}$ & $(0.4, 0.9)$ \\
$\kappa/({\rm cm^2\,g^{-1}})$ & $0.73_{-0.06}^{+0.53}$ & $(0.5, 1.5)$ \\
$M_{\rm ej}/{\rm M_{\odot}}$ & $0.037_{-0.004}^{+0.008}$ & $(0.02, 0.05)$ \\
$v_{\rm ej}/{\rm c}$ & $0.24_{-0.02}^{+0.06}$ & $(0.1, 0.4)$ \\
\hline
log$E(B-V)$ & $-2.1_{-0.5}^{+0.5}$ & $(-3, -1)$ \\
log$(v/{\rm mag})$ & $-0.58_{-0.02}^{+0.10}$ & $(-1.5, 0.0)$ \\
\hline
\end{tabular}
\end{table*}

\clearpage

\begin{figure}
\includegraphics[width=160 mm]{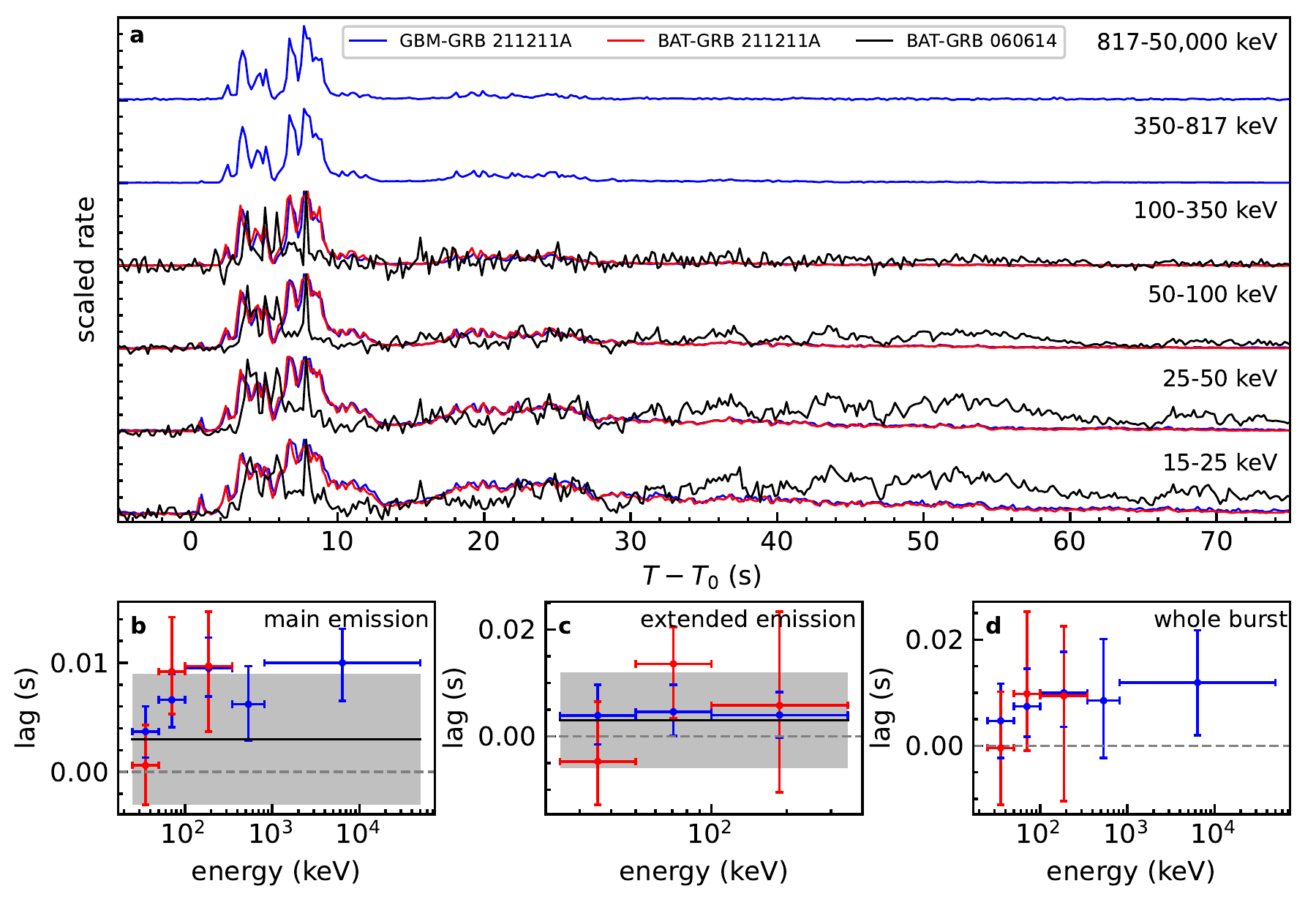}
\caption{\noindent\textbf{The multi-wavelength light curves of prompt emissions and spectral lags.} \textbf{a}, The scaled multi-wavelength light curves obtained from {\it Fermi}/GBM (blue) and {\it Swift}/BAT (red) data for GRB 211211A, and {\it Swift}/BAT (black) data for GRB 060614. For comparison, we reduce the trigger time of GRB 060614 by 5 seconds. \textbf{b}, \textbf{c} and \textbf{d} show the spectral lags between each of the higher energy bands and the lowest energy band calculated for the main emission, extended emission, and whole burst of GRB 211211A. The lags in blue and red are derived from {\it Fermi}/GBM and {\it Swift}/BAT data, respectively. The horizontal black lines and grey shaded areas in \textbf{b} and \textbf{c} show the spectral lags and their uncertainties of GRB 060614, respectively. All error bars on data points represent their 1$\sigma$ confidence level.}
\label{fig:wavelag}
\end{figure}

\begin{figure*}
\centering
\includegraphics[width=160 mm]{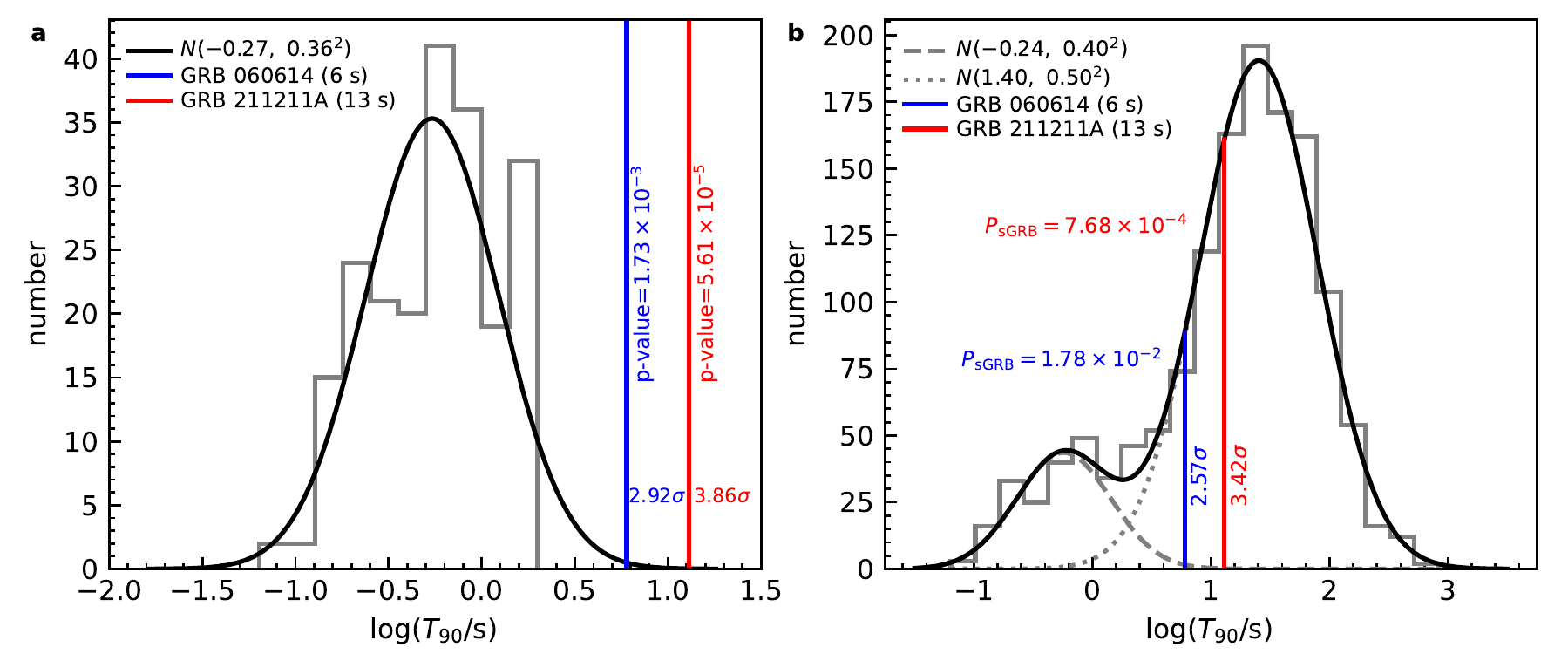}
\caption{\noindent\textbf{The $T_{\rm 90}$ distributions.} \textbf{a}, The $T_{\rm 90}$ distribution (grey hist) of the short GRB ($T_{\rm 90} < 2~{\rm s}$) sample from the fourth {\it Fermi}/GBM catalog \cite{von_2020ApJ} fit with a single log-normal distribution (solid black line). GRB 211211A (with 13-s main emission) and GRB 060614 (with 6-s main emission) are highlighted in red and blue, respectively. \textbf{b}, The $T_{\rm 90}$ distribution (grey hist) of the whole GRB sample from the fourth {\it Fermi}/GBM catalog fit with a two-component log-normal mixture model (solid black line). The two components responsible for short and long GRB populations are shown with grey dashed and dotted lines, respectively. GRB 211211A (with 13-s main emission) and GRB 060614 (with 6-s main emission) are highlighted in red and blue, respectively.}
\label{fig:t90pdf}
\end{figure*}

\begin{figure*}
\includegraphics[width=160 mm]{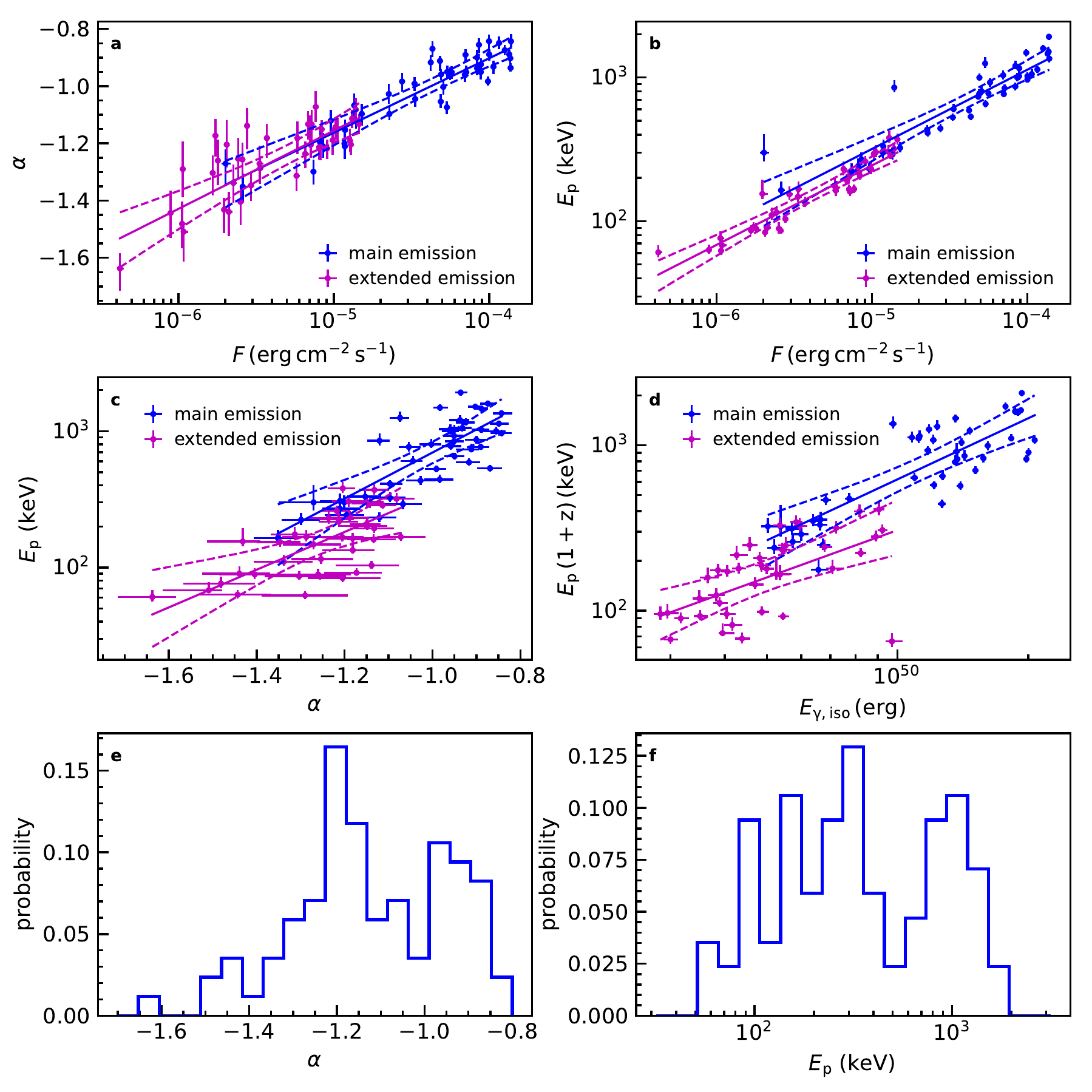}
\caption{\noindent\textbf{The $\alpha$, $E_{\rm p}$, and flux $F$ correlation diagrams.} \textbf{a}, \textbf{b}, \textbf{c} and \textbf{d} show the linear fits to $\alpha-{\rm log}F$, ${\rm log}E_{\rm p}-{\rm log}F$, ${\rm log}E_{\rm p}-\alpha$ and ${\rm log}E_{\rm p,z}-{\rm log}E_{\rm \gamma,iso}$ relations during main emission phase (blue) and extended emission (magenta) phase. The solid and dashed lines show the best-fit correlations and 3$\sigma$ error bands, respectively. All error bars on data points represent their 1$\sigma$ confidence level. \textbf{e} and \textbf{f} show the distributions of the best-fit $\alpha$ and $E_{\rm p}$ obtained from time-resolved spectral fits.}
\label{fig:alpha_Ep_flux}
\end{figure*}

\begin{figure*}
\centering
\includegraphics[width=150 mm]{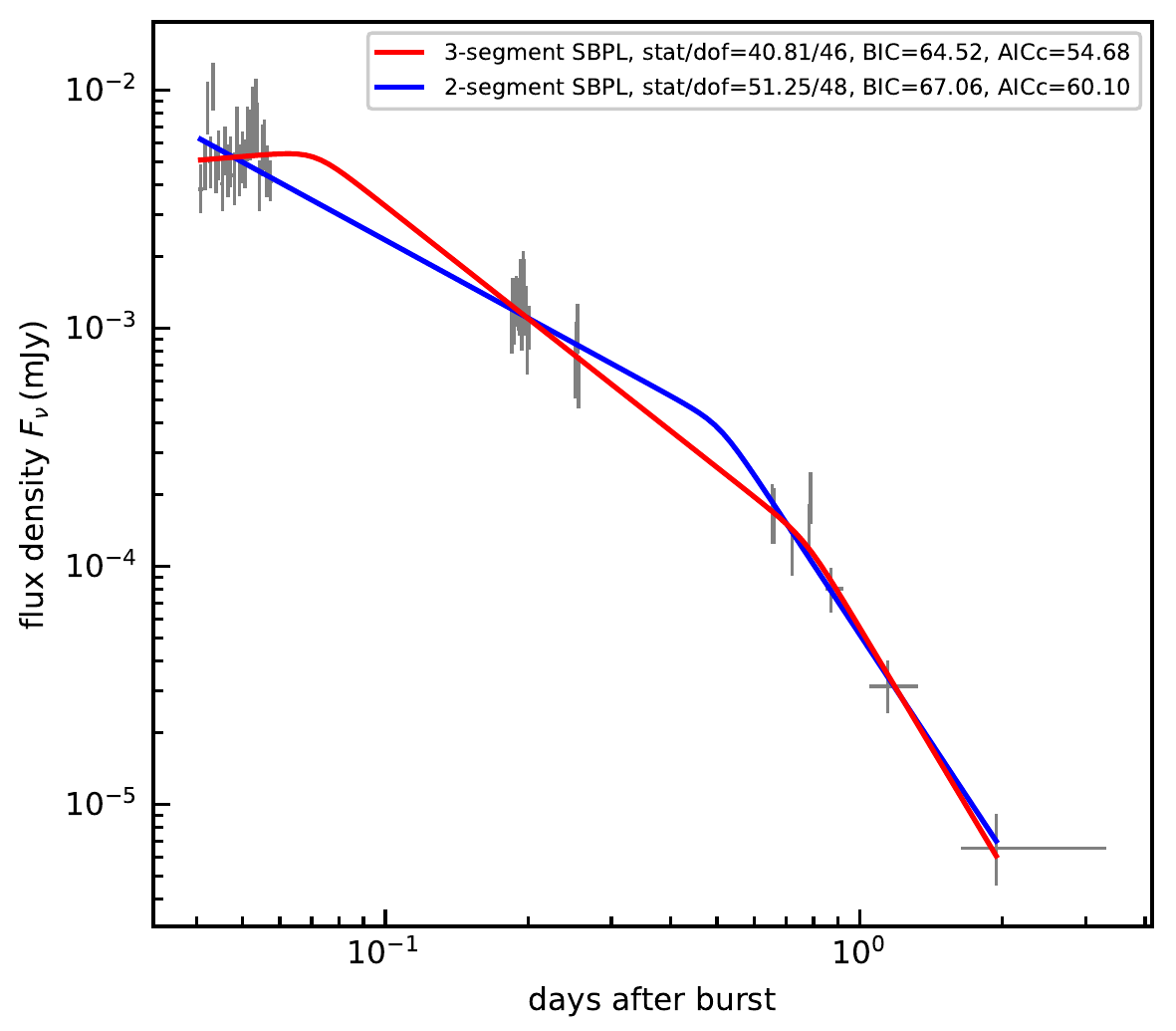}
\caption{\noindent\textbf{The X-ray afterglow of GRB 211211A fit with smoothed broken power-law (SBPL) function.} The X-ray afterglow in PC mode is shown with grey error bars. The 2-segment and 3-segment SBPL models are represented by blue and red lines, respectively. Based on the statistics (stat/dof, BIC, and AICc), the data favors the 3-segment more than the 2-segment SBPL model. The differences in BIC and AICc between the 2-segment and 3-segment SBPL models are $\Delta{\rm BIC}=2.54$ and $\Delta{\rm AICc}=5.42$, respectively. Such ``strength of evidence'' (i.e., $\Delta{\rm BIC}$ and $\Delta{\rm AICc}$) also positively supports the 3-segment SBPL model.}
\label{fig:xrtfit}
\end{figure*}

\begin{figure*}
\centering
\includegraphics[width=160 mm]{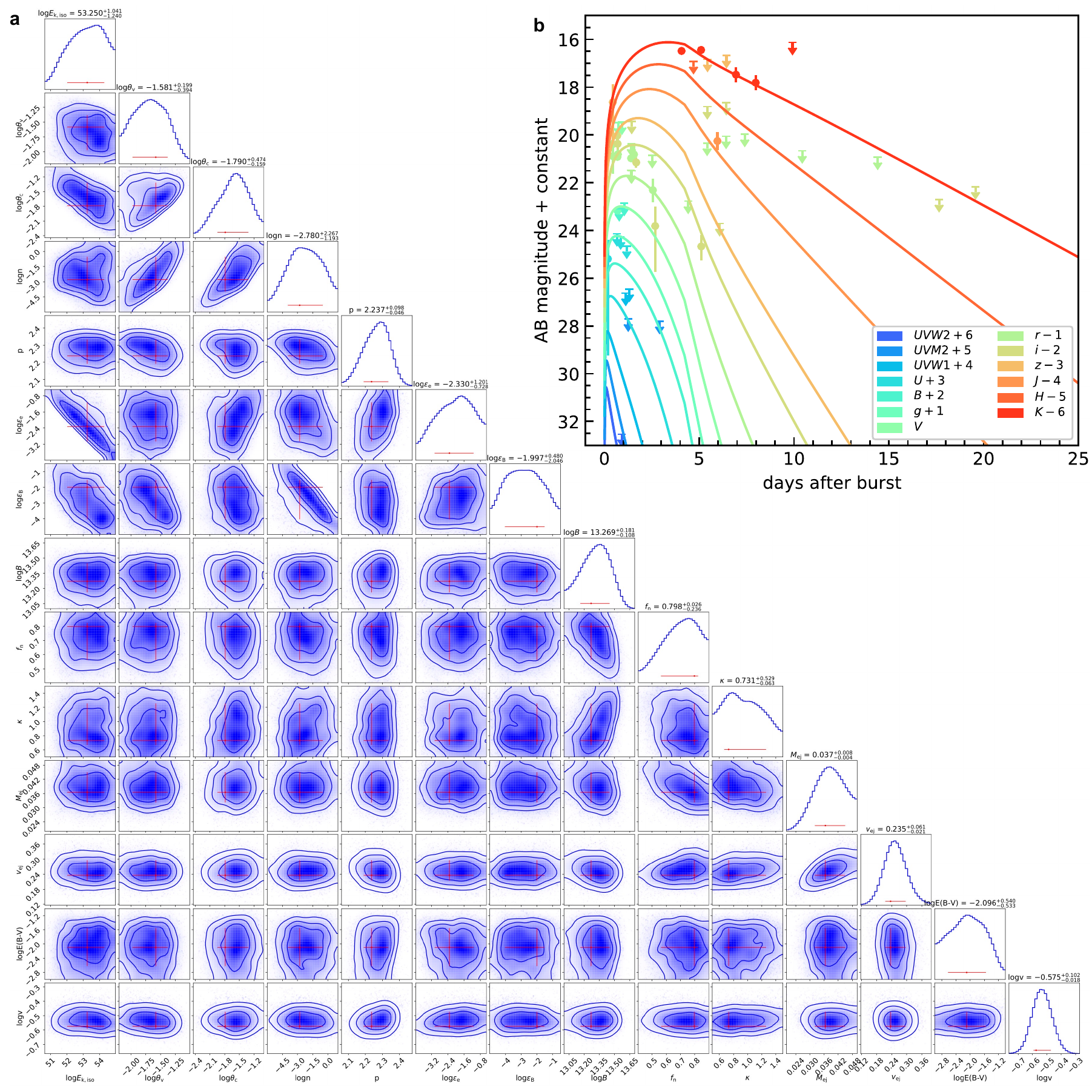}
\caption{\noindent\textbf{The fit of the afterglow plus engine-fed kilonova model to multi-wavelength data.} \textbf{a}, Corner plot of the posterior probability distributions of the parameters. The red error bars represent the 1$\sigma$ uncertainties. \textbf{b}, Afterglow-subtracted observations and best-fitting engine-fed kilonova model. The detections and upper limits of the afterglow-subtracted observations are shown with solid circles and downward arrows, respectively. The best-fit models in different bands are presented with solid lines.}
\label{fig:mnmodeling}
\end{figure*}


\begin{thebibliography}{10}
\expandafter\ifx\csname url\endcsname\relax
  \def\url#1{\texttt{#1}}\fi
\expandafter\ifx\csname urlprefix\endcsname\relax\def\urlprefix{URL }\fi
\providecommand{\bibinfo}[2]{#2}
\providecommand{\eprint}[2][]{\url{#2}}

\bibitem[1]{Woosley_2006ARAA}
\bibinfo{author}{{Woosley}, S.~E.} \& \bibinfo{author}{{Bloom}, J.~S.}
\newblock \bibinfo{title}{{The Supernova Gamma-Ray Burst Connection}}.
\newblock \emph{\bibinfo{journal}{\araa}} \textbf{\bibinfo{volume}{44}},
  \bibinfo{pages}{507--556} (\bibinfo{year}{2006}).

\bibitem[2]{Berger_2014ARAA}
\bibinfo{author}{{Berger}, E.}
\newblock \bibinfo{title}{{Short-Duration Gamma-Ray Bursts}}.
\newblock \emph{\bibinfo{journal}{\araa}} \textbf{\bibinfo{volume}{52}},
  \bibinfo{pages}{43--105} (\bibinfo{year}{2014}).

\bibitem[3]{Gehrels_2006Natur}
\bibinfo{author}{{Gehrels}, N.} \emph{et~al.}
\newblock \bibinfo{title}{{A new {\ensuremath{\gamma}}-ray burst classification
  scheme from GRB060614}}.
\newblock \emph{\bibinfo{journal}{\nat}} \textbf{\bibinfo{volume}{444}},
  \bibinfo{pages}{1044--1046} (\bibinfo{year}{2006}).

\bibitem[4]{Della_2006Natur}
\bibinfo{author}{{Della Valle}, M.} \emph{et~al.}
\newblock \bibinfo{title}{{An enigmatic long-lasting {\ensuremath{\gamma}}-ray
  burst not accompanied by a bright supernova}}.
\newblock \emph{\bibinfo{journal}{\nat}} \textbf{\bibinfo{volume}{444}},
  \bibinfo{pages}{1050--1052} (\bibinfo{year}{2006}).

\bibitem[5]{Zhang_2021NatAs}
\bibinfo{author}{{Zhang}, B.-B.} \emph{et~al.}
\newblock \bibinfo{title}{{A peculiarly short-duration gamma-ray burst from
  massive star core collapse}}.
\newblock \emph{\bibinfo{journal}{Nature Astronomy}}
  \textbf{\bibinfo{volume}{5}}, \bibinfo{pages}{911--916}
  (\bibinfo{year}{2021}).

\bibitem[6]{Ahumada_2021NatAs}
\bibinfo{author}{{Ahumada}, T.} \emph{et~al.}
\newblock \bibinfo{title}{{Discovery and confirmation of the shortest gamma-ray
  burst from a collapsar}}.
\newblock \emph{\bibinfo{journal}{Nature Astronomy}}
  \textbf{\bibinfo{volume}{5}}, \bibinfo{pages}{917--927}
  (\bibinfo{year}{2021}).

\bibitem[7]{Zhang_2009ApJ}
\bibinfo{author}{{Zhang}, B.} \emph{et~al.}
\newblock \bibinfo{title}{{Discerning the Physical Origins of Cosmological
  Gamma-ray Bursts Based on Multiple Observational Criteria: The Cases of z =
  6.7 GRB 080913, z = 8.2 GRB 090423, and Some Short/Hard GRBs}}.
\newblock \emph{\bibinfo{journal}{\apj}} \textbf{\bibinfo{volume}{703}},
  \bibinfo{pages}{1696--1724} (\bibinfo{year}{2009}).

\bibitem[8]{Zhang_2007ApJ}
\bibinfo{author}{{Zhang}, B.} \emph{et~al.}
\newblock \bibinfo{title}{{Making a Short Gamma-Ray Burst from a Long One:
  Implications for the Nature of GRB 060614}}.
\newblock \emph{\bibinfo{journal}{\apjl}} \textbf{\bibinfo{volume}{655}},
  \bibinfo{pages}{L25--L28} (\bibinfo{year}{2007}).

\bibitem[9]{Yang_2015NatCo}
\bibinfo{author}{{Yang}, B.} \emph{et~al.}
\newblock \bibinfo{title}{{A possible macronova in the late afterglow of the
  long-short burst GRB 060614}}.
\newblock \emph{\bibinfo{journal}{Nature Communications}}
  \textbf{\bibinfo{volume}{6}}, \bibinfo{pages}{7323} (\bibinfo{year}{2015}).

\bibitem[10]{Rastinejad_2022arXiv}
\bibinfo{author}{{Rastinejad}, J.~C.} \emph{et~al.}
\newblock \bibinfo{title}{{A Kilonova Following a Long-Duration Gamma-Ray Burst
  at 350 Mpc}}.
\newblock \emph{\bibinfo{journal}{arXiv e-prints}}
  \bibinfo{pages}{arXiv:2204.10864} (\bibinfo{year}{2022}).

\bibitem[11]{Yu_2013ApJ}
\bibinfo{author}{{Yu}, Y.-W.}, \bibinfo{author}{{Zhang}, B.} \&
  \bibinfo{author}{{Gao}, H.}
\newblock \bibinfo{title}{{Bright ``Merger-nova'' from the Remnant of a Neutron
  Star Binary Merger: A Signature of a Newly Born, Massive, Millisecond
  Magnetar}}.
\newblock \emph{\bibinfo{journal}{\apjl}} \textbf{\bibinfo{volume}{776}},
  \bibinfo{pages}{L40} (\bibinfo{year}{2013}).

\bibitem[12]{Ai_2022arXiv}
\bibinfo{author}{{Ai}, S.}, \bibinfo{author}{{Zhang}, B.} \&
  \bibinfo{author}{{Zhu}, Z.}
\newblock \bibinfo{title}{{Engine-fed Kilonovae (Mergernovae) -- I. Dynamical
  Evolution and Energy Injection / Heating Efficiencies}}.
\newblock \emph{\bibinfo{journal}{arXiv e-prints}}
  \bibinfo{pages}{arXiv:2203.03045} (\bibinfo{year}{2022}).

\bibitem[13]{Meegan_2009ApJ}
\bibinfo{author}{{Meegan}, C.} \emph{et~al.}
\newblock \bibinfo{title}{{The Fermi Gamma-ray Burst Monitor}}.
\newblock \emph{\bibinfo{journal}{\apj}} \textbf{\bibinfo{volume}{702}},
  \bibinfo{pages}{791--804} (\bibinfo{year}{2009}).

\bibitem[14]{Barthelmy_2005SSRv}
\bibinfo{author}{{Barthelmy}, S.~D.} \emph{et~al.}
\newblock \bibinfo{title}{{The Burst Alert Telescope (BAT) on the SWIFT Midex
  Mission}}.
\newblock \emph{\bibinfo{journal}{\ssr}} \textbf{\bibinfo{volume}{120}},
  \bibinfo{pages}{143--164} (\bibinfo{year}{2005}).

\bibitem[15]{Xiao2022arXiv}
\bibinfo{author}{{Xiao}, S.} \emph{et~al.}
\newblock \bibinfo{title}{{The quasi-periodically oscillating precursor of a
  long gamma-ray burst from a binary neutron star merger}}.
\newblock \emph{\bibinfo{journal}{arXiv e-prints}}
  \bibinfo{pages}{arXiv:2205.02186} (\bibinfo{year}{2022}).

\bibitem[16]{Band_1993ApJ}
\bibinfo{author}{{Band}, D.} \emph{et~al.}
\newblock \bibinfo{title}{{BATSE Observations of Gamma-Ray Burst Spectra. I.
  Spectral Diversity}}.
\newblock \emph{\bibinfo{journal}{\apj}} \textbf{\bibinfo{volume}{413}},
  \bibinfo{pages}{281} (\bibinfo{year}{1993}).

\bibitem[17]{Amati_2002A&A}
\bibinfo{author}{{Amati}, L.} \emph{et~al.}
\newblock \bibinfo{title}{{Intrinsic spectra and energetics of BeppoSAX
  Gamma-Ray Bursts with known redshifts}}.
\newblock \emph{\bibinfo{journal}{\aap}} \textbf{\bibinfo{volume}{390}},
  \bibinfo{pages}{81--89} (\bibinfo{year}{2002}).

\bibitem[18]{Li_1998ApJ}
\bibinfo{author}{{Li}, L.-X.} \& \bibinfo{author}{{Paczy{\'n}ski}, B.}
\newblock \bibinfo{title}{{Transient Events from Neutron Star Mergers}}.
\newblock \emph{\bibinfo{journal}{\apjl}} \textbf{\bibinfo{volume}{507}},
  \bibinfo{pages}{L59--L62} (\bibinfo{year}{1998}).

\bibitem[19]{Metzger_2010MNRAS}
\bibinfo{author}{{Metzger}, B.~D.} \emph{et~al.}
\newblock \bibinfo{title}{{Electromagnetic counterparts of compact object
  mergers powered by the radioactive decay of r-process nuclei}}.
\newblock \emph{\bibinfo{journal}{\mnras}} \textbf{\bibinfo{volume}{406}},
  \bibinfo{pages}{2650--2662} (\bibinfo{year}{2010}).

\bibitem[20]{Lu_2014MNRAS}
\bibinfo{author}{{L{\"u}}, H.-J.}, \bibinfo{author}{{Zhang}, B.},
  \bibinfo{author}{{Liang}, E.-W.}, \bibinfo{author}{{Zhang}, B.-B.} \&
  \bibinfo{author}{{Sakamoto}, T.}
\newblock \bibinfo{title}{{The `amplitude' parameter of gamma-ray bursts and
  its implications for GRB classification}}.
\newblock \emph{\bibinfo{journal}{\mnras}} \textbf{\bibinfo{volume}{442}},
  \bibinfo{pages}{1922--1929} (\bibinfo{year}{2014}).

\bibitem[21]{zhang_2018book}
\bibinfo{author}{{Zhang}, B.}
\newblock \emph{\bibinfo{title}{{The Physics of Gamma-Ray Bursts}}}
  (\bibinfo{year}{2018}).

\bibitem[22]{Kluzniak_1998ApJ}
\bibinfo{author}{{Klu{\'z}niak}, W.} \& \bibinfo{author}{{Ruderman}, M.}
\newblock \bibinfo{title}{{The Central Engine of Gamma-Ray Bursters}}.
\newblock \emph{\bibinfo{journal}{\apjl}} \textbf{\bibinfo{volume}{505}},
  \bibinfo{pages}{L113--L117} (\bibinfo{year}{1998}).

\bibitem[23]{Ruderman_2000ApJ}
\bibinfo{author}{{Ruderman}, M.~A.}, \bibinfo{author}{{Tao}, L.} \&
  \bibinfo{author}{{Klu{\'z}niak}, W.}
\newblock \bibinfo{title}{{A Central Engine for Cosmic Gamma-Ray Burst
  Sources}}.
\newblock \emph{\bibinfo{journal}{\apj}} \textbf{\bibinfo{volume}{542}},
  \bibinfo{pages}{243--250} (\bibinfo{year}{2000}).

\bibitem[24]{Dai_2006Sci}
\bibinfo{author}{{Dai}, Z.~G.}, \bibinfo{author}{{Wang}, X.~Y.},
  \bibinfo{author}{{Wu}, X.~F.} \& \bibinfo{author}{{Zhang}, B.}
\newblock \bibinfo{title}{{X-ray Flares from Postmerger Millisecond Pulsars}}.
\newblock \emph{\bibinfo{journal}{Science}} \textbf{\bibinfo{volume}{311}},
  \bibinfo{pages}{1127--1129} (\bibinfo{year}{2006}).

\bibitem[25]{Toonen_2018AA}
\bibinfo{author}{{Toonen}, S.}, \bibinfo{author}{{Perets}, H.~B.},
  \bibinfo{author}{{Igoshev}, A.~P.}, \bibinfo{author}{{Michaely}, E.} \&
  \bibinfo{author}{{Zenati}, Y.}
\newblock \bibinfo{title}{{The demographics of neutron star - white dwarf
  mergers. Rates, delay-time distributions, and progenitors}}.
\newblock \emph{\bibinfo{journal}{\aap}} \textbf{\bibinfo{volume}{619}},
  \bibinfo{pages}{A53} (\bibinfo{year}{2018}).

\bibitem[26]{Buikema_2020PhRvD}
\bibinfo{author}{{Buikema}, A.} \emph{et~al.}
\newblock \bibinfo{title}{{Sensitivity and performance of the Advanced LIGO
  detectors in the third observing run}}.
\newblock \emph{\bibinfo{journal}{\prd}} \textbf{\bibinfo{volume}{102}},
  \bibinfo{pages}{062003} (\bibinfo{year}{2020}).

\bibitem[27]{Bersanetti_2021Univ}
\bibinfo{author}{{Bersanetti}, D.} \emph{et~al.}
\newblock \bibinfo{title}{{Advanced Virgo: Status of the Detector, Latest
  Results and Future Prospects}}.
\newblock \emph{\bibinfo{journal}{Universe}} \textbf{\bibinfo{volume}{7}},
  \bibinfo{pages}{322} (\bibinfo{year}{2021}).

\bibitem[28]{Kagra_2019NatAs}
\bibinfo{author}{{Kagra Collaboration}} \emph{et~al.}
\newblock \bibinfo{title}{{KAGRA: 2.5 generation interferometric gravitational
  wave detector}}.
\newblock \emph{\bibinfo{journal}{Nature Astronomy}}
  \textbf{\bibinfo{volume}{3}}, \bibinfo{pages}{35--40} (\bibinfo{year}{2019}).

\bibitem[29]{Amaro_2017arXiv}
\bibinfo{author}{{Amaro-Seoane}, P.} \emph{et~al.}
\newblock \bibinfo{title}{{Laser Interferometer Space Antenna}}.
\newblock \emph{\bibinfo{journal}{arXiv e-prints}}
  \bibinfo{pages}{arXiv:1702.00786} (\bibinfo{year}{2017}).

\bibitem[30]{Luo_2020ResPh}
\bibinfo{author}{{Luo}, Z.}, \bibinfo{author}{{Guo}, Z.},
  \bibinfo{author}{{Jin}, G.}, \bibinfo{author}{{Wu}, Y.} \&
  \bibinfo{author}{{Hu}, W.}
\newblock \bibinfo{title}{{A brief analysis to Taiji: Science and technology}}.
\newblock \emph{\bibinfo{journal}{Results in Physics}}
  \textbf{\bibinfo{volume}{16}}, \bibinfo{pages}{102918}
  (\bibinfo{year}{2020}).

\bibitem[31]{Luo_2016CQGra}
\bibinfo{author}{{Luo}, J.} \emph{et~al.}
\newblock \bibinfo{title}{{TianQin: a space-borne gravitational wave
  detector}}.
\newblock \emph{\bibinfo{journal}{Classical and Quantum Gravity}}
  \textbf{\bibinfo{volume}{33}}, \bibinfo{pages}{035010}
  (\bibinfo{year}{2016}).

\bibitem[32]{GCN5264}
\bibinfo{author}{{Golenetskii}, S.} \emph{et~al.}
\newblock \bibinfo{title}{{Konus-wind observation of GRB 060614.}}
\newblock \emph{\bibinfo{journal}{GRB Coordinates Network}}
  \textbf{\bibinfo{volume}{5264}}, \bibinfo{pages}{1} (\bibinfo{year}{2006}).

\bibitem[33]{Amati_2007AA}
\bibinfo{author}{{Amati}, L.} \emph{et~al.}
\newblock \bibinfo{title}{{On the consistency of peculiar GRBs 060218 and
  060614 with the E\_p,i - E\_iso correlation}}.
\newblock \emph{\bibinfo{journal}{\aap}} \textbf{\bibinfo{volume}{463}},
  \bibinfo{pages}{913--919} (\bibinfo{year}{2007}).

\bibitem[34]{Blanchard_2016ApJ}
\bibinfo{author}{{Blanchard}, P.~K.}, \bibinfo{author}{{Berger}, E.} \&
  \bibinfo{author}{{Fong}, W.-f.}
\newblock \bibinfo{title}{{The Offset and Host Light Distributions of Long
  Gamma-Ray Bursts: A New View From HST Observations of Swift Bursts}}.
\newblock \emph{\bibinfo{journal}{\apj}} \textbf{\bibinfo{volume}{817}},
  \bibinfo{pages}{144} (\bibinfo{year}{2016}).
\end{thebibliography}

\begin{thebibliography}{10}
\expandafter\ifx\csname url\endcsname\relax
  \def\url#1{\texttt{#1}}\fi
\expandafter\ifx\csname urlprefix\endcsname\relax\def\urlprefix{URL }\fi
\providecommand{\bibinfo}[2]{#2}
\providecommand{\eprint}[2][]{\url{#2}}

\bibitem[35]{von_2020ApJ}
\bibinfo{author}{{von Kienlin}, A.} \emph{et~al.}
\newblock \bibinfo{title}{{The Fourth Fermi-GBM Gamma-Ray Burst Catalog: A
  Decade of Data}}.
\newblock \emph{\bibinfo{journal}{\apj}} \textbf{\bibinfo{volume}{893}},
  \bibinfo{pages}{46} (\bibinfo{year}{2020}).

\bibitem[36]{Scargle_2013ApJ}
\bibinfo{author}{{Scargle}, J.~D.}, \bibinfo{author}{{Norris}, J.~P.},
  \bibinfo{author}{{Jackson}, B.} \& \bibinfo{author}{{Chiang}, J.}
\newblock \bibinfo{title}{{Studies in Astronomical Time Series Analysis. VI.
  Bayesian Block Representations}}.
\newblock \emph{\bibinfo{journal}{\apj}} \textbf{\bibinfo{volume}{764}},
  \bibinfo{pages}{167} (\bibinfo{year}{2013}).

\bibitem[37]{Vianello_2018ApJ}
\bibinfo{author}{{Vianello}, G.} \emph{et~al.}
\newblock \bibinfo{title}{{The Bright and the Slow{\textemdash}GRBs 100724B and
  160509A with High-energy Cutoffs at {\ensuremath{\lesssim}}100 MeV}}.
\newblock \emph{\bibinfo{journal}{\apj}} \textbf{\bibinfo{volume}{864}},
  \bibinfo{pages}{163} (\bibinfo{year}{2018}).

\bibitem[38]{Zhang_2018NatAs}
\bibinfo{author}{{Zhang}, B.~B.} \emph{et~al.}
\newblock \bibinfo{title}{{Transition from fireball to Poynting-flux-dominated
  outflow in the three-episode GRB 160625B}}.
\newblock \emph{\bibinfo{journal}{Nature Astronomy}}
  \textbf{\bibinfo{volume}{2}}, \bibinfo{pages}{69--75} (\bibinfo{year}{2018}).

\bibitem[39]{Burgess_2018MNRAS}
\bibinfo{author}{{Burgess}, J.~M.}, \bibinfo{author}{{Yu}, H.-F.},
  \bibinfo{author}{{Greiner}, J.} \& \bibinfo{author}{{Mortlock}, D.~J.}
\newblock \bibinfo{title}{{Awakening the BALROG: BAyesian Location
  Reconstruction Of GRBs}}.
\newblock \emph{\bibinfo{journal}{\mnras}} \textbf{\bibinfo{volume}{476}},
  \bibinfo{pages}{1427--1444} (\bibinfo{year}{2018}).

\bibitem[40]{Berlato_2019ApJ}
\bibinfo{author}{{Berlato}, F.}, \bibinfo{author}{{Greiner}, J.} \&
  \bibinfo{author}{{Burgess}, J.~M.}
\newblock \bibinfo{title}{{Improved Fermi-GBM GRB Localizations Using BALROG}}.
\newblock \emph{\bibinfo{journal}{\apj}} \textbf{\bibinfo{volume}{873}},
  \bibinfo{pages}{60} (\bibinfo{year}{2019}).

\bibitem[41]{Feroz_2008MNRAS}
\bibinfo{author}{{Feroz}, F.} \& \bibinfo{author}{{Hobson}, M.~P.}
\newblock \bibinfo{title}{{Multimodal nested sampling: an efficient and robust
  alternative to Markov Chain Monte Carlo methods for astronomical data
  analyses}}.
\newblock \emph{\bibinfo{journal}{\mnras}} \textbf{\bibinfo{volume}{384}},
  \bibinfo{pages}{449--463} (\bibinfo{year}{2008}).

\bibitem[42]{Feroz_2009MNRAS}
\bibinfo{author}{{Feroz}, F.}, \bibinfo{author}{{Hobson}, M.~P.} \&
  \bibinfo{author}{{Bridges}, M.}
\newblock \bibinfo{title}{{MULTINEST: an efficient and robust Bayesian
  inference tool for cosmology and particle physics}}.
\newblock \emph{\bibinfo{journal}{\mnras}} \textbf{\bibinfo{volume}{398}},
  \bibinfo{pages}{1601--1614} (\bibinfo{year}{2009}).

\bibitem[43]{Buchner_2014A&A}
\bibinfo{author}{{Buchner}, J.} \emph{et~al.}
\newblock \bibinfo{title}{{X-ray spectral modelling of the AGN obscuring region
  in the CDFS: Bayesian model selection and catalogue}}.
\newblock \emph{\bibinfo{journal}{\aap}} \textbf{\bibinfo{volume}{564}},
  \bibinfo{pages}{A125} (\bibinfo{year}{2014}).

\bibitem[44]{Feroz_2019OJAp}
\bibinfo{author}{{Feroz}, F.}, \bibinfo{author}{{Hobson}, M.~P.},
  \bibinfo{author}{{Cameron}, E.} \& \bibinfo{author}{{Pettitt}, A.~N.}
\newblock \bibinfo{title}{{Importance Nested Sampling and the MultiNest
  Algorithm}}.
\newblock \emph{\bibinfo{journal}{The Open Journal of Astrophysics}}
  \textbf{\bibinfo{volume}{2}}, \bibinfo{pages}{10} (\bibinfo{year}{2019}).

\bibitem[45]{Arnaud_1996ASPC}
\bibinfo{author}{{Arnaud}, K.~A.}
\newblock \bibinfo{title}{{XSPEC: The First Ten Years}}.
\newblock In \bibinfo{editor}{{Jacoby}, G.~H.} \& \bibinfo{editor}{{Barnes},
  J.} (eds.) \emph{\bibinfo{booktitle}{Astronomical Data Analysis Software and
  Systems V}}, vol. \bibinfo{volume}{101} of
  \emph{\bibinfo{series}{Astronomical Society of the Pacific Conference
  Series}}, \bibinfo{pages}{17} (\bibinfo{year}{1996}).

\bibitem[46]{Lu_2012ApJ}
\bibinfo{author}{{Lu}, R.-J.} \emph{et~al.}
\newblock \bibinfo{title}{{A Comprehensive Analysis of Fermi Gamma-Ray Burst
  Data. II. E $_{p}$ Evolution Patterns and Implications for the Observed
  Spectrum-Luminosity Relations}}.
\newblock \emph{\bibinfo{journal}{\apj}} \textbf{\bibinfo{volume}{756}},
  \bibinfo{pages}{112} (\bibinfo{year}{2012}).

\bibitem[47]{Li_2019ApJ}
\bibinfo{author}{{Li}, L.} \emph{et~al.}
\newblock
  \bibinfo{title}{{{\textquotedblleft}Double-tracking{\textquotedblright}
  Characteristics of the Spectral Evolution of GRB 131231A: Synchrotron
  Origin?}}
\newblock \emph{\bibinfo{journal}{\apj}} \textbf{\bibinfo{volume}{884}},
  \bibinfo{pages}{109} (\bibinfo{year}{2019}).

\bibitem[48]{Foreman_2013PASP}
\bibinfo{author}{{Foreman-Mackey}, D.}, \bibinfo{author}{{Hogg}, D.~W.},
  \bibinfo{author}{{Lang}, D.} \& \bibinfo{author}{{Goodman}, J.}
\newblock \bibinfo{title}{{emcee: The MCMC Hammer}}.
\newblock \emph{\bibinfo{journal}{\pasp}} \textbf{\bibinfo{volume}{125}},
  \bibinfo{pages}{306} (\bibinfo{year}{2013}).

\bibitem[49]{Preece_1998ApJ}
\bibinfo{author}{{Preece}, R.~D.} \emph{et~al.}
\newblock \bibinfo{title}{{The Synchrotron Shock Model Confronts a ``Line of
  Death'' in the BATSE Gamma-Ray Burst Data}}.
\newblock \emph{\bibinfo{journal}{\apjl}} \textbf{\bibinfo{volume}{506}},
  \bibinfo{pages}{L23--L26} (\bibinfo{year}{1998}).

\bibitem[50]{Meszaros_2000ApJ}
\bibinfo{author}{{M{\'e}sz{\'a}ros}, P.} \& \bibinfo{author}{{Rees}, M.~J.}
\newblock \bibinfo{title}{{Steep Slopes and Preferred Breaks in Gamma-Ray Burst
  Spectra: The Role of Photospheres and Comptonization}}.
\newblock \emph{\bibinfo{journal}{\apj}} \textbf{\bibinfo{volume}{530}},
  \bibinfo{pages}{292--298} (\bibinfo{year}{2000}).

\bibitem[51]{Deng_2014ApJ}
\bibinfo{author}{{Deng}, W.} \& \bibinfo{author}{{Zhang}, B.}
\newblock \bibinfo{title}{{Low Energy Spectral Index and E$_{p}$ Evolution of
  Quasi-thermal Photosphere Emission of Gamma-Ray Bursts}}.
\newblock \emph{\bibinfo{journal}{\apj}} \textbf{\bibinfo{volume}{785}},
  \bibinfo{pages}{112} (\bibinfo{year}{2014}).

\bibitem[52]{Zhang_2002ApJ}
\bibinfo{author}{{Zhang}, B.} \& \bibinfo{author}{{M{\'e}sz{\'a}ros}, P.}
\newblock \bibinfo{title}{{An Analysis of Gamma-Ray Burst Spectral Break
  Models}}.
\newblock \emph{\bibinfo{journal}{\apj}} \textbf{\bibinfo{volume}{581}},
  \bibinfo{pages}{1236--1247} (\bibinfo{year}{2002}).

\bibitem[53]{Uhm_2014NatPh}
\bibinfo{author}{{Uhm}, Z.~L.} \& \bibinfo{author}{{Zhang}, B.}
\newblock \bibinfo{title}{{Fast-cooling synchrotron radiation in a decaying
  magnetic field and {\ensuremath{\gamma}}-ray burst emission mechanism}}.
\newblock \emph{\bibinfo{journal}{Nature Physics}}
  \textbf{\bibinfo{volume}{10}}, \bibinfo{pages}{351--356}
  (\bibinfo{year}{2014}).

\bibitem[54]{Zhang_2011ApJ}
\bibinfo{author}{{Zhang}, B.} \& \bibinfo{author}{{Yan}, H.}
\newblock \bibinfo{title}{{The Internal-collision-induced Magnetic Reconnection
  and Turbulence (ICMART) Model of Gamma-ray Bursts}}.
\newblock \emph{\bibinfo{journal}{\apj}} \textbf{\bibinfo{volume}{726}},
  \bibinfo{pages}{90} (\bibinfo{year}{2011}).

\bibitem[55]{Yi_2006MNRAS}
\bibinfo{author}{{Yi}, T.}, \bibinfo{author}{{Liang}, E.},
  \bibinfo{author}{{Qin}, Y.} \& \bibinfo{author}{{Lu}, R.}
\newblock \bibinfo{title}{{On the spectral lags of the short gamma-ray
  bursts}}.
\newblock \emph{\bibinfo{journal}{\mnras}} \textbf{\bibinfo{volume}{367}},
  \bibinfo{pages}{1751--1756} (\bibinfo{year}{2006}).

\bibitem[56]{Bernardini_2015MNRAS}
\bibinfo{author}{{Bernardini}, M.~G.} \emph{et~al.}
\newblock \bibinfo{title}{{Comparing the spectral lag of short and long
  gamma-ray bursts and its relation with the luminosity}}.
\newblock \emph{\bibinfo{journal}{\mnras}} \textbf{\bibinfo{volume}{446}},
  \bibinfo{pages}{1129--1138} (\bibinfo{year}{2015}).

\bibitem[57]{Norris_2000ApJ}
\bibinfo{author}{{Norris}, J.~P.}, \bibinfo{author}{{Marani}, G.~F.} \&
  \bibinfo{author}{{Bonnell}, J.~T.}
\newblock \bibinfo{title}{{Connection between Energy-dependent Lags and Peak
  Luminosity in Gamma-Ray Bursts}}.
\newblock \emph{\bibinfo{journal}{\apj}} \textbf{\bibinfo{volume}{534}},
  \bibinfo{pages}{248--257} (\bibinfo{year}{2000}).

\bibitem[58]{Ukwatta_2010ApJ}
\bibinfo{author}{{Ukwatta}, T.~N.} \emph{et~al.}
\newblock \bibinfo{title}{{Spectral Lags and the Lag-Luminosity Relation: An
  Investigation with Swift BAT Gamma-ray Bursts}}.
\newblock \emph{\bibinfo{journal}{\apj}} \textbf{\bibinfo{volume}{711}},
  \bibinfo{pages}{1073--1086} (\bibinfo{year}{2010}).

\bibitem[59]{Zhang_2012ApJ}
\bibinfo{author}{{Zhang}, B.-B.} \emph{et~al.}
\newblock \bibinfo{title}{{Unusual Central Engine Activity in the Double Burst
  GRB 110709B}}.
\newblock \emph{\bibinfo{journal}{\apj}} \textbf{\bibinfo{volume}{748}},
  \bibinfo{pages}{132} (\bibinfo{year}{2012}).

\bibitem[60]{Shao_2017ApJ}
\bibinfo{author}{{Shao}, L.} \emph{et~al.}
\newblock \bibinfo{title}{{A New Measurement of the Spectral Lag of Gamma-Ray
  Bursts and its Implications for Spectral Evolution Behaviors}}.
\newblock \emph{\bibinfo{journal}{\apj}} \textbf{\bibinfo{volume}{844}},
  \bibinfo{pages}{126} (\bibinfo{year}{2017}).

\bibitem[61]{GCN31221}
\bibinfo{author}{{Malesani}, D.~B.} \emph{et~al.}
\newblock \bibinfo{title}{{GRB 211211A: NOT optical spectroscopy}}.
\newblock \emph{\bibinfo{journal}{GRB Coordinates Network}}
  \textbf{\bibinfo{volume}{31221}}, \bibinfo{pages}{1} (\bibinfo{year}{2021}).

\bibitem[62]{GCN31230}
\bibinfo{author}{{Minaev}, P.}, \bibinfo{author}{{Pozanenko}, A.} \&
  \bibinfo{author}{{GRB IKI FuN}}.
\newblock \bibinfo{title}{{GRB 211211A: redshift estimation and
  SPI-ACS/INTEGRAL detection}}.
\newblock \emph{\bibinfo{journal}{GRB Coordinates Network}}
  \textbf{\bibinfo{volume}{31230}}, \bibinfo{pages}{1} (\bibinfo{year}{2021}).

\bibitem[63]{GCN31235}
\bibinfo{author}{{Levan}, A.~J.} \emph{et~al.}
\newblock \bibinfo{title}{{GRB 211211A - Gemini K-band detection}}.
\newblock \emph{\bibinfo{journal}{GRB Coordinates Network}}
  \textbf{\bibinfo{volume}{31235}}, \bibinfo{pages}{1} (\bibinfo{year}{2021}).

\bibitem[64]{GCN31203}
\bibinfo{author}{{Zheng}, W.}, \bibinfo{author}{{Filippenko}, A.~V.} \&
  \bibinfo{author}{{KAIT GRB Team}}.
\newblock \bibinfo{title}{{GRB 211211A: KAIT Optical Afterglow Candidate}}.
\newblock \emph{\bibinfo{journal}{GRB Coordinates Network}}
  \textbf{\bibinfo{volume}{31203}}, \bibinfo{pages}{1} (\bibinfo{year}{2021}).

\bibitem[65]{Adelman_2008ApJS}
\bibinfo{author}{{Adelman-McCarthy}, J.~K.} \emph{et~al.}
\newblock \bibinfo{title}{{The Sixth Data Release of the Sloan Digital Sky
  Survey}}.
\newblock \emph{\bibinfo{journal}{\apjs}} \textbf{\bibinfo{volume}{175}},
  \bibinfo{pages}{297--313} (\bibinfo{year}{2008}).

\bibitem[66]{Bloom_2002AJ}
\bibinfo{author}{{Bloom}, J.~S.}, \bibinfo{author}{{Kulkarni}, S.~R.} \&
  \bibinfo{author}{{Djorgovski}, S.~G.}
\newblock \bibinfo{title}{{The Observed Offset Distribution of Gamma-Ray Bursts
  from Their Host Galaxies: A Robust Clue to the Nature of the Progenitors}}.
\newblock \emph{\bibinfo{journal}{\aj}} \textbf{\bibinfo{volume}{123}},
  \bibinfo{pages}{1111--1148} (\bibinfo{year}{2002}).

\bibitem[67]{Stalder_2017ApJ}
\bibinfo{author}{{Stalder}, B.} \emph{et~al.}
\newblock \bibinfo{title}{{Observations of the GRB Afterglow ATLAS17aeu and Its
  Possible Association with GW 170104}}.
\newblock \emph{\bibinfo{journal}{\apj}} \textbf{\bibinfo{volume}{850}},
  \bibinfo{pages}{149} (\bibinfo{year}{2017}).

\bibitem[68]{Lien_2016ApJ}
\bibinfo{author}{{Lien}, A.} \emph{et~al.}
\newblock \bibinfo{title}{{The Third Swift Burst Alert Telescope Gamma-Ray
  Burst Catalog}}.
\newblock \emph{\bibinfo{journal}{\apj}} \textbf{\bibinfo{volume}{829}},
  \bibinfo{pages}{7} (\bibinfo{year}{2016}).

\bibitem[69]{Bhat_2016ApJS}
\bibinfo{author}{{Narayana Bhat}, P.} \emph{et~al.}
\newblock \bibinfo{title}{{The Third Fermi GBM Gamma-Ray Burst Catalog: The
  First Six Years}}.
\newblock \emph{\bibinfo{journal}{\apjs}} \textbf{\bibinfo{volume}{223}},
  \bibinfo{pages}{28} (\bibinfo{year}{2016}).

\bibitem[70]{GCN31201}
\bibinfo{author}{{Fermi GBM Team}}.
\newblock \bibinfo{title}{{GRB 211211A: Fermi GBM Final Real-time
  Localization}}.
\newblock \emph{\bibinfo{journal}{GRB Coordinates Network}}
  \textbf{\bibinfo{volume}{31201}}, \bibinfo{pages}{1} (\bibinfo{year}{2021}).

\bibitem[71]{Kann_2011ApJ}
\bibinfo{author}{{Kann}, D.~A.} \emph{et~al.}
\newblock \bibinfo{title}{{The Afterglows of Swift-era Gamma-Ray Bursts. II.
  Type I GRB versus Type II GRB Optical Afterglows}}.
\newblock \emph{\bibinfo{journal}{\apj}} \textbf{\bibinfo{volume}{734}},
  \bibinfo{pages}{96} (\bibinfo{year}{2011}).

\bibitem[72]{Berger_2010ApJ}
\bibinfo{author}{{Berger}, E.}
\newblock \bibinfo{title}{{A Short Gamma-ray Burst ``No-host'' Problem?
  Investigating Large Progenitor Offsets for Short GRBs with Optical
  Afterglows}}.
\newblock \emph{\bibinfo{journal}{\apj}} \textbf{\bibinfo{volume}{722}},
  \bibinfo{pages}{1946--1961} (\bibinfo{year}{2010}).

\bibitem[73]{Hogg_1997MNRAS}
\bibinfo{author}{{Hogg}, D.~W.} \emph{et~al.}
\newblock \bibinfo{title}{{Counts and colours of faint galaxies in the U and R
  bands}}.
\newblock \emph{\bibinfo{journal}{\mnras}} \textbf{\bibinfo{volume}{288}},
  \bibinfo{pages}{404--410} (\bibinfo{year}{1997}).

\bibitem[74]{Beckwith_2006AJ}
\bibinfo{author}{{Beckwith}, S. V.~W.} \emph{et~al.}
\newblock \bibinfo{title}{{The Hubble Ultra Deep Field}}.
\newblock \emph{\bibinfo{journal}{\aj}} \textbf{\bibinfo{volume}{132}},
  \bibinfo{pages}{1729--1755} (\bibinfo{year}{2006}).

\bibitem[75]{Fong_2013ApJ}
\bibinfo{author}{{Fong}, W.} \& \bibinfo{author}{{Berger}, E.}
\newblock \bibinfo{title}{{The Locations of Short Gamma-Ray Bursts as Evidence
  for Compact Object Binary Progenitors}}.
\newblock \emph{\bibinfo{journal}{\apj}} \textbf{\bibinfo{volume}{776}},
  \bibinfo{pages}{18} (\bibinfo{year}{2013}).

\bibitem[76]{Burrows_2005SSRv}
\bibinfo{author}{{Burrows}, D.~N.} \emph{et~al.}
\newblock \bibinfo{title}{{The Swift X-Ray Telescope}}.
\newblock \emph{\bibinfo{journal}{\ssr}} \textbf{\bibinfo{volume}{120}},
  \bibinfo{pages}{165--195} (\bibinfo{year}{2005}).

\bibitem[77]{GCN31205}
\bibinfo{author}{{Beardmore}, A.~P.}, \bibinfo{author}{{Evans}, P.~A.},
  \bibinfo{author}{{Goad}, M.~R.}, \bibinfo{author}{{Osborne}, J.~P.} \&
  \bibinfo{author}{{Swift-XRT Team.}}
\newblock \bibinfo{title}{{GRB 211211A: Enhanced Swift-XRT position}}.
\newblock \emph{\bibinfo{journal}{GRB Coordinates Network}}
  \textbf{\bibinfo{volume}{31205}}, \bibinfo{pages}{1} (\bibinfo{year}{2021}).

\bibitem[78]{Liang_2007ApJ}
\bibinfo{author}{{Liang}, E.-W.}, \bibinfo{author}{{Zhang}, B.-B.} \&
  \bibinfo{author}{{Zhang}, B.}
\newblock \bibinfo{title}{{A Comprehensive Analysis of Swift XRT Data. II.
  Diverse Physical Origins of the Shallow Decay Segment}}.
\newblock \emph{\bibinfo{journal}{\apj}} \textbf{\bibinfo{volume}{670}},
  \bibinfo{pages}{565--583} (\bibinfo{year}{2007}).

\bibitem[79]{Xiao_2018ApJ}
\bibinfo{author}{{Xiao}, D.}, \bibinfo{author}{{Peng}, Z.-k.},
  \bibinfo{author}{{Zhang}, B.-B.} \& \bibinfo{author}{{Dai}, Z.-G.}
\newblock \bibinfo{title}{{Prompt Emission of Gamma-Ray Bursts from the Wind of
  Newborn Millisecond Magnetars: A Case Study of GRB 160804A}}.
\newblock \emph{\bibinfo{journal}{\apj}} \textbf{\bibinfo{volume}{867}},
  \bibinfo{pages}{52} (\bibinfo{year}{2018}).

\bibitem[80]{Schwarz_1978AnSta}
\bibinfo{author}{{Schwarz}, G.}
\newblock \bibinfo{title}{{Estimating the Dimension of a Model}}.
\newblock \emph{\bibinfo{journal}{Annals of Statistics}}
  \textbf{\bibinfo{volume}{6}}, \bibinfo{pages}{461--464}
  (\bibinfo{year}{1978}).

\bibitem[81]{Akaike1974}
\bibinfo{author}{{Akaike}, H.}
\newblock \bibinfo{title}{{A New Look at the Statistical Model
  Identification}}.
\newblock \emph{\bibinfo{journal}{IEEE Transactions on Automatic Control}}
  \textbf{\bibinfo{volume}{19}}, \bibinfo{pages}{716--723}
  (\bibinfo{year}{1974}).

\bibitem[82]{sugiura1978}
\bibinfo{author}{Sugiura, N.}
\newblock \bibinfo{title}{Further analysts of the data by akaike's information
  criterion and the finite corrections: Further analysts of the data by
  akaike's}.
\newblock \emph{\bibinfo{journal}{Communications in Statistics-theory and
  Methods}} \textbf{\bibinfo{volume}{7}}, \bibinfo{pages}{13--26}
  (\bibinfo{year}{1978}).

\bibitem[83]{Zhang_2006ApJ}
\bibinfo{author}{{Zhang}, B.} \emph{et~al.}
\newblock \bibinfo{title}{{Physical Processes Shaping Gamma-Ray Burst X-Ray
  Afterglow Light Curves: Theoretical Implications from the Swift X-Ray
  Telescope Observations}}.
\newblock \emph{\bibinfo{journal}{\apj}} \textbf{\bibinfo{volume}{642}},
  \bibinfo{pages}{354--370} (\bibinfo{year}{2006}).

\bibitem[84]{Nousek_2006ApJ}
\bibinfo{author}{{Nousek}, J.~A.} \emph{et~al.}
\newblock \bibinfo{title}{{Evidence for a Canonical Gamma-Ray Burst Afterglow
  Light Curve in the Swift XRT Data}}.
\newblock \emph{\bibinfo{journal}{\apj}} \textbf{\bibinfo{volume}{642}},
  \bibinfo{pages}{389--400} (\bibinfo{year}{2006}).

\bibitem[85]{Kumar_2000ApJ}
\bibinfo{author}{{Kumar}, P.} \& \bibinfo{author}{{Panaitescu}, A.}
\newblock \bibinfo{title}{{Afterglow Emission from Naked Gamma-Ray Bursts}}.
\newblock \emph{\bibinfo{journal}{\apjl}} \textbf{\bibinfo{volume}{541}},
  \bibinfo{pages}{L51--L54} (\bibinfo{year}{2000}).

\bibitem[86]{Dermer_2004ApJ}
\bibinfo{author}{{Dermer}, C.~D.}
\newblock \bibinfo{title}{{Curvature Effects in Gamma-Ray Burst Colliding
  Shells}}.
\newblock \emph{\bibinfo{journal}{\apj}} \textbf{\bibinfo{volume}{614}},
  \bibinfo{pages}{284--292} (\bibinfo{year}{2004}).

\bibitem[87]{BBZhang_2007ApJ}
\bibinfo{author}{{Zhang}, B.-B.}, \bibinfo{author}{{Liang}, E.-W.} \&
  \bibinfo{author}{{Zhang}, B.}
\newblock \bibinfo{title}{{A Comprehensive Analysis of Swift XRT Data. I.
  Apparent Spectral Evolution of Gamma-Ray Burst X-Ray Tails}}.
\newblock \emph{\bibinfo{journal}{\apj}} \textbf{\bibinfo{volume}{666}},
  \bibinfo{pages}{1002--1011} (\bibinfo{year}{2007}).

\bibitem[88]{BBZhang_2009ApJ}
\bibinfo{author}{{Zhang}, B.-B.}, \bibinfo{author}{{Zhang}, B.},
  \bibinfo{author}{{Liang}, E.-W.} \& \bibinfo{author}{{Wang}, X.-Y.}
\newblock \bibinfo{title}{{Curvature Effect of a Non-Power-Law Spectrum and
  Spectral Evolution of GRB X-Ray Tails}}.
\newblock \emph{\bibinfo{journal}{\apjl}} \textbf{\bibinfo{volume}{690}},
  \bibinfo{pages}{L10--L13} (\bibinfo{year}{2009}).

\bibitem[89]{gompertz2022}
\bibinfo{author}{{Gompertz}, B.~P.} \emph{et~al.}
\newblock \bibinfo{title}{{A minute-long merger-driven gamma-ray burst from
  fast-cooling synchrotron emission}}.
\newblock \emph{\bibinfo{journal}{arXiv e-prints}}
  \bibinfo{pages}{arXiv:2205.05008} (\bibinfo{year}{2022}).

\bibitem[90]{Dai_1998PhRvL}
\bibinfo{author}{{Dai}, Z.~G.} \& \bibinfo{author}{{Lu}, T.}
\newblock \bibinfo{title}{{{\ensuremath{\gamma}}-Ray Bursts and Afterglows from
  Rotating Strange Stars and Neutron Stars}}.
\newblock \emph{\bibinfo{journal}{\prl}} \textbf{\bibinfo{volume}{81}},
  \bibinfo{pages}{4301--4304} (\bibinfo{year}{1998}).

\bibitem[91]{Zhang_2001ApJ}
\bibinfo{author}{{Zhang}, B.} \& \bibinfo{author}{{M{\'e}sz{\'a}ros}, P.}
\newblock \bibinfo{title}{{Gamma-Ray Burst Afterglow with Continuous Energy
  Injection: Signature of a Highly Magnetized Millisecond Pulsar}}.
\newblock \emph{\bibinfo{journal}{\apjl}} \textbf{\bibinfo{volume}{552}},
  \bibinfo{pages}{L35--L38} (\bibinfo{year}{2001}).

\bibitem[92]{Rees_1998ApJ}
\bibinfo{author}{{Rees}, M.~J.} \& \bibinfo{author}{{M{\'e}sz{\'a}ros}, P.}
\newblock \bibinfo{title}{{Refreshed Shocks and Afterglow Longevity in
  Gamma-Ray Bursts}}.
\newblock \emph{\bibinfo{journal}{\apjl}} \textbf{\bibinfo{volume}{496}},
  \bibinfo{pages}{L1--L4} (\bibinfo{year}{1998}).

\bibitem[93]{Sari_2000ApJ}
\bibinfo{author}{{Sari}, R.} \& \bibinfo{author}{{M{\'e}sz{\'a}ros}, P.}
\newblock \bibinfo{title}{{Impulsive and Varying Injection in Gamma-Ray Burst
  Afterglows}}.
\newblock \emph{\bibinfo{journal}{\apjl}} \textbf{\bibinfo{volume}{535}},
  \bibinfo{pages}{L33--L37} (\bibinfo{year}{2000}).

\bibitem[94]{Troja_2007ApJ}
\bibinfo{author}{{Troja}, E.} \emph{et~al.}
\newblock \bibinfo{title}{{Swift Observations of GRB 070110: An Extraordinary
  X-Ray Afterglow Powered by the Central Engine}}.
\newblock \emph{\bibinfo{journal}{\apj}} \textbf{\bibinfo{volume}{665}},
  \bibinfo{pages}{599--607} (\bibinfo{year}{2007}).

\bibitem[95]{Lyons_2010MNRAS}
\bibinfo{author}{{Lyons}, N.} \emph{et~al.}
\newblock \bibinfo{title}{{Can X-ray emission powered by a spinning-down
  magnetar explain some gamma-ray burst light-curve features?}}
\newblock \emph{\bibinfo{journal}{\mnras}} \textbf{\bibinfo{volume}{402}},
  \bibinfo{pages}{705--712} (\bibinfo{year}{2010}).

\bibitem[96]{Racusin_2009ApJ}
\bibinfo{author}{{Racusin}, J.~L.} \emph{et~al.}
\newblock \bibinfo{title}{{Jet Breaks and Energetics of Swift Gamma-Ray Burst
  X-Ray Afterglows}}.
\newblock \emph{\bibinfo{journal}{\apj}} \textbf{\bibinfo{volume}{698}},
  \bibinfo{pages}{43--74} (\bibinfo{year}{2009}).

\bibitem[97]{Roming_2005SSRv}
\bibinfo{author}{{Roming}, P. W.~A.} \emph{et~al.}
\newblock \bibinfo{title}{{The Swift Ultra-Violet/Optical Telescope}}.
\newblock \emph{\bibinfo{journal}{\ssr}} \textbf{\bibinfo{volume}{120}},
  \bibinfo{pages}{95--142} (\bibinfo{year}{2005}).

\bibitem[98]{GCN31222}
\bibinfo{author}{{Belles}, A.}, \bibinfo{author}{{D'Ai}, A.} \&
  \bibinfo{author}{{Swift/UVOT Team}}.
\newblock \bibinfo{title}{{GRB 211211A: Swift/UVOT detection}}.
\newblock \emph{\bibinfo{journal}{GRB Coordinates Network}}
  \textbf{\bibinfo{volume}{31222}}, \bibinfo{pages}{1} (\bibinfo{year}{2021}).

\bibitem[99]{GCN31217}
\bibinfo{author}{{Ito}, N.} \emph{et~al.}
\newblock \bibinfo{title}{{GRB 211211A: MITSuME Akeno optical observation}}.
\newblock \emph{\bibinfo{journal}{GRB Coordinates Network}}
  \textbf{\bibinfo{volume}{31217}}, \bibinfo{pages}{1} (\bibinfo{year}{2021}).

\bibitem[100]{GCN31227}
\bibinfo{author}{{Kumar}, H.} \emph{et~al.}
\newblock \bibinfo{title}{{GRB 211211A: HCT and GIT optical follow up
  observations}}.
\newblock \emph{\bibinfo{journal}{GRB Coordinates Network}}
  \textbf{\bibinfo{volume}{31227}}, \bibinfo{pages}{1} (\bibinfo{year}{2021}).

\bibitem[101]{GCN31214}
\bibinfo{author}{{Strausbaugh}, R.} \& \bibinfo{author}{{Cucchiara}, A.}
\newblock \bibinfo{title}{{GRB 211211A: LCO Optical Observations}}.
\newblock \emph{\bibinfo{journal}{GRB Coordinates Network}}
  \textbf{\bibinfo{volume}{31214}}, \bibinfo{pages}{1} (\bibinfo{year}{2021}).

\bibitem[102]{GCN31232}
\bibinfo{author}{{Mao}, J.}, \bibinfo{author}{{Xin}, Y.~X.} \&
  \bibinfo{author}{{Bai}, J.~M.}
\newblock \bibinfo{title}{{GRB 211211A: GMG upper limit}}.
\newblock \emph{\bibinfo{journal}{GRB Coordinates Network}}
  \textbf{\bibinfo{volume}{31232}}, \bibinfo{pages}{1} (\bibinfo{year}{2021}).

\bibitem[103]{GCN31299}
\bibinfo{author}{{Gupta}, R.} \emph{et~al.}
\newblock \bibinfo{title}{{GRB 211211A: observations with the 3.6m Devasthal
  Optical Telescope}}.
\newblock \emph{\bibinfo{journal}{GRB Coordinates Network}}
  \textbf{\bibinfo{volume}{31299}}, \bibinfo{pages}{1} (\bibinfo{year}{2021}).

\bibitem[104]{GCN31233}
\bibinfo{author}{{Pankov}, N.} \emph{et~al.}
\newblock \bibinfo{title}{{GRB 211211A: AbAO optical observations}}.
\newblock \emph{\bibinfo{journal}{GRB Coordinates Network}}
  \textbf{\bibinfo{volume}{31233}}, \bibinfo{pages}{1} (\bibinfo{year}{2021}).

\bibitem[105]{GCN31234}
\bibinfo{author}{{Moskvitin}, A.} \emph{et~al.}
\newblock \bibinfo{title}{{GRB 211211A: SAO RAS optical observations}}.
\newblock \emph{\bibinfo{journal}{GRB Coordinates Network}}
  \textbf{\bibinfo{volume}{31234}}, \bibinfo{pages}{1} (\bibinfo{year}{2021}).

\bibitem[106]{GCN31242}
\bibinfo{author}{{D'Avanzo}, P.} \emph{et~al.}
\newblock \bibinfo{title}{{GRB 211211A: TNG NIR observations}}.
\newblock \emph{\bibinfo{journal}{GRB Coordinates Network}}
  \textbf{\bibinfo{volume}{31242}}, \bibinfo{pages}{1} (\bibinfo{year}{2021}).

\bibitem[107]{Gal-Yam_2006Natur}
\bibinfo{author}{{Gal-Yam}, A.} \emph{et~al.}
\newblock \bibinfo{title}{{A novel explosive process is required for the
  {\ensuremath{\gamma}}-ray burst GRB 060614}}.
\newblock \emph{\bibinfo{journal}{\nat}} \textbf{\bibinfo{volume}{444}},
  \bibinfo{pages}{1053--1055} (\bibinfo{year}{2006}).

\bibitem[108]{Fynbo_2006Natur}
\bibinfo{author}{{Fynbo}, J. P.~U.} \emph{et~al.}
\newblock \bibinfo{title}{{No supernovae associated with two long-duration
  {\ensuremath{\gamma}}-ray bursts}}.
\newblock \emph{\bibinfo{journal}{\nat}} \textbf{\bibinfo{volume}{444}},
  \bibinfo{pages}{1047--1049} (\bibinfo{year}{2006}).

\bibitem[109]{Galama_1998Natur}
\bibinfo{author}{{Galama}, T.~J.} \emph{et~al.}
\newblock \bibinfo{title}{{An unusual supernova in the error box of the
  {\ensuremath{\gamma}}-ray burst of 25 April 1998}}.
\newblock \emph{\bibinfo{journal}{\nat}} \textbf{\bibinfo{volume}{395}},
  \bibinfo{pages}{670--672} (\bibinfo{year}{1998}).

\bibitem[110]{Reeves_2002Natur}
\bibinfo{author}{{Reeves}, J.~N.} \emph{et~al.}
\newblock \bibinfo{title}{{The signature of supernova ejecta in the X-ray
  afterglow of the {\ensuremath{\gamma}}-ray burst 011211}}.
\newblock \emph{\bibinfo{journal}{\nat}} \textbf{\bibinfo{volume}{416}},
  \bibinfo{pages}{512--515} (\bibinfo{year}{2002}).

\bibitem[111]{Hjorth_2003Natur}
\bibinfo{author}{{Hjorth}, J.} \emph{et~al.}
\newblock \bibinfo{title}{{A very energetic supernova associated with the
  {\ensuremath{\gamma}}-ray burst of 29 March 2003}}.
\newblock \emph{\bibinfo{journal}{\nat}} \textbf{\bibinfo{volume}{423}},
  \bibinfo{pages}{847--850} (\bibinfo{year}{2003}).

\bibitem[112]{Clocchiatti_2011AJ}
\bibinfo{author}{{Clocchiatti}, A.}, \bibinfo{author}{{Suntzeff}, N.~B.},
  \bibinfo{author}{{Covarrubias}, R.} \& \bibinfo{author}{{Candia}, P.}
\newblock \bibinfo{title}{{The Ultimate Light Curve of SN 1998bw/GRB 980425}}.
\newblock \emph{\bibinfo{journal}{\aj}} \textbf{\bibinfo{volume}{141}},
  \bibinfo{pages}{163} (\bibinfo{year}{2011}).

\bibitem[113]{Cano_2013MNRAS}
\bibinfo{author}{{Cano}, Z.}
\newblock \bibinfo{title}{{A new method for estimating the bolometric
  properties of Ibc supernovae}}.
\newblock \emph{\bibinfo{journal}{\mnras}} \textbf{\bibinfo{volume}{434}},
  \bibinfo{pages}{1098--1116} (\bibinfo{year}{2013}).

\bibitem[114]{Ryan_2020ApJ}
\bibinfo{author}{{Ryan}, G.}, \bibinfo{author}{{van Eerten}, H.},
  \bibinfo{author}{{Piro}, L.} \& \bibinfo{author}{{Troja}, E.}
\newblock \bibinfo{title}{{Gamma-Ray Burst Afterglows in the Multimessenger
  Era: Numerical Models and Closure Relations}}.
\newblock \emph{\bibinfo{journal}{\apj}} \textbf{\bibinfo{volume}{896}},
  \bibinfo{pages}{166} (\bibinfo{year}{2020}).

\bibitem[115]{Fitzpatrick_1999PASP}
\bibinfo{author}{{Fitzpatrick}, E.~L.}
\newblock \bibinfo{title}{{Correcting for the Effects of Interstellar
  Extinction}}.
\newblock \emph{\bibinfo{journal}{\pasp}} \textbf{\bibinfo{volume}{111}},
  \bibinfo{pages}{63--75} (\bibinfo{year}{1999}).

\bibitem[116]{Astropy_2013AA}
\bibinfo{author}{{Astropy Collaboration}} \emph{et~al.}
\newblock \bibinfo{title}{{Astropy: A community Python package for astronomy}}.
\newblock \emph{\bibinfo{journal}{\aap}} \textbf{\bibinfo{volume}{558}},
  \bibinfo{pages}{A33} (\bibinfo{year}{2013}).

\bibitem[117]{Kasen_2013ApJ}
\bibinfo{author}{{Kasen}, D.}, \bibinfo{author}{{Badnell}, N.~R.} \&
  \bibinfo{author}{{Barnes}, J.}
\newblock \bibinfo{title}{{Opacities and Spectra of the r-process Ejecta from
  Neutron Star Mergers}}.
\newblock \emph{\bibinfo{journal}{\apj}} \textbf{\bibinfo{volume}{774}},
  \bibinfo{pages}{25} (\bibinfo{year}{2013}).

\bibitem[118]{Wang_2015ApJS}
\bibinfo{author}{{Wang}, X.-G.} \emph{et~al.}
\newblock \bibinfo{title}{{How Bad or Good Are the External Forward Shock
  Afterglow Models of Gamma-Ray Bursts?}}
\newblock \emph{\bibinfo{journal}{\apjs}} \textbf{\bibinfo{volume}{219}},
  \bibinfo{pages}{9} (\bibinfo{year}{2015}).

\bibitem[119]{rosswog07}
\bibinfo{author}{{Rosswog}, S.}
\newblock \bibinfo{title}{{Fallback accretion in the aftermath of a compact
  binary merger}}.
\newblock \emph{\bibinfo{journal}{\mnras}} \textbf{\bibinfo{volume}{376}},
  \bibinfo{pages}{L48--L51} (\bibinfo{year}{2007}).

\bibitem[120]{lu2022}
\bibinfo{author}{{Lu}, W.} \& \bibinfo{author}{{Quataert}, E.}
\newblock \bibinfo{title}{{Late-time accretion in neutron star mergers:
  implications for short gamma-ray bursts and kilonovae}}.
\newblock \emph{\bibinfo{journal}{arXiv e-prints}}
  \bibinfo{pages}{arXiv:2208.04293} (\bibinfo{year}{2022}).

\bibitem[121]{vanputten14}
\bibinfo{author}{{van Putten}, M.~H.~P.~M.}, \bibinfo{author}{{Lee}, G.~M.},
  \bibinfo{author}{{Della Valle}, M.}, \bibinfo{author}{{Amati}, L.} \&
  \bibinfo{author}{{Levinson}, A.}
\newblock \bibinfo{title}{{On the origin of short GRBs with extended emission
  and long GRBs without associated SN.}}
\newblock \emph{\bibinfo{journal}{\mnras}} \textbf{\bibinfo{volume}{444}},
  \bibinfo{pages}{L58--L62} (\bibinfo{year}{2014}).

\bibitem[122]{vanputten12}
\bibinfo{author}{{van Putten}, M.~H.~P.~M.}
\newblock \bibinfo{title}{{Discovery of Black Hole Spindown in the BATSE
  Catalogue of Long GRBs}}.
\newblock \emph{\bibinfo{journal}{Progress of Theoretical Physics}}
  \textbf{\bibinfo{volume}{127}}, \bibinfo{pages}{331--354}
  (\bibinfo{year}{2012}).

\bibitem[123]{metzger08}
\bibinfo{author}{{Metzger}, B.~D.}, \bibinfo{author}{{Quataert}, E.} \&
  \bibinfo{author}{{Thompson}, T.~A.}
\newblock \bibinfo{title}{{Short-duration gamma-ray bursts with extended
  emission from protomagnetar spin-down}}.
\newblock \emph{\bibinfo{journal}{\mnras}} \textbf{\bibinfo{volume}{385}},
  \bibinfo{pages}{1455--1460} (\bibinfo{year}{2008}).

\bibitem[124]{shapiro83}
\bibinfo{author}{{Shapiro}, S.~L.}, \bibinfo{author}{{Teukolsky}, S.~A.} \&
  \bibinfo{author}{{Lightman}, A.~P.}
\newblock \bibinfo{title}{{Black Holes, White Dwarfs, and Neutron Stars: The
  Physics of Compact Objects}}.
\newblock \emph{\bibinfo{journal}{Physics Today}}
  \textbf{\bibinfo{volume}{36}}, \bibinfo{pages}{89} (\bibinfo{year}{1983}).

\bibitem[125]{Xiao_2019ApJ}
\bibinfo{author}{{Xiao}, D.} \& \bibinfo{author}{{Dai}, Z.-G.}
\newblock \bibinfo{title}{{Determining the Efficiency of Converting Magnetar
  Spindown Energy into Gamma-Ray Burst X-Ray Afterglow Emission and Its
  Possible Implications}}.
\newblock \emph{\bibinfo{journal}{\apj}} \textbf{\bibinfo{volume}{878}},
  \bibinfo{pages}{62} (\bibinfo{year}{2019}).

\bibitem[126]{Xiao_2019ApJL}
\bibinfo{author}{{Xiao}, D.}, \bibinfo{author}{{Zhang}, B.-B.} \&
  \bibinfo{author}{{Dai}, Z.-G.}
\newblock \bibinfo{title}{{On the Properties of a Newborn Magnetar Powering the
  X-Ray Transient CDF-S XT2}}.
\newblock \emph{\bibinfo{journal}{\apjl}} \textbf{\bibinfo{volume}{879}},
  \bibinfo{pages}{L7} (\bibinfo{year}{2019}).

\bibitem[127]{fryer99}
\bibinfo{author}{{Fryer}, C.}, \bibinfo{author}{{Benz}, W.},
  \bibinfo{author}{{Herant}, M.} \& \bibinfo{author}{{Colgate}, S.~A.}
\newblock \bibinfo{title}{{What Can the Accretion-induced Collapse of White
  Dwarfs Really Explain?}}
\newblock \emph{\bibinfo{journal}{\apj}} \textbf{\bibinfo{volume}{516}},
  \bibinfo{pages}{892--899} (\bibinfo{year}{1999}).

\bibitem[128]{Paczynski1986}
\bibinfo{author}{{Paczynski}, B.}
\newblock \bibinfo{title}{{Gamma-ray bursters at cosmological distances}}.
\newblock \emph{\bibinfo{journal}{\apjl}} \textbf{\bibinfo{volume}{308}},
  \bibinfo{pages}{L43--L46} (\bibinfo{year}{1986}).

\bibitem[129]{Eichler1989}
\bibinfo{author}{{Eichler}, D.}, \bibinfo{author}{{Livio}, M.},
  \bibinfo{author}{{Piran}, T.} \& \bibinfo{author}{{Schramm}, D.~N.}
\newblock \bibinfo{title}{{Nucleosynthesis, neutrino bursts and
  {\ensuremath{\gamma}}-rays from coalescing neutron stars}}.
\newblock \emph{\bibinfo{journal}{\nat}} \textbf{\bibinfo{volume}{340}},
  \bibinfo{pages}{126--128} (\bibinfo{year}{1989}).

\bibitem[130]{Paczynski1991}
\bibinfo{author}{{Paczynski}, B.}
\newblock \bibinfo{title}{{Cosmological gamma-ray bursts.}}
\newblock \emph{\bibinfo{journal}{\actaa}} \textbf{\bibinfo{volume}{41}},
  \bibinfo{pages}{257--267} (\bibinfo{year}{1991}).

\bibitem[131]{Rueda2018}
\bibinfo{author}{{Rueda}, J.~A.} \emph{et~al.}
\newblock \bibinfo{title}{{GRB 170817A-GW170817-AT 2017gfo and the observations
  of NS-NS, NS-WD and WD-WD mergers}}.
\newblock \emph{\bibinfo{journal}{\jcap}} \textbf{\bibinfo{volume}{2018}},
  \bibinfo{pages}{006} (\bibinfo{year}{2018}).

\bibitem[132]{Siegel2019}
\bibinfo{author}{{Siegel}, D.~M.}, \bibinfo{author}{{Barnes}, J.} \&
  \bibinfo{author}{{Metzger}, B.~D.}
\newblock \bibinfo{title}{{Collapsars as a major source of r-process
  elements}}.
\newblock \emph{\bibinfo{journal}{\nat}} \textbf{\bibinfo{volume}{569}},
  \bibinfo{pages}{241--244} (\bibinfo{year}{2019}).

\bibitem[133]{waxman22}
\bibinfo{author}{{Waxman}, E.}, \bibinfo{author}{{Ofek}, E.~O.} \&
  \bibinfo{author}{{Kushnir}, D.}
\newblock \bibinfo{title}{{Strong NIR emission following the long duration GRB
  211211A: Dust heating as an alternative to a kilonova}}.
\newblock \emph{\bibinfo{journal}{arXiv e-prints}}
  \bibinfo{pages}{arXiv:2206.10710} (\bibinfo{year}{2022}).

\end{thebibliography}
\end{document}